\documentclass[acmlarge]{acmart}
\usepackage{pifont}
\usepackage{algorithm}
\usepackage{algorithmicx}
\usepackage{algpseudocode}
\usepackage{color} 
\usepackage[utf8]{inputenc} 
\usepackage{multicol}
\usepackage{multirow}
\usepackage{xcolor}
\usepackage{tikz}
\usepackage{longtable}
\usepackage{graphicx}  
\usepackage{subcaption}  
\usepackage{supertabular}
\usepackage{etoolbox}
\usepackage{bbding}
\usepackage{tcolorbox}
\usepackage{wasysym}
\usepackage{threeparttable}
\newcommand{\sqbox}[1]{\textcolor{#1}{\SquareSolid}}
\newcommand{\sqboxEmpty}[1]{%
\textcolor{#1}{\Square}%
}%

\makeatletter

\newcommand{\Rmnum}[1]{\expandafter\@slowromancap\romannumeral #1@}
\makeatother

\newcommand{\customtnote}[1]{%
    \textcolor{blue}{\raisebox{0.1ex}{\scalebox{1.1}{#1}}}%
}
\newcommand{\customfootnote}[1]{%
    \textsuperscript{\customtnote{#1}}%
}

\newcommand{\blackdiamond}{%
  \begin{tikzpicture}[scale=0.12]
    \filldraw[fill=black, draw=black] 
    (0,1) -- (1,2) -- (2,1) -- (1,0) -- cycle;
  \end{tikzpicture}%
}

\AtBeginDocument{%
  }

\setcopyright{acmlicensed}
\copyrightyear{2025}
\acmYear{2025}
\acmDOI{XXXXXXX.XXXXXXX}

\acmJournal{POMACS}
\acmVolume{37}
\acmNumber{4}
\acmArticle{111}
\acmMonth{8}

\begin{document}

\title{Coverage-Guided Testing for Deep Learning Models: A Comprehensive Survey}

\author{Hongjing Guo}
\email{guohongjing@nuaa.edu.cn}
\orcid{0000-0001-5286-874X}

\author{Chuanqi Tao}
\email{taochuanqi@nuaa.edu.cn}
\authornote{The authors are with the Key Laboratory for Safety-Critical Software Development and Verification, Ministry of Industry and Information Technology (Nanjing University of Aeronautics and Astronautics), Nanjing, China, 211106. They are with the Collaborative Innovation Center of Novel Software Technology and Industrialization, Nanjing, China, 210023. Chuanqi Tao is also affiliated with the State Key Laboratory for Novel Software Technology, Nanjing University, Nanjing, China, 210023. Zhiqiu Huang and Chuanqi Tao are corresponding authors.}
\orcid{0000-0002-0698-7307}

\author{Zhiqiu Huang}
\authornotemark[1]
\email{zqhuang@nuaa.edu.cn}
\orcid{0000-0001-6843-1892}

\author{Weiqin Zou}
\authornotemark[1]
\email{weiqin@nuaa.edu.cn}
\orcid{0000-0002-0913-1539}

\affiliation{
  \institution{Nanjing University of Aeronautics and Astronautics}
  \streetaddress{No. 29 Jiangjun Avenue, Jiangning District}
  \city{Nanjing}
  \country{China}
  \postcode{211106}
}

\renewcommand{\shortauthors}{Guo et al.}

\begin{abstract}
As Deep Learning (DL) models are increasingly applied in safety-critical domains, ensuring their quality has emerged as a pressing challenge in modern software engineering. Among emerging validation paradigms, coverage-guided testing (CGT) has gained prominence as a systematic framework for identifying erroneous or unexpected model behaviors. Despite growing research attention, existing CGT studies remain methodologically fragmented, limiting the understanding of current advances and emerging trends. This work addresses that gap through a comprehensive review of state-of-the-art CGT methods for DL models, including test coverage analysis, coverage-guided test input generation, and coverage-guided test input optimization. This work provides detailed taxonomies to organize these methods based on methodological characteristics and application scenarios. We also investigate evaluation practices adopted in existing studies, including the use of benchmark datasets, model architectures, and evaluation aspects. Finally, open challenges and future directions are highlighted in terms of the correlation between structural coverage and testing objectives, method generalizability across tasks and models, practical deployment concerns, and the need for standardized evaluation and tool support. This work aims to provide a roadmap for future academic research and engineering practice in DL model quality assurance.

\end{abstract}


\ccsdesc[500]{Software and its engineering}
\ccsdesc[300]{Software testing and debugging}
\ccsdesc[100]{Neural networks}

\keywords{Coverage-Guided Testing, Deep Learning Testing, Deep Neural Network Testing, Coverage Criteria}

\received{20 February 2007}
\received[revised]{12 March 2009}
\received[accepted]{5 June 2009}

\maketitle

\section{Introduction}
Deep learning (DL) has achieved remarkable success across a wide range of real-world applications, with growing adoption in safety- and mission-critical domains such as autonomous driving \cite{DBLP:journals/tiv/ChibS24, DBLP:journals/tits/MuhammadULSA21}, medical diagnostics \cite{groh2024deep}, and financial decision-making \cite{Mallikarjuna}. At the core of these DL-enabled systems lie DL models. However, despite their impressive capabilities, DL models have exhibited erroneous or unexpected behaviors under certain conditions \cite{yu2024surveyfailureanalysisfault, DBLP:conf/chi/BalaynR0B23}, raising serious concerns about their reliability and trustworthiness. These challenges highlight an urgent need for quality assurance of DL models, drawing increasing attention from both academia and industry.

Software testing has long been a cornerstone of traditional software engineering, providing systematic methodologies to assess software behavior against expected specifications. Over decades of evolution, the field has matured into a rich body of techniques, such as mutation testing \cite{DBLP:conf/icse/PetrovicIFJ21, DBLP:journals/tse/JiaH11}, fuzz testing \cite{DBLP:journals/csur/MallisseryW24, DBLP:journals/csur/LiLTY15}, and metamorphic testing \cite{DBLP:journals/tse/SeguraFSC16, DBLP:journals/csur/ChenKLPTTZ18}. In a similar vein, DL models also require rigorous testing to detect inference errors, defined as deviations between model outputs and expected results for given inputs under specific task specifications. However, directly applying traditional testing methodologies to DL models remains challenging. Unlike conventional software with explicitly defined control logic, DL models operate under a data-driven paradigm, where decision-making is implicitly encoded in non-linear architectures and high-dimensional parameters \cite{10.1145/3672446, DBLP:journals/tosem/XieLWMGJL22}. This inherent opacity, commonly referred to as the black-box nature, introduces uncertainty in prediction outputs \cite{DBLP:journals/tse/ZhangHML22}. Moreover, DL models operate on semantically complex input domains such as images, natural language, and sensor data, and produce diverse outputs ranging from discrete labels to continuous or structured results. Effective testing of such models often demands large volumes of high-quality, task-specific ground-truth data, the acquisition of which is both costly and labor-intensive \cite{DBLP:journals/tosem/HuGXCMPT24, DBLP:journals/tosem/HuGCXMPT22}. These fundamental differences necessitate the development of dedicated testing methodologies tailored to the unique characteristics of DL models.

The research community has increasingly embraced coverage-guided testing (CGT) as a central paradigm. CGT leverages coverage criteria to guide the testing process, aiming to maximize the exploration of a model’s behavior space. In software testing, coverage has long served as a fundamental indicator of test adequacy, traditionally measured through control flow elements such as branches, statements, and paths. Inspired by these structural coverage principles, researchers have redefined the notion of coverage for DL models by introducing a variety of metrics that capture the extent to which a model’s internal states, input domain, and output behaviors are exercised during testing \cite{DBLP:journals/tosem/XieLWMGJL22, DBLP:conf/icse/GerasimouE0C20, DBLP:conf/icse/JiMYW23, DBLP:conf/icse/YuanPW23, DBLP:conf/wcre/MaJXLLLZ19, DBLP:conf/sosp/PeiCYJ17, DBLP:journals/tosem/DolaDS23}. Coverage criteria serve a dual purpose: they not only quantify test adequacy but also provide actionable feedback to steer the testing process. Guided by coverage signals, two primary research directions have emerged. The first focuses on test input generation, aiming to synthesize inputs that increase model coverage and reveal erroneous behaviors \cite{DBLP:journals/tosem/ZhiXSSZG24, DBLP:conf/issta/LeeCLO20, DBLP:conf/issta/XieMJXCLZLYS19}. The underlying assumption is that broader exploration of a model’s behavior space, reflected by increased test coverage, enhances the likelihood of uncovering errors. The second direction focuses on test input optimization, aiming to alleviate the high cost of ground-truth labeling in DL testing. This includes test prioritization, selection, and minimization strategies that aim to retain testing effectiveness while reducing the number of labeled samples required \cite{DBLP:journals/tosem/HuGCXMPT22, DBLP:journals/tosem/HuGXCMPT24, DBLP:journals/tosem/0003WWYZY20}.

Although various CGT approaches have been proposed, the sheer volume and diversity of existing work pose challenges for researchers and practitioners to grasp recent advances. While several surveys have been conducted on DL testing, few offer an in-depth and focused analysis of CGT techniques. Table \ref{Tab-survey} presents a comparative summary of representative surveys. Studies by Zhang et al. \cite{DBLP:journals/tse/ZhangHML22}, Riccio et al. \cite{DBLP:journals/ese/RiccioJSHWT20}, and Huang et al. \cite{DBLP:journals/csr/HuangKRSSTWY20}, were published in 2020, and thus do not reflect the rapid methodological advancements and emerging trends of recent years. These works focus on broader topics such as general ML or DL testing, with CGT treated as a subtopic. They lack detailed discussions on the specific methodologies and evaluation practices unique to CGT. Hu et al. (2024) \cite{DBLP:journals/tosem/HuGXCMPT24} provide a review on DL test optimization, but do not specifically address approaches that leverage coverage feedback. Zhang et al. \cite{DBLP:journals/csur/ZhangJSLWLG25} focus on DL Library testing rather than model-level techniques. Across all these surveys, the evaluation dimension, particularly how CGT methods are experimentally validated, is insufficiently discussed. These gaps point to the need for a dedicated and up-to-date review of CGT methodologies, evaluation practices, and research trends to support ongoing and future work in this field.

\begin{table}
\begin{threeparttable}
  \caption{Comparison of Our Work with Related Surveys}
  \small
  \label{Tab-survey}
  \fontsize{9.3}{9.8}\selectfont
  \setlength{\tabcolsep}{0.5 mm}{
  \begin{tabular}{cccccccc}
    \toprule
\multirow{3}{*}{\textbf{Paper}} & \multirow{3}{*}{\textbf{Year}} &  \multirow{3}{*}{\textbf{Scope}} &  \multicolumn{3}{c}{\textbf{Covered Aspect} }  & \multicolumn{2}{c}{\textbf{Evaluation Design} } \\ \cmidrule(lr) {4-6} \cmidrule(lr) {7-8} 
&& & Coverage & Test Input  & Test & Dataset & Aspect \\ 
&&& Criteria & Generation   & Optimization & $\&$ Model & $\&$ Metric \\ \midrule
Zhang et al. \cite{DBLP:journals/tse/ZhangHML22} & 2020 & ML Testing &  \ding{51} & \ding{52}\rotatebox[origin=c]{-9.2}{\kern-0.7em\ding{55}}  &  \ding{52}\rotatebox[origin=c]{-9.2}{\kern-0.7em\ding{55}} & \ding{51} & \\

Huang et al. \cite{DBLP:journals/csr/HuangKRSSTWY20} & 2020 & DL Safety and Trustworthiness  & \ding{51} & \ding{52}\rotatebox[origin=c]{-9.2}{\kern-0.7em\ding{55}}  & \\

Riccio et al. \cite{DBLP:journals/ese/RiccioJSHWT20} & 2020 & ML Testing & \ding{51} & \ding{52}\rotatebox[origin=c]{-9.2}{\kern-0.7em\ding{55}} & & \ding{51} & \ding{51} \\

Hu et al. \cite{DBLP:conf/icmlca/HuWLL23} & 2023 & DL Model Testing &  \ding{51} & \ding{52}\rotatebox[origin=c]{-9.2}{\kern-0.7em\ding{55}}  & \\ 

Hu et al. \cite{DBLP:journals/tosem/HuGXCMPT24} &  2024 & DL Test Optimization & &  & \ding{52}\rotatebox[origin=c]{-9.2}{\kern-0.7em\ding{55}} & \ding{51} & \ding{51} \\ 

Zhang et al. \cite{DBLP:journals/csur/ZhangJSLWLG25} & 2025 & DL Library Testing & \\  \midrule

Our Survey & 2025 & DL Model CGT &   \ding{51} & \ding{51} & \ding{51} & \ding{51} & \ding{51} \\
\bottomrule
\end{tabular}}
\begin{tablenotes}
\small
\item  The symbol \ding{51} indicates support; \ding{52}\rotatebox[origin=c]{-9.2}{\kern-0.7em\ding{55}} implies partial support, without an explicit focus on coverage-driven methods.
\end{tablenotes}
\end{threeparttable}
\end{table}

Therefore, this work presents a comprehensive retrospective on the development of CGT for DL models. We survey and analyze 89 relevant papers, capturing key advancements in coverage analysis, test input generation, and test optimization. This work presents a comprehensive survey of CGT approaches, systematically categorizing each research aspect into a taxonomy based on methodological characteristics and application scenarios. Furthermore, we provide an overview of experimental designs commonly employed to evaluate CGT methods. We also pinpoint open challenges and outline future directions to inspire continued innovation in this important field of DL quality assurance. This work aims to offer a systematic introduction for readers who are new to DL testing practices, while also delivering an in-depth analysis for experienced researchers by highlighting the strengths and limitations of current approaches. 

The main contribution can be summarized as follows:
\begin{itemize}
    \item We present a comprehensive survey dedicated to CGT approaches for DL models, covering three aspects: coverage analysis, coverage-guided test input generation, and coverage-guided test optimization. It complements existing DL testing surveys by providing a focused perspective on coverage-driven methods. 
    \item We develop taxonomies that organize existing CGT approaches based on their underlying methodologies and application scenarios, and offer detailed discussions of their design principles, strengths, and limitations.
    \item We analyze experimental designs reported in the literature, including choices of benchmark datasets, model architectures, dataset treatments, and evaluation aspects. This analysis highlights the diversity in evaluation practices, emphasizing the pressing need for more standardized benchmarking in CGT research.
    \item We provide open challenges and future research directions of CGT, aiming to foster research efforts toward quality assurance and improvement of DL models.
\end{itemize}

The remainder of this paper is organized as follows. Section \ref{Sec-framework} introduces the CGT framework. Section \ref{Sec-SurveyMethodology} details the methodology used for our systematic literature review. The subsequent sections summarize existing approaches, including coverage analysis (Section \ref{Sec-coverage}), coverage-guided test input generation (Section \ref{Sec-generation}), and coverage-guided test optimization (Section \ref{Sec-opt}). Section \ref{Sec-evaluation} presents empirical evaluation practices in CGT research, including experimental designs and existing empirical studies. Section \ref{Sec-challenges} discusses open challenges and outlines promising directions for future work. Section \ref{Sec-conclusion} concludes this article.

\section{PRELIMINARIES}
\label{Sec-framework}
\subsection{DL Model Testing}
Testing DL models is a critical process in the development of DL-enabled software and systems, serving to ensure their quality and reliability. It aims to detect erroneous inferences made by DL models. An inference error is typically defined as a deviation between the model’s output and the expected outcome for a given input, under specific task specifications and operational conditions. 

Unlike traditional software testing, which focuses on verifying deterministic logic against explicitly defined rules, DL model testing deals with the probabilistic and data-driven nature of learned representations. As such, it involves a range of testing objectives, such as assessing robustness to natural input variations or adversarial perturbations, and evaluating fairness across demographic groups or other sensitive attributes. 

\subsection{CGT Framework}
CGT refers to a testing paradigm that leverages coverage criteria to guide the testing process. Based on our survey and analysis, coverage information is primarily employed in three core activities—test input generation, test optimization, and test adequacy analysis, to support the generation, optimization, and evaluation of test inputs.  

\begin{figure}  
    \centering  
    \includegraphics[width = 14.7 cm]{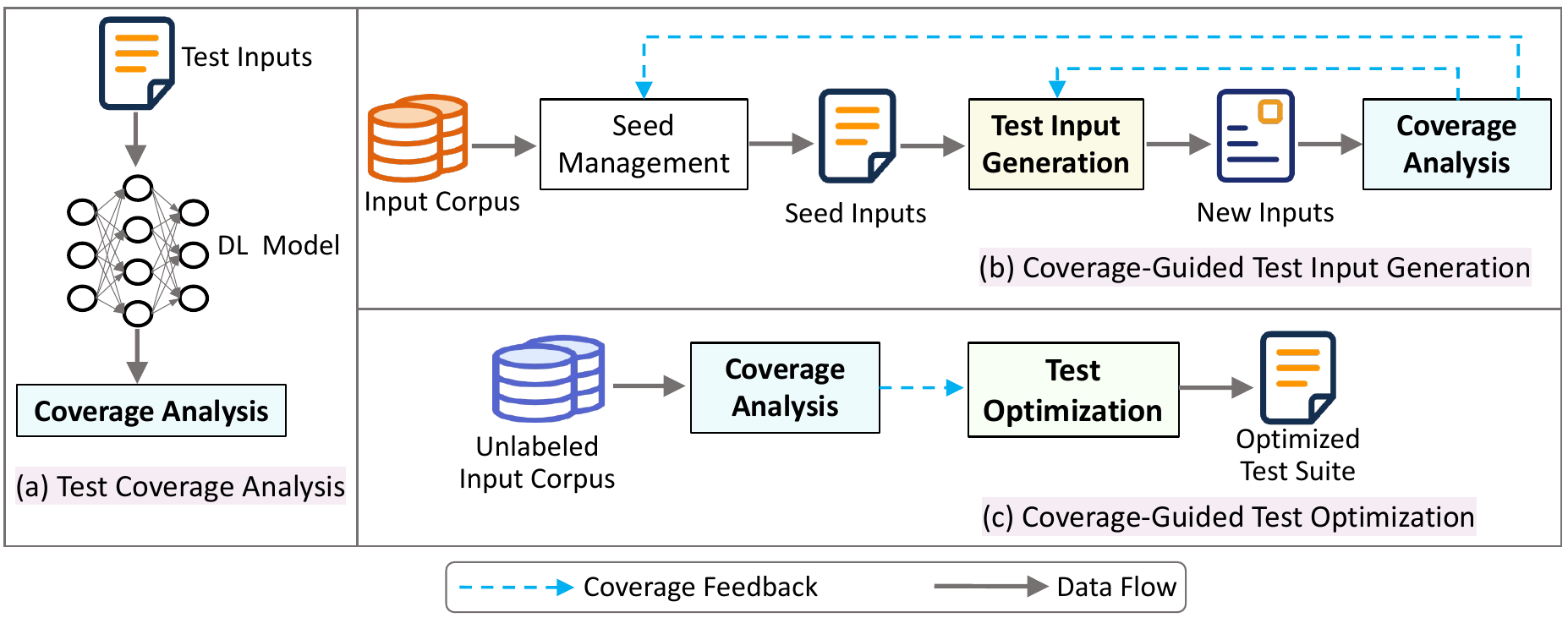} 
    \caption{Workflow of Coverage-Guided Testing for DL Models}  
    \label{WorkflowCGT}
\end{figure}

Figure \ref{WorkflowCGT} illustrates the typical CGT workflow for DL models. At the core of CGT lies coverage analysis, which adopts coverage criteria to evaluate how thoroughly a DL model's behavior space has been exercised by test inputs. The test input generation process begins with a set of initial seed inputs sampled from an input corpus. These seeds are iteratively mutated to synthesize new test inputs, with the objective of maximizing coverage. The underlying assumption is that greater coverage correlates with an increased likelihood of uncovering erroneous behaviors. Throughout this process, test coverage serves as feedback to guide the testing loop by (1)  identifying promising seeds for mutation, and (2) informing mutation strategies or directions. This iterative procedure continues until a stopping condition is met, such as achieving a predefined coverage threshold or exhausting computational resources. To improve testing efficiency, CGT integrates a test suite optimization phase. Given a potentially large and unlabeled input corpus, optimization techniques are employed to sample or prioritize test inputs based on their contribution to coverage improvement. Notably, recent DL testing efforts commonly reuse the optimized test suite as the seed pool for subsequent test input generation. Overall, through its use of coverage-driven feedback, CGT facilitates systematic exploration and efficient test design for DL models.

\section{Survey Methodology}
\label{Sec-SurveyMethodology}
\begin{figure}  
    \centering  
    \includegraphics[scale = 0.75, width = 15 cm]{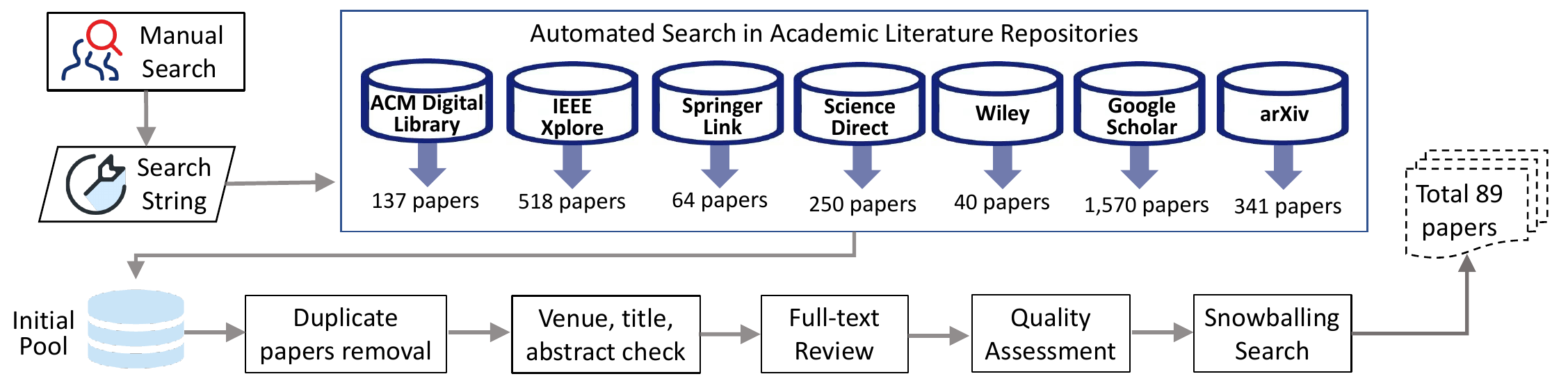} 
    \caption{Study Collection and Selection Process}  
    \label{StudySearch}
\end{figure}

This work adopts the `Quasi-Gold Standard' (QGS) method \cite{ZHANG2011625} to systematically collect relevant studies. The process begins with a manual search to gather initial studies from high-quality publication venues. These papers are carefully reviewed to construct search strings, as described in Section \ref{SearchItems}. The refined search strings are then applied in an automated search across digital repositories. Retrieved results are screened using inclusion and exclusion criteria to retain the most relevant studies. To ensure a comprehensive survey, we adopt a snowballing strategy, detailed in Section \ref{CollectionSelection}. Given the vast body of literature in fields such as Software Engineering (SE) and Artificial Intelligence (AI), the QGS method ensures the identification of highly relevant studies with improved efficiency and precision compared to fully manual search processes. Following this multi-stage procedure, a total of 89 studies were selected for analysis. The workflow of the collection and selection process is illustrated in Figure~\ref{StudySearch}, while publication trends across conferences and journals are analyzed in Section~\ref{TrendObservation}.

\begin{table}
\begin{threeparttable}
  \caption{Search Keywords}
  \small
  \label{Tab-keywords}
  \fontsize{8.3}{8}\selectfont
  \setlength{\tabcolsep}{0.1 mm}{
  \begin{tabular}{c|l}
    \toprule
\textbf{Group} & \multicolumn{1}{c}{\textbf{Keywords}}   \\
    \midrule
\multirow{3}{*}{\textbf{A}} & "\textit{Deep Learning System*}" OR "\textit{DL System*}" OR "\textit{Deep Learning Model*}" OR "\textit{DL Model*}" OR "\textit{Deep Neural Network*}"  \\
&  OR "\textit{DNN*}" OR "\textit{Neural Network*}" OR "\textit{Deep Neural Network Model*}" OR "\textit{DNN Model*}" OR "\textit{Deep Learning}" OR "\textit{DL}"   \\
&  OR "\textit{Deep Reinforcement Learning}" OR "\textit{DRL}"  \\ \midrule  
\multirow{4}{*}{\textbf{B}} & "\textit{Test Coverage}" OR "\textit{Coverage Analysis}" OR "\textit{Coverage Criteria}"  OR "\textit{Coverage Criterion}" OR "\textit{Testing Criteria}"  \\ 
& OR "\textit{Coverage Metric*}" OR "Neuron Coverage" OR "\textit{Coverage-Guided Testing}" OR "\textit{Coverage-Guided Fuzz*}" OR "\textit{Testing}" \\
&  OR "\textit{Fuzz*}" OR "\textit{Test Generat*}" OR "\textit{Combinatorial Testing}" OR "\textit{Concolic Testing}" OR "\textit{Test Optimization}" OR "\textit{Test Selection}" \\ 
&  OR "\textit{Test Prioritization}" OR "\textit{Test Minimization}" OR "\textit{Input Prioritization}" OR "\textit{Input Reduction}" OR "\textit{Input Selection}"  \\ \midrule
\multicolumn{2}{l}{\textbf{Search String: } \textit{Group A} \textbf{AND} \textit{Group B} } \\
\bottomrule
\end{tabular}}
\begin{tablenotes}
\small
  \item[*] is a wildcard used to match zero or more characters.
\end{tablenotes}
\end{threeparttable}
\end{table}

\subsection{Search Items}
\label{SearchItems}
Given that our survey focuses on CGT, a specialized domain within SE, we primarily target SE publication venues for the manual search. We select four top-tier SE conferences, ICSE, ESEC/FSE, ASE, and ISSTA, along with three leading journals, TOSEM, TSE, and EMSE, to ensure a search of state-of-the-art advancements in CGT. In our literature survey, we found that the earliest research \cite{DBLP:conf/sosp/PeiCYJ17} can be traced back to 2017. As a result, we limit the literature search to the period October 1, 2017, to April 30, 2025. Through manual inspection, we identified a total of 44 papers within this time frame. These selected papers serve as the foundation for designing the search string. To enable automated searches, we systematically categorize the search terms into two distinct groups. 

\begin{tcolorbox}
    [colback=gray!10, rounded corners]
\small
\textbf{A. DL-related group:} This group comprises commonly used keywords associated with DL systems or DL models.  \\
\textbf{B. Testing-related group:} This group contains keywords frequently used to describe testing activities for DL models, including test coverage analysis, test input generation, and test optimization. 
\end{tcolorbox}
Table \ref{Tab-keywords} presents the complete search string. The two groups are combined using the `AND' operator to ensure that the retrieved CGT studies are those applied to DL models.

\begin{table}
\begin{threeparttable}
  \caption{Inclusion Criteria and Exclusion Criteria}
  \small
  \label{Tab-Inclusion-Exclusion}
  \fontsize{9.3}{7.7}\selectfont
  \setlength{\tabcolsep}{0.1 mm}{
  \begin{tabular}{c|l}
    \toprule
\textbf{Index} & \multicolumn{1}{c}{\textbf{Inclusion Criteria}}   \\
    \midrule
\ding{172} & The paper addresses the challenges of testing DL models, with an explicit contribution to CGT techniques. \\
\multirow{2}{*}{\ding{173}} & The papers, including those from arXiv and short-paper venues, have demonstrated impact, each receiving \\ 
& at least 10 citations, and showing high relevance to our research focus. \\
\ding{174}  & The full text of the study is publicly accessible.\\
\ding{175} & The paper is written exclusively in English. \\ \midrule
\textbf{Index} & \multicolumn{1}{c}{\textbf{Exclusion Criteria}}   \\ \midrule
\ding{186} & The paper is published as a literature review or survey. \\
\ding{187} & The paper applies DL models to test traditional software, or applies CGT outside the context of DL models. \\ 
\ding{188} & The paper is published as books, keynotes, tool demonstrations, technical reports, or editorials. \\ 
\ding{189} & Duplicate papers or similar studies that appear in different venues but are authored by the same researchers. \\
\ding{190} & The paper lacks a clearly defined methodology for validating the proposed technique. \\

\bottomrule
\end{tabular}}
\end{threeparttable}
\end{table}

\subsection{Study Collection and Selection}
\label{CollectionSelection}
To conduct an automated literature search, we collect relevant studies from 7 academic repositories, \textit{ACM Digital Library}, \textit{IEEE Xplore Digital Library}, \textit{SpringerLink}, \textit{ScienceDirect}, \textit{Wiley}, \textit{Google Scholar repository}, and \textit{arXiv} using the search string. Using the predefined search string, we retrieved a total of 137, 518, 64, 250, 40, 1570, and 341 from these repositories, respectively. 

\subsubsection{Inclusion and Exclusion Criteria}
After collecting the studies through the automated search strategy, we first removed duplicate records retrieved from different digital libraries (criterion \ding{189}). Second, we conducted a manual screening of the titles, abstracts, keywords, and publication venues to assess the relevance of each paper based on predefined inclusion and exclusion criteria, as detailed in Table \ref{Tab-Inclusion-Exclusion}. For instance, following the inclusion criteria in the survey by Xie et al. \cite{DBLP:journals/tosem/XieLDZW24}, we adopted criterion \ding{173} to include influential short papers published in workshops and special tracks, such as Demonstrations, New Ideas and Emerging Results, and Companion sessions. Considering the rapidly evolving nature of this research field, we also included relevant preprints from open-access repositories arXiv, especially those focusing on DL testing that may not yet have been formally peer-reviewed. To ensure quality, we retained papers that have received at least 10 citations and demonstrate high relevance to the topic. Furthermore, we excluded similar versions of the same work published across multiple venues, such as conference papers later extended to journals, or arXiv preprints later formally published, when authored by the same research group. According to the exclusion criterion \ding{187}, we also removed studies that utilize DL models to test traditional software, or that apply CGT techniques outside the context of DL systems. Finally, we conducted full-text reviews of papers to confirm their relevance to the scope and quality standards of this survey. At this stage, papers lacking a clear methodology for validating the proposed technique were excluded (criterion \ding{190}). Through this process, a total of 82 papers were identified as directly relevant to our research focus.

\subsubsection{Snowballing Search}
To mitigate the risk of overlooking relevant papers and to ensure comprehensive coverage of the DL model testing field, we supplement our search with a lightweight backward and forward snowballing process. This entails reviewing the references cited in each of our selected primary studies, as well as identifying and examining subsequent works that have cited these studies. Through this manual inspection, we manually incorporated 7 additional papers into our survey

\begin{table}
\begin{threeparttable}
  \caption{Overview of Publication Venues on Coverage-Guided Testing Studies}
  \small
  \label{Tab-Venue}
  \fontsize{7.7}{7.8}\selectfont
  \setlength{\tabcolsep}{0.02 mm}{
  \begin{tabular}{cllcc}
    \toprule
\textbf{Venue} & \textbf{Acronym} & \textbf{Full Name}  &  \textbf{\# Publications}  &  \textbf{Proportion}  \\ \midrule
\multirow{22}{*}{\rotatebox{90} {\textbf{Conference (61.8\%)}}}  & ICSE & ACM/IEEE International Conference on Software Engineering & 15 & 16.85\% \\
& ISSTA & ACM SIGSOFT International Symposium on Software Testing and Analysis & 5 & 5.62\%\\
& \multirow{2}{*}{ESEC/FSE} & ACM Joint European Software Engineering Conference and Symposium on the Foundations & \multirow{2}{*}{5} & \multirow{2}{*}{5.62\%} \\
&&  of Software Engineering \\
& ASE & IEEE/ACM International Conference Automated Software Engineering & 4 & 4.49\% \\
& ISSRE & IEEE International Symposium on Software Reliability & 3 & 3.37\% \\ 
& SANER & IEEE International Conference on Software Analysis, Evolution and Reengineering & 3 & 3.37\% \\ 
& QRS & IEEE International Conference on Software Quality, Reliability and Security & 3 & 3.37\% \\
& ICML & International Conference on Machine Learning & 2  & 2.25\%\\ 
& IJCNN & International Joint Conference on Neural Networks & 2 & 2.25\% \\
& SOSP & ACM Symposium on Operating Systems Principles & 1 & 1.12\% \\
& ESEM & ACM/IEEE International Symposium on Empirical Software Engineering and Measurement & 1 & 1.12\%  \\
& ICSME & IEEE International Conference on Software Maintenance and Evolution & 1 & 1.12\%  \\ 
& ICST & IEEE International Conference on Software Testing, Verification and Validation & 1 & 1.12\%  \\ 
& ICLR &  International Conference on Learning Representations & 1 & 1.12\%  \\
& IJCAI & International Joint Conference on Artificial Intelligence & 1  & 1.12\% \\
& NAACL & Conference of the North American Chapter of the Association for Computational Linguistics & 1 & 1.12\%  \\
& AITest & IEEE International Conference on Artificial Intelligence Testing & 1 & 1.12\%  \\
& SEKE &  International Conference on Software Engineering and Knowledge Engineering & 1 & 1.12\%  \\
& STTT & International Journal on Software Tools for Technology Transfer  & 1 &1.12\% \\
& SETTA & International Symposium on Dependable Software Engineering: Theories, Tools, and Applications & 1 &1.12\% \\
& ICECCS & International Conference on Engineering of Complex Computer Systems & 1 & 1.12\% \\
& HPCC & International Conference on High Performance Computing and Communications & 1 & 1.12\%  \\
 \midrule
 
\multirow{14}{*}{\rotatebox{90} {\textbf{Journal (38.2\%)} }}  & TSE & IEEE Transactions on Software Engineering & 8 & 8.99 \% \\ 
& TOSEM & ACM Transactions on Software Engineering and Methodology & 5 & 5.62\%\\
& IST & Information and Software Technology & 5 & 5.62\% \\
& EMSE & Empirical Software Engineering & 2 & 2.25\% \\
& JSS & Journal of Systems and Software & 2 & 2.25\% \\
& TR & IEEE Transactions on Reliability & 2 & 2.25\% \\
& JSEP & Journal of Software: Evolution and Process & 1 & 1.12\% \\
& TNNLS & IEEE Transactions on Neural Networks and Learning Systems & 1 &  1.12\% \\
& TECS & ACM Transactions on Embedded Computing Systems & 1 & 1.12\%\\
& JSA & Journal of Systems Architecture & 1 & 1.12\%\\
& SQJ & Software Quality Journal & 1 & 1.12\%\\
& APIN & Applied Intelligence & 1 & 1.12\% \\
& IJON & Neurocomputing & 1 & 1.12\% \\
& ASC & Applied Soft Computing & 1 & 1.12\%\\
\midrule
\textbf{Other} & arXiv & N.A & 2 & 2.25\% \\
\midrule
{\textbf{Total}}  & N.A  & N.A  & 89 & 100\% \\ \bottomrule
\end{tabular}}
\end{threeparttable}
\end{table}

\subsection{Publication Venues and Trends}
\label{TrendObservation}
Table \ref{Tab-Venue} presents a summary of the publication venues for CGT-related studies. Conferences serve as the primary dissemination channel in this field, accounting for 55 publications (61.8\% of the total), while journal articles are comparatively fewer, with 34 identified. This distribution reflects the fast-paced nature of DL testing research, where new advancements are often introduced through conference proceedings. CGT has emerged as a prominent topic within the SE community, with over two-thirds of the studies published in SE venues such as TOSEM and ICSE. The topic has also garnered increasing attention from the AI community, with publications in venues like ICML and TNNLS. Among conference venues, ICSE leads with 15 publications, followed by ISSTA and ESEC/FSE, each featuring 5 papers, and ASE with 4. These venues represent top-tier conferences in the SE domain, underscoring the ongoing development of CGT research. In terms of journal publications, TSE is the most prominent venue, with 8 papers, followed by other SE journals such as TOSEM, IST, EMSE, and JSS. Additionally, two arXiv preprints, each cited over 10 times, were included, highlighting the rapid emergence of CGT research.

\begin{figure}  
    \centering  
    \includegraphics[width = 14.5 cm]{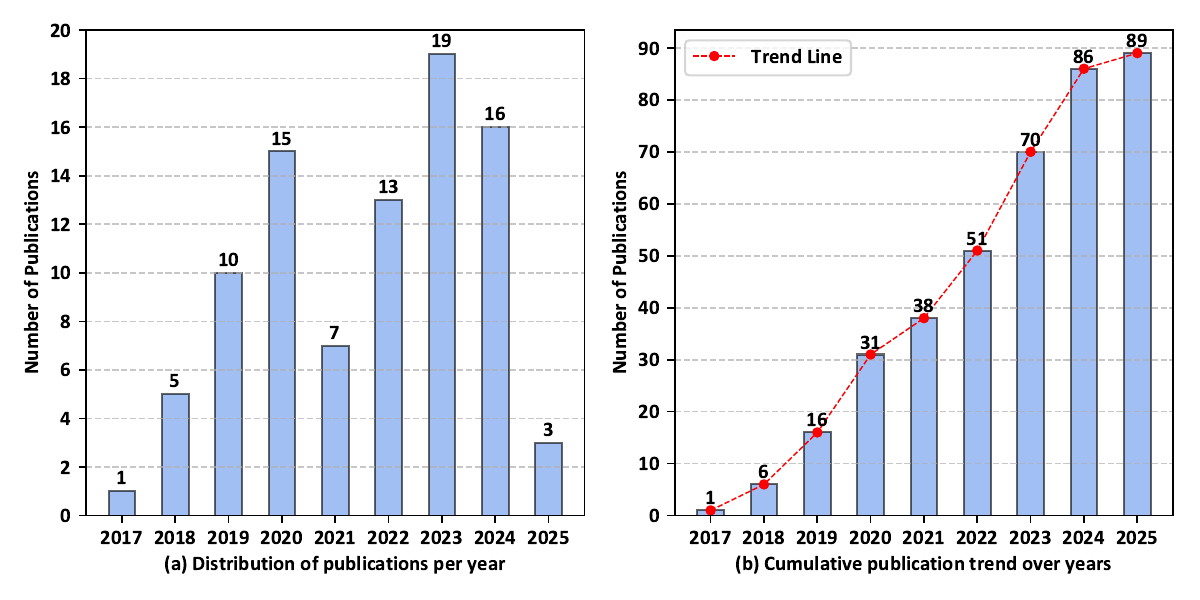} 
    \caption{Publication Trends of Coverage-Guided Testing Studies}  
    \label{PublicationTrends}
\end{figure}

The publication trends presented in Figure \ref{PublicationTrends} highlight the evolving research landscape of CGT. Figure \ref{PublicationTrends}a illustrates the annual number of publications from October 2017 to April 2025, while Figure \ref{PublicationTrends}b shows the cumulative growth over this period, indicating increasing academic interest. Since the first CGT study was introduced in 2017, research activity has generally increased. A slight decline observed in 2021 corresponds with increasing skepticism regarding the reliability of existing coverage criteria, leading to critical analyses and a temporary slowdown. The field regained momentum from 2022 onward, driven by innovative methodologies that addressed these concerns from alternative perspectives. It highlights continued interest, with a growing body of research contributing to its theoretical foundations and practical applications.	

\section{Coverage Criteria}
\label{Sec-coverage}
Depending on the level of access and knowledge the tester has about the model’s internal structure, coverage criteria can be broadly categorized into white-box, gray-box, and black-box metrics. These three types of coverage criteria are complementary, collectively enhancing confidence that the DNN has undergone thorough testing. 
\begin{tcolorbox}
    [colback=gray!10, rounded corners]
\small
\textbf{(1) White-box Coverage Criteria} rely on full access to the internal structure and learned parameters of a DL model, including its architecture, intermediate representations, connection weights, and gradients. \\
\textbf{(2) Black-box Coverage Criteria} assess test adequacy based on the input and output domain, without requiring any knowledge of the model’s internal architecture or parameters.\\
\textbf{(3) Gray-box Coverage Criteria} balance white-box and black-box criteria by utilizing partial internal information. Internal parameters such as weights and gradients are inaccessible to testers. 
\end{tcolorbox}

\begin{figure*}[htbp]
  \centering
  \begin{subfigure}{0.45\textwidth}
    \centering
    \includegraphics[width = 6.2 cm]{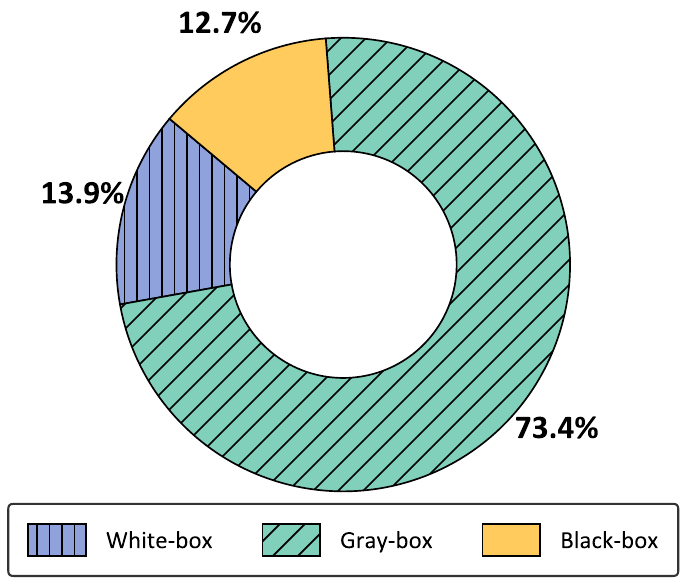}
    \caption{Distribution of White-box, Gray-box, and Black-box Coverage Criteria}
    \label{dis-cov1}
  \end{subfigure}
    \hspace{0.4cm}  
  \begin{subfigure}{0.45\textwidth}
    \centering
    \includegraphics[width = 5.3 cm]{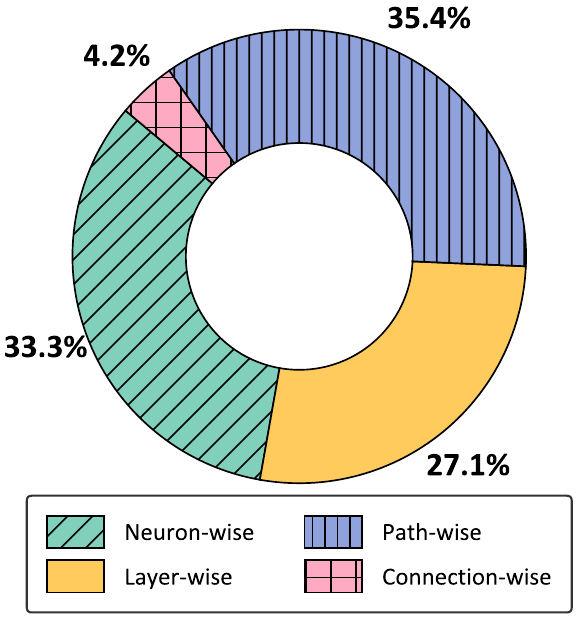}
    \caption{Distribution of White-box and Gray-box Coverage Criteria Across Different Granularity Levels for FNNs}
    \label{dis-cov2}
  \end{subfigure}
  \caption{Distribution of Coverage Criteria for DL Models}
  \label{fig:coverage}
\end{figure*}

Figure \ref{dis-cov1} illustrates their distribution: gray-box criteria dominate, accounting for 73.4\%, while white-box and black-box criteria comprise 13.9\% and 12.7\%, respectively. The prevalence of gray-box criteria may be attributed to their adaptability, particularly when internal model details are partially accessible. Black-box criteria primarily focus on partitioning the input/output space and are model architecture-agnostic.

\subsection{White-box and Gray-box Coverage Criteria}
\label{WhiteAndGrayCC}
White-box and gray-box criteria quantify the extent to which the internal components of a DL model are exercised by a given test suite. Both types of criteria rely on access to model internals, and are thus introduced together in this section. To reflect the architectural diversity of DL models and the corresponding design of coverage criteria, we organize the subsequent discussion by model architectures. Section \ref{CC4FNN} reviews criteria designed for Feedforward Neural Networks (FNNs), including Convolutional Neural Networks (CNNs) and customized FNNs variants. Section \ref{CC4Other} turns to coverage criteria designed for other architectures, including Recurrent Neural Networks (RNNs), Transformer-based models, and Deep Reinforcement Learning (DRL) models.

\subsubsection{Coverage Criteria for FNN Models}
\label{CC4FNN}
\begin{figure}  
    \centering  
    \includegraphics[width = 12.5 cm]{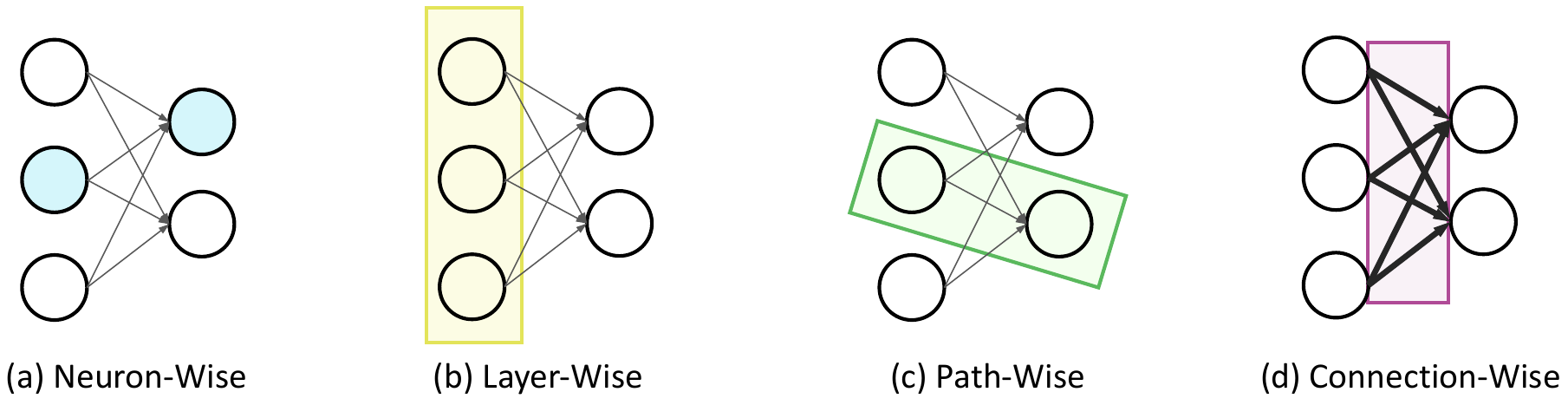} 
    \caption{High-Level Illustrations of White-box and Gray-box Coverage Criteria for FNNs}  
    \label{CCWise}
\end{figure}

White-box and gray-box coverage criteria for FNNs can be categorized by granularity into neuron-wise, layer-wise, path-wise, and connection-wise levels. Figure \ref{CCWise} presents high-level illustrations of these categories, highlighting internal components within DL models to indicate their coverage domain. \textbf{Neuron-wise criteria} operate at the most fundamental level, focusing on individual neurons as basic computational units. \textbf{Layer-wise criteria} examine the aggregation of neurons at the layer level, capturing intra-layer dynamics and interactions. \textbf{Path-wise criteria} follow information flow across multiple layers, offering a global view of network functionality. \textbf{Connection-wise criteria} target neuron connections between adjacent layers. Both path-wise and connection-wise capture inter-layer interactions. 

Figure~\ref{dis-cov2} presents the distribution of coverage criteria across different granularity levels. Path-wise coverage is the most commonly adopted, representing 35.4\% of the total, followed closely by neuron-wise coverage at 33.3\%. Layer-wise coverage accounts for 27.1\%, while connection-wise coverage remains relatively underexplored, comprising only 4.2\%. Table \ref{Tab-CC4FNN} summarizes white-box and gray-box criteria for FNN models. The `Training Inps' column reflects whether a criterion utilizes prior knowledge extracted from training inputs. The `Percentage' column indicates whether the criterion reports coverage as a percentage. The `Task-Agnostic' column indicates whether a criterion could be directly applied to both classification and regression tasks.

\textbf{(1) Neuron-wise.} Neuron-wise criteria trace how thoroughly individual neurons are exercised by a test suite. DeepXplore \cite{DBLP:conf/sosp/PeiCYJ17} pioneered this direction by drawing an analogy between neurons in DNNs and statements in traditional software programs, introducing Neuron Coverage, the first coverage criterion for DNNs. NC discretizes each neuron's continuous output into binary states and considers a neuron activated if its output exceeds a predefined threshold (\textit{e.g.}, 0.5 or 0.75). It computes the ratio of activated neurons across a DNN.

DeepGauge \cite{DBLP:conf/kbse/MaJZSXLCSLLZW18} proposes finer-grained metrics by analyzing each neuron's output distribution over the training data. The distribution is partitioned into three regions: the major function region $[low_n, high_n]$, and two corner-case regions $( -\infty, low_n)$ and $(hign_n, + \infty) $, where $low_n$ and $high_n$ are the observed output bounds. The intuition is that different output ranges correspond to distinct functional behaviors. $k$-Multisection Neuron Coverage divides the major region into $k$ equal intervals and considers a section covered when a neuron's output, after executing a test input, falls within that section. It measures how many of these sections are covered. Neuron Boundary Coverage measures the proportion of neurons activated in the corner regions, while Strong Neuron Activation Coverage focuses on activations in the upper corner region. Extending Neuron Boundary Coverage and $k$-Multisection Neuron Coverage, Rossolini et al. \cite{DBLP:journals/tse/RossoliniBB23} perform coverage analysis to monitor input behavior and assess trustworthiness during inference. They introduce a mapping that converts coverage deviations, quantified as the degree of neuron activations falling outside the established training bounds, into a confidence score.

Several studies have advanced these concepts by incorporating neuron semantics. CriticalFuzz \cite{DBLP:journals/infsof/BaiHHWXQY24} introduces a `criticality level' to quantify neuron importance, defining a neuron as critical if it exhibits high activation across a large portion of the training data. Class-critical neurons are those strongly activated by inputs from a specific class. Critical Neuron Coverage and Class-Critical Neuron Coverage are proposed to measure the extent to which a test suite covers these neurons. DANCe  \cite{DBLP:conf/ijcnn/YeWZK22} leverages an influence factor to quantify how frequently a neuron's output falls within its major function region as defined by DeepGauge \cite{DBLP:conf/kbse/MaJZSXLCSLLZW18}. Themis \cite{DBLP:conf/issre/0005LXC24} quantifies neuron sensitivity by computing the absolute difference in activation values between original and perturbed inputs.  Neurons exhibiting high sensitivity variance are sampled, and their sensitivity distributions are estimated via Markov Chain Monte Carlo sampling. Sensitivity Convergence Coverage is defined as the proportion of sampled neurons whose sensitivity distributions have statistically converged.

FalsifyAI \cite{DBLP:journals/tse/ZhangHML22} designs coverage criteria for DNN-enabled Cyber-Physical Systems (CPSs), where DNN-based controllers process continuous input signals received from the environment and execute consecutively over a period of time. Instant Neuron Coverage quantifies the ratio of neurons activated at least once within a specified time window. Timed Neuron Coverage captures the proportion of neurons that remain continuously activated for a given duration within a time window.

Neuron Activation Coverage (NAC) \cite{DBLP:conf/iclr/LiuTLM024} characterizes the statistical activation behavior of neurons over in-distribution (ID) data by leveraging the probability density function. The core insight is that neuron states rarely activated by ID inputs are more likely to be associated with erroneous behaviors under out-of-distribution (OOD) scenarios, thus requiring targeted testing. Two derived metrics are introduced: NAC–Uncertainty Estimation, which estimates the uncertainty of inputs by averaging NAC scores over all neurons, and NAC–Model Evaluation, which assesses model generalization by integrating the NAC distribution \textit{w.r.t} all neurons.

\begin{table}
\begin{threeparttable}
  \caption{Overview of White-box and Gray-box Coverage Criteria for FNNs Organized by Publication Year}
  \small
  \label{Tab-CC4FNN}
  \fontsize{7.2}{7.4}\selectfont
  \setlength{\tabcolsep}{0.05 mm}{
  \begin{tabular}{cllcccccc}
    \toprule
\multirow{2}{*} {\textbf{Year}} &   \multirow{2}{*} {\textbf{Study}} &    \multirow{2}{*} {\textbf{Criterion} } &   \multirow{2}{*} {\textbf{Accessed} \customfootnote{*}} &    \multirow{2}{*} {\textbf{Granularity} \customfootnote{**}} & \textbf{Training}  &  \textbf{Hyperparameter}  &   \multirow{2}{*} {\textbf{Percentage} } &   \textbf{Task} \\
&&&&&\textbf{Inps}  & \textbf{-Free}  & & \textbf{-Agnostic}  \\
    \midrule
2017& DeepXplore \cite{DBLP:conf/sosp/PeiCYJ17}& Neuron Coverage & \scalebox{0.80}{\sqbox{gray}}& Neuron & \scriptsize  \ding{55} &  \scriptsize \ding{55}& \scriptsize \ding{51}& \scriptsize \ding{51}\\ \midrule
    
\multirow{5}{*} {2018}& \multirow{5}{*} {DeepGauge \cite{DBLP:conf/kbse/MaJZSXLCSLLZW18} }  & $k$-Multisection Neuron Coverage & \scalebox{0.80}{\sqbox{gray}} & Neuron & \scriptsize \ding{51} & \footnotesize \scriptsize \ding{55}& \scriptsize \ding{51}& \scriptsize \ding{51}\\
  && Neuron Boundary Coverage&  \scalebox{0.80}{\sqbox{gray}} & Neuron & \scriptsize \ding{51} & \scriptsize \ding{51}& \scriptsize \ding{51}& \scriptsize \ding{51}\\
 && Strong Neuron Activation Coverage & \scalebox{0.80}{\sqbox{gray}} & Neuron &\scriptsize \ding{51} & \scriptsize \ding{51} & \scriptsize \ding{51}& \scriptsize \ding{51}\\
  && Top-$k$ Neuron Coverage& \scalebox{0.80}{\sqbox{gray}} & Layer  & \scriptsize \ding{55} &  \scriptsize \ding{55} & \scriptsize \ding{51}& \scriptsize \ding{51}\\
 && Top-$k$ Neuron Pattern Coverage & \scalebox{0.80}{\sqbox{gray}} & Layer  &  \scriptsize \ding{55} & \scriptsize \ding{55} &  \scriptsize \ding{55} & \scriptsize \ding{51}\\\midrule
 
  \multirow{4}{*} {2018} &\multirow{4}{*} {Sun et al. \cite{DBLP:journals/tecs/SunHKSHA19}}& Sign-Sign Coverage & \scalebox{0.80}{\sqbox{gray}} & Path & \scriptsize \ding{55}&  \scriptsize \ding{55}& \scriptsize \ding{51}& \scriptsize \ding{51}\\
   && Value-Sign Coverage & \scalebox{0.80}{\sqbox{gray}} & Path & \scriptsize \ding{55}& \scriptsize \ding{55}& \scriptsize \ding{51}& \scriptsize \ding{51}\\
   && Sign-Value Coverage & \scalebox{0.80}{\sqbox{gray}} & Path & \scriptsize \ding{55}& \scriptsize \ding{55}& \scriptsize \ding{51}& \footnotesize \ding{51}\\
  &&  Value-Value Coverage & \scalebox{0.80}{\sqbox{gray}}  & Path & \scriptsize \ding{55}& \scriptsize \ding{55}& \scriptsize \ding{51}& \ding{51}\\ \midrule
  
\multirow{3}{*} {2019}  & \multirow{3}{*} {DeepPath \cite{DBLP:conf/aitest/WangWFCC19}} & $l$-length Activated Path Coverage & \scalebox{0.80}{\sqbox{gray}} & Path  &  \scriptsize \ding{55}&  \scriptsize \ding{55}& \scriptsize \ding{51}& \scriptsize \ding{51}\\
 && $l$-length Output Activated Path Coverage & \scalebox{0.80}{\sqbox{gray}} & Path  & \footnotesize \scriptsize \ding{55}& \footnotesize \scriptsize \ding{55}& \scriptsize \ding{51}& \scriptsize \ding{51}\\
  && $l$-length Full State Path Coverage & \scalebox{0.80}{\sqbox{gray}} & Path  & \footnotesize \scriptsize \ding{55}& \footnotesize \scriptsize \ding{55}& \scriptsize \ding{51}& \footnotesize \ding{51}\\ \midrule
  
  2019 &TensorFuzz \cite{DBLP:conf/icml/OdenaOAG19} & Cluster-based Coverage & \scalebox{0.80}{\sqbox{gray}}  & Layer & \scriptsize \ding{51}& \footnotesize \scriptsize \ding{55} & \footnotesize \scriptsize \ding{55}& \footnotesize \ding{51}\\ \midrule
  
 \multirow{3}{*} {2019}  & \multirow{3}{*} {SADL \cite{DBLP:conf/icse/KimFY19, DBLP:conf/sigsoft/KimJFY20}} & Distance-based Surprise Coverage & \scalebox{0.80}{\sqbox{gray}} & Path & \scriptsize \ding{51}& \scriptsize \ding{55}& \scriptsize \ding{51}&  \scriptsize \ding{55}\\
    & &Likelihood-based Surprise Coverage& \scalebox{0.80}{\sqbox{gray}} & Path & \scriptsize \ding{51}& \scriptsize \ding{55}& \scriptsize \ding{51}& \scriptsize \ding{51}\\
   & & Mahalanobis Distance-based Surprise Coverage & \scalebox{0.80}{\sqbox{gray}} & Path & \scriptsize \ding{51}& \scriptsize \ding{55}& \scriptsize \ding{51}& \ding{51}\\ \midrule
   
\multirow{3}{*} {2019} & \multirow{3}{*} {DeepCT \cite{DBLP:conf/wcre/MaJXLLLZ19} } &  $t$-way Combination Sparse Coverage & \scalebox{0.80}{\sqbox{gray}}  & Layer & \scriptsize \ding{55}&  \scriptsize \ding{55}& \scriptsize \ding{51}& \scriptsize \ding{51}\\
  &&  $t$-way Combination Dense Coverage & \scalebox{0.80}{\sqbox{gray}} & Layer  & \scriptsize \ding{55}&  \scriptsize \ding{55}& \scriptsize \ding{51}& \scriptsize \ding{51}\\
    && $(p,t)$-Completeness & \scalebox{0.80}{\sqbox{gray}} & Layer & \scriptsize \ding{55}&  \scriptsize \ding{55}& \scriptsize \ding{51}& \scriptsize \ding{51}\\ \midrule

2019 & Mani et al. \cite{DBLP:journals/corr/abs-1911-07309} & Centroid Positioning & \scalebox{0.80}{\sqbox{gray}} & Layer & \scriptsize \ding{55}& \scriptsize \ding{55}& \scriptsize \ding{51}& \scriptsize \ding{55}  \\ \midrule

2020 & DeepImportance \cite{DBLP:conf/icse/GerasimouE0C20} & Importance-Driven Coverage & \small \sqboxEmpty{black} & Layer & \scriptsize \ding{51}& \scriptsize \ding{55}& \scriptsize \ding{51}& \scriptsize \ding{51}  \\ \midrule

2021  & DeepCon \cite{DBLP:conf/wcre/ZhouDLZ0Y21}& Contribution Coverage & \small \sqboxEmpty{black} & Path & \scriptsize \ding{55}& \scriptsize \ding{55}& \scriptsize \ding{51}&  \ding{51}\\ \midrule

2021  & FilterFuzz \cite{DBLP:conf/qrs/WeiC21} & Filter Coverage & \scalebox{0.80}{\sqbox{gray}} & Layer & \scriptsize \ding{55}&  \scriptsize \ding{55}& \scriptsize \ding{51} & \scriptsize \ding{51}\\ \midrule
 
\multirow{2}{*} {2022}   &\multirow{2}{*} {NPC \cite{DBLP:journals/tosem/XieLWMGJL22}} & Structure-based Neuron Path Coverage & \small \sqboxEmpty{black} & Path & \scriptsize \ding{51}&  \scriptsize \ding{55}&   \scriptsize \ding{51}& \scriptsize \ding{55} \\
  &&  Activation-based Neuron Path Coverage & \small \sqboxEmpty{black} & Path & \scriptsize \ding{51}&  \scriptsize \ding{55}& \scriptsize \ding{51}& \scriptsize \ding{55}\\  \midrule 
  
2022  & DANCe \cite{DBLP:conf/ijcnn/YeWZK22}& Influence Coverage &  \scalebox{0.80}{\sqbox{gray}} & Neuron & \scriptsize \ding{51}& \scriptsize \ding{55}& \scriptsize \ding{51}&  \ding{51}\\ \midrule
   
2023  & GradFuzz \cite{DBLP:journals/ijon/ParkCKK23}& Gradient Vector Coverage & \small \sqboxEmpty{black} & Connection & \scriptsize \ding{55}&  \scriptsize \ding{55}& \scriptsize \ding{55}&  \scriptsize \ding{55}  \\ \midrule

\multirow{3}{*} {2023}  & \multirow{3}{*} {Rossolini et al. \cite{DBLP:journals/tse/RossoliniBB23}} & Single-Range Coverage & \scalebox{0.80}{\sqbox{gray}} & Neuron & \scriptsize \ding{51} & \scriptsize \ding{55} & \scriptsize \ding{55}  & \scriptsize \ding{55} \\
&& Multi-Range Coverage & \scalebox{0.80}{\sqbox{gray}} & Neuron & \scriptsize \ding{51} & \scriptsize \ding{55} & \scriptsize \ding{55}  & \scriptsize \ding{55} \\ 
&& $k$-Nearest Neighbors Coverage & \scalebox{0.80}{\sqbox{gray}}  & Layer & \scriptsize \ding{51} & \scriptsize \ding{55} & \scriptsize \ding{55}  & \scriptsize \ding{55}  \\ \midrule
  
\multirow{6}{*} {2023}   &\multirow{6}{*} {FalsifyAI \cite{DBLP:journals/tse/ZhangLAMHZ23}}  &Instant Neuron Coverage & \scalebox{0.80}{\sqbox{gray}} & Neuron & \scriptsize \ding{55}& \scriptsize \ding{55}& \scriptsize \ding{51}& \scriptsize \ding{51}\\
&& Instant Top-$k$ Neuron Coverage & \scalebox{0.80}{\sqbox{gray}} & Layer &  \scriptsize \ding{55}& \scriptsize \ding{55} & \scriptsize \ding{51}& \scriptsize \ding{51}\\
&&  Timed Neuron Coverage & \scalebox{0.80}{\sqbox{gray}}  & Neuron & \scriptsize \ding{55}& \scriptsize \ding{55}& \scriptsize \ding{51}& \ding{51}\\
 && Timed Top-$k$ Neuron Coverage & \scalebox{0.80}{\sqbox{gray}} & Layer & \scriptsize \ding{55}&  \scriptsize \ding{55}& \scriptsize \ding{51}& \ding{51}\\
&&  Positive/Negative Differential Neuron Coverage & \scalebox{0.80}{\sqbox{gray}} & Neuron & \scriptsize \ding{55}& \scriptsize \ding{55}& \scriptsize \ding{51}& \scriptsize \ding{51}\\
&&  Monotonic Increase/Decrease Neuron Coverage & \scalebox{0.80}{\sqbox{gray}} & Neuron & \scriptsize \ding{55}&  \scriptsize \ding{55}& \scriptsize \ding{51}& \scriptsize \ding{51}\\  \midrule

2023 & NLC \cite{DBLP:conf/icse/YuanPW23} & NeuraL Coverage & \scalebox{0.80}{\sqbox{gray}} & Layer & \scriptsize \ding{51}& \scriptsize \ding{51}& \scriptsize \ding{55}& \scriptsize \ding{51}\\ \midrule
    
2023  & Ji et al. \cite{DBLP:conf/icse/JiMYW23}& Causality-Aware Coverage & \scalebox{0.80}{\sqbox{gray}} & Connection & \scriptsize \ding{55}& \scriptsize \ding{55}&  \scriptsize \ding{51}& \scriptsize \ding{51}\\ \midrule

2023  & HeatC \cite{DBLP:conf/setta/SunLLS23}& HeatC Coverage &  \small \sqboxEmpty{black} & Path & \scriptsize \ding{51}& \scriptsize \ding{55}&  \scriptsize \ding{51}& \scriptsize \ding{51}\\ \midrule

2023  & HashC \cite{DBLP:journals/jsa/SunXLZS23} & HashC Coverage &  \small \sqboxEmpty{black} & Neuron & \scriptsize \ding{51} & \scriptsize \ding{55} & \scriptsize \ding{51} & \scriptsize \ding{51}\\ \midrule

\multirow{2}{*} {2024}  & \multirow{2}{*} {CriticalFuzz \cite{DBLP:journals/infsof/BaiHHWXQY24}}  & Critical Neuron Coverage & \scalebox{0.80}{\sqbox{gray}} & Neuron & \scriptsize \ding{51}& \scriptsize \ding{55}&\scriptsize \ding{51} & \scriptsize \ding{51}\\
 &&  Class-Critical Neuron Coverage & \scalebox{0.80}{\sqbox{gray}} & Neuron & \scriptsize \ding{51}& \scriptsize \ding{55}& \scriptsize \ding{51}& \scriptsize \ding{55}\\
 \midrule

\multirow{2}{*} {2024}   &\multirow{2}{*} {DeepLID \cite{DBLP:journals/ese/GuoTH24}} & LID-based Surprise Coverage &  \small \sqboxEmpty{black} & Path & \scriptsize \ding{51}& \scriptsize \ding{55}& \scriptsize \ding{51}& \scriptsize \ding{51}\\
  &&  Top-$n$ Surprise Coverage & \small \sqboxEmpty{black} & Path & \scriptsize \ding{51}& \scriptsize \ding{55}& \scriptsize \ding{51}& \scriptsize \ding{51}\\ \midrule
  
2024  & Liu et al. \cite{DBLP:conf/iclr/LiuTLM024} & Neuron Activation Coverage & \scalebox{0.80}{\sqbox{gray}} & Neuron & \scriptsize \ding{51}& \scriptsize \ding{55}&  \scriptsize \ding{51}& \scriptsize \ding{55}\\ \midrule

2024  & Themis \cite{DBLP:conf/issre/0005LXC24} & Sensitivity Convergence Coverage & \scalebox{0.80}{\sqbox{gray}} &  Neuron & \scriptsize \ding{55} & \scriptsize \ding{55} & \scriptsize \ding{51} & \scriptsize \ding{51} \\\midrule

2025  & Shi et al. \cite{DBLP:journals/infsof/ShiYS25} & Markov Chain Coverage & \small \sqboxEmpty{black} & Path & \scriptsize \ding{51} & \scriptsize \ding{55} & \scriptsize \ding{51} & \scriptsize \ding{51} \\ 

\bottomrule
\end{tabular}}
\begin{tablenotes}
\footnotesize
\item[\customtnote{*}] This column denotes the tester's level of knowledge regarding the DL model under test. \scalebox{1.5}{\sqboxEmpty{black}} : white-box. \scalebox{0.85}{\sqbox{gray}}: gray-box. \item[\customtnote{**}] This column denotes the granularity of the coverage criteria. The symbol \ding{51} indicates support, while \ding{55} indicates lack of support. 
\end{tablenotes}
\end{threeparttable}
\end{table}

\textbf{(2) Layer-wise.} 
In FNNs, neurons are organized in a layer-wise fashion, with each layer performing specific computations on the output of the preceding one. Layer-wise coverage aggregates neurons within a layer as a collective computational unit. It captures higher-level behavioral patterns beyond individual activations, such as neuron combinations \cite{DBLP:conf/kbse/MaJZSXLCSLLZW18, DBLP:journals/tse/ZhangHML22, DBLP:conf/wcre/MaJXLLLZ19, DBLP:conf/qrs/WeiC21}, neuron output clusters \cite{DBLP:conf/icse/GerasimouE0C20, DBLP:conf/icml/OdenaOAG19}, or intra-layer neuron outputs \cite{DBLP:conf/icse/YuanPW23}. 

DeepGuage \cite{DBLP:conf/kbse/MaJZSXLCSLLZW18} introduces Top-$k$ Neuron Coverage (TKNC) and Top-$k$ Neuron Pattern Coverage (TKNPC), to capture the most active neurons within each layer. A neuron is considered top-$k$ if its activation ranks among the $k$ highest in its layer for a given input. TKNC captures how many distinct neurons have ever appeared in the top-$k$ across a test suite. TKNPC counts the number of unique combinations of top-$k$ activated neurons. The basic assumption is that neurons with higher values are more influential in determining the model's output. FalsifyAI \cite{DBLP:journals/tse/ZhangLAMHZ23} extends TKNC to consecutive DNN executions in CPSs. It introduces Instant Top-$k$ Neuron Coverage and Time Top-$k$ Neuron Coverage to quantify the proportion of neurons appearing in the top-$k$ at an individual timestamp and across multiple timestamps, respectively. 

Filter Coverage \cite{DBLP:conf/qrs/WeiC21} targets the behavior of filters, groups of spatially structured neurons, in CNNs. A filter is deemed activated if more than half of its neurons yield positive outputs for an input. Coverage is achieved when each filter is activated by at least one input. DeepCT \cite{DBLP:conf/wcre/MaJXLLLZ19} applies $t$-way combinatorial testing to group neurons within a layer, measuring how well a test suite explores interactions among different neuron combinations.

The aforementioned layer-wise criteria commonly treat neurons as isolated units whose activation states are represented using Boolean values. However, neuron outputs in DNNs are continuous-valued, expressed as floating-point numbers. Several criteria have shifted toward modeling neuron outputs in the continuous space. DeepImportance \cite{DBLP:conf/icse/GerasimouE0C20} and TensorFuzz \cite{DBLP:conf/icml/OdenaOAG19} define coverage based on clusters of layer-wise neuron outputs. DeepImportance leverages the model interpretability technique Layer-wise Relevance Propagation to identify neurons critical to decision-making, and clusters their outputs across training data. It measures coverage as the proportion of cluster combinations triggered by the test suite. TensorFuzz dynamically forms clusters based on Euclidean distance: neuron outputs are grouped into separate clusters if their pairwise distance exceeds a predefined threshold, with coverage quantified by the number of unique clusters observed during testing.

NeuraL Coverage \cite{DBLP:conf/icse/YuanPW23} defines the coverage domain over neuron outputs within a layer. It quantifies how extensively each neuron is exercised by computing the variance of its activations across a test suite and evaluates pairwise correlations to capture inter-neuron dependencies. These statistics are organized into a covariance matrix, serving as a unified representation of intra-layer activation dynamics. Besides, $k$-Nearest Neighbors Coverage \cite{DBLP:journals/tse/RossoliniBB23} measures how closely the new input’s layer-wise neuron outputs match those in the training inputs, quantifying confidence based on the frequency of neighbors belonging to the predicted class. 

\textbf{(3) Path-wise.} 
Path-wise coverage tracks neuron behaviors across multiple layers. This is conceptually analogous to executable paths in traditional software programs. In the context of FNNs, a path is defined as a sequence of neurons across connected layers, modeling the information flow of the model.

Inspired by the Modified Condition/Decision Coverage (MC/DC) in traditional software verification, Sun et al. \cite{DBLP:journals/tecs/SunHKSHA19} draw an analogy between features learned by a layer and decisions in a traditional software program, where features from one layer serve as conditions influencing neuron outputs (decisions) in the next. They define sign-change and value-change to capture how variations in condition activations impact subsequent outputs, and propose four MC/DC variants for DNNs to model different types of condition-decision interactions.

DeepPath \cite{DBLP:conf/aitest/WangWFCC19} conceptualizes the sequences of neurons across layers as a path. Each neuron’s state is categorized as either activated or inactivated, forming state paths that track activation behaviors of neurons along these paths. A state path is considered covered if a test input triggers that exact sequence of activations. DeepPath quantifies the extent to which these possible state paths are covered by a test suite.

Beyond syntactic paths, several white-box criteria explore semantic dependencies between neurons by leveraging weights or gradients. DeepCon \cite{DBLP:conf/wcre/ZhouDLZ0Y21} defines inter-neuron contribution as the product of a neuron's output and its connection weight to the next layer, measuring coverage as the ratio of contributions exceeding a predefined threshold. NPC \cite{DBLP:journals/tosem/XieLWMGJL22} adopts Layer-wise Relevance Propagation, a technique designed for model interpretability, to identify critical neurons responsible for the model's decision. It connects critical neurons across layers to construct Critical Decision Paths (CDPs) that represent semantic information flows. It introduces two metrics: Structure-based Neuron Path Coverage, which measures the proportion of unique CDPs exercised by a test suite, and Activation-based Neuron Path Coverage, which evaluates how thoroughly test inputs explore neuron outputs along each CDP. Shi et al. \cite{DBLP:journals/infsof/ShiYS25} abstract CDPs into a Markov chain. KL divergence is adopted to quantify the discrepancy between transition matrices from training and test data. Markov Chain Coverage discretizes these divergence values across layers and computes the proportion of covered buckets to assess test adequacy.

Kim et al. \cite{DBLP:conf/icse/KimFY19, DBLP:journals/tosem/KimFY23} argue that an effective test suite should exhibit high diversity, including inputs ranging from similar to the training set to significantly different ones. To quantify this diversity, they introduce the concept of Surprise Adequacy (SA), which quantifies the novelty of a test input relative to the training set by analyzing the similarity of neuron output traces across layers. Surprise Coverage (SC) divides the continuous SA value space into equal-sized buckets and defines coverage as the ratio of buckets covered by test inputs, \textit{i.e.}, a bucket is covered if at least one input’s SA value falls within its range. Several SC variants have been proposed using different distance metrics, including Likelihood-based SC \cite{DBLP:conf/icse/KimFY19}, Distance-based SC \cite{DBLP:conf/icse/KimFY19}, Mahalanobis Distance-based SC \cite{DBLP:conf/sigsoft/KimJFY20}, and Local Intrinsic Dimensionality-based SC (LDSC) \cite{DBLP:journals/ese/GuoTH24}. DeepLID \cite{DBLP:journals/ese/GuoTH24} further introduces Top-$n$ SC (TNSC) to quantify the coverage of SA buckets associated with higher values. Notably, LDSC and TNSC leverage gradients to identify critical neurons, making them white-box criteria.

\textbf{(4) Connection-wise.} 
Connections between neurons are crucial in shaping how information flows across layers. Recent research has shifted attention toward connections, leading to the development of connection-wise criteria that evaluate whether possible links between neurons have been adequately exercised by test inputs.

Gradient Vector Coverage \cite{DBLP:journals/ijon/ParkCKK23} defines the coverage domain using gradients of connection weights between the penultimate and logit layers. For each input, a gradient vector is formed. If its distance from all previously observed vectors exceeds a predefined threshold, the input is considered to enhance coverage.

Ji et al. \cite{DBLP:conf/icse/JiMYW23} propose a causality-aware connection coverage based on the causal structure of DNNs. Using Structural Causal Models, they infer causal graphs from input executions, where each edge denotes a causal relationship between neurons. Coverage is assessed by comparing the inferred causal graph to a ground-truth reference. It examines which causal edges can be falsified during testing, ensuring that key causal structures are sufficiently covered. It provides a semantic abstraction over DNN structural connections. The causal edges reflect semantic influence, whether one neuron's activation causally affects another.

\begin{table}
\begin{threeparttable}
  \caption{Overview of Coverage Criteria for RNN, Transformer, and DRL Models}
  \small
  \label{Tab-CC4RNN}
  \fontsize{8.1}{8}\selectfont
  \setlength{\tabcolsep}{0.4 mm}{
  \begin{tabular}{cllcccccc}
    \toprule
   \multirow{2}{*} {\textbf{Year}} &    \multirow{2}{*} {\textbf{Study}} &    \multirow{2}{*} {\textbf{Criterion}}  &    \multirow{2}{*} {\textbf{Architecture} \customfootnote{*}} &    \multirow{2}{*} {\textbf{Accessed} \customfootnote{**}} & \textbf{Training}  &  \textbf{Hyperparameter} &     \multirow{2}{*} {\textbf{Percentage}}  \\
   &&&&& \textbf{Inps} & \textbf{-Free} &  \\
    \midrule
   \multirow{5}{*} {2019} &  \multirow{5}{*} {DeepStellar \cite{DBLP:conf/sigsoft/DuXLM0Z19}} &  Basic State Coverage & \multirow{5}{*} {RNN} & \multirow{5}{*} {\scalebox{0.85}{\sqbox{gray}}} & \footnotesize \ding{51}& \footnotesize \ding{55}& \footnotesize \ding{51}&  \\
   && Weighted State Coverage  &  & & \footnotesize \ding{51}& \footnotesize \ding{55}& \footnotesize \ding{51}& \\
   && n-step State Boundary Coverage & & & \footnotesize \ding{51}& \footnotesize \ding{55}& \footnotesize \ding{51}& \\
  && Basic Transition Coverage & & & \footnotesize \ding{51}& \footnotesize \ding{55}& \footnotesize \ding{51}& \\
  &&  Weighted Transition Coverage & & & \footnotesize \ding{51}& \footnotesize \ding{55}& \footnotesize \ding{51}& \\ \midrule
  
 \multirow{3}{*} {2022}  &  \multirow{3}{*} {TESTRNN \cite{DBLP:journals/tr/HuangSZSRMH22}} & Boundary Coverage & \multirow{3}{*} {RNN} & \multirow{3}{*} {\scalebox{0.85}{\sqbox{gray}}} & \footnotesize \ding{51} &  \footnotesize \ding{55} &  \small \footnotesize \ding{51} \\
  && Step-Wise Coverage & & &  \footnotesize \ding{55}&  \footnotesize \ding{55}& \footnotesize \ding{51}& \\
   && Temporal Coverage & & & \footnotesize \ding{51}&  \footnotesize \ding{55}&  \footnotesize \ding{51}& \\ \midrule
   
 \multirow{2}{*} {2022} &  \multirow{2}{*} {RNN-Test \cite{DBLP:journals/tse/GuoZZSJS22}} &   Hidden State Coverage & \multirow{2}{*} {RNN} & \multirow{2}{*} {\scalebox{0.85}{\sqbox{gray}}} & \footnotesize \ding{55} & \footnotesize \ding{51} & \footnotesize \ding{51}&  \\
  && Cell State Coverage & & & \footnotesize \ding{55}& \footnotesize \ding{55} & \footnotesize \ding{51}& \\  \midrule
  
 2022 &  MNCOVER \cite{DBLP:conf/naacl/SekhonJDQ22} & Mask Neuron Coverage & Transformer &   \scalebox{1.5}{\sqboxEmpty{black}} & \footnotesize \ding{51}& \footnotesize \ding{51}& \footnotesize \ding{51} \\\midrule
 
\multirow{4}{*} {2020}  &  \multirow{4}{*} {Trujillo et al. \cite{DBLP:conf/icse/TrujilloLEDC20}} & Neuron Coverage & \multirow{4}{*} {DRL} & \multirow{4}{*} {\scalebox{0.85}{\sqbox{gray}}} &  \footnotesize \ding{55}& \footnotesize \ding{51}& \footnotesize \ding{51} \\
&& Neuron Layered Coverage  & & &  \footnotesize \ding{55}& \footnotesize \ding{51}&  \footnotesize \ding{51}\\
&& Cumulative Neuron Coverage & & & \footnotesize \ding{55}& \footnotesize \ding{51}& \footnotesize \ding{51}\\ 
&& Cumulative Neuron Layered Coverage & & & \footnotesize \ding{55}& \footnotesize \ding{51}& \footnotesize \ding{51}\\ \midrule

\multirow{6}{*} {2024}  &  \multirow{6}{*} {Shi et al. \cite{DBLP:journals/jss/ShiYZ24}} & State boundary coverage & \multirow{6}{*} {DRL} &  \multirow{6}{*} {\scalebox{0.85}{\sqbox{gray}}} & \footnotesize \ding{51}& \footnotesize \ding{51}& \footnotesize \ding{51} \\
&& $k$-Multisection State Coverage  & & & \footnotesize \ding{51}& \footnotesize \ding{55}& \footnotesize \ding{51}\\
&& Action Boundary Coverage  & & &  \footnotesize \ding{51}& \footnotesize \ding{51}& \footnotesize \ding{51}\\ 
&& $k$-Multisection Action Coverage  & & & \footnotesize \ding{51}& \footnotesize \ding{55}& \footnotesize \ding{52}\rotatebox[origin=c]{-9.2}{\kern-0.7em\ding{55}} \\ 
&& State-Action Coverage  & & & \footnotesize \ding{51}& \footnotesize \ding{55}& \footnotesize \ding{51}\\ 
&& Optimized State Coverage  & & & \footnotesize \ding{51}& \footnotesize \ding{55}& \footnotesize \ding{51}\\ 
\bottomrule
  
\end{tabular}}
\begin{tablenotes}
\small
  \item[\customtnote{*}] This column shows the targeted neural network architecture.
  \item[\customtnote{**}] This column denotes the tester's level of knowledge regarding the DL model under test. \scalebox{1.5}{\sqboxEmpty{black}}: white-box. \scalebox{0.85}{\sqbox{gray}}: gray-box.
\item[] The symbol \ding{51} indicates support, \ding{55} indicates lack of support, and \ding{52}\rotatebox[origin=c]{-9.2}{\kern-0.7em\ding{55}} implies partial support. 
\end{tablenotes}
\end{threeparttable}
\end{table}

\subsubsection{Coverage Criteria for RNN, Transformer, and DRL Models}
\label{CC4Other}
Table \ref{Tab-CC4RNN} presents an overview of white-box and gray-box criteria designed for RNN, Transformer, and DRL models. 

\textbf{(1) RNN-specific coverage criteria.}
RNNs are designed to model temporal dependencies through internal loops and memory mechanisms. At each time step, an RNN processes both the current input and a hidden state carrying information from previous steps, capturing the sequential structure of time-series data. Unlike FNNs, where each layer serves a fixed role in feature extraction, a layer in an unrolled RNN does not maintain the same latent feature space across different input sequences. While unrolling creates a feedforward-like architecture, it fails to preserve the model’s intrinsic recursive behavior. Coverage criteria developed for FNNs are inadequate to capture the temporal dynamics of RNNs. Therefore, RNN-specific coverage criteria have been proposed. 

DeepStellar \cite{DBLP:conf/sigsoft/DuXLM0Z19} abstracts the behavior of RNNs into a state transition model. It reduces the dimensionality of hidden states to construct abstract states and aggregates transitions between concrete states into abstract transitions. The abstraction forms a Discrete-Time Markov Chain, where states and transitions observed during training define the major function regions, while unvisited states form corner-case regions. DeepStellar defines state-level criteria to assess how thoroughly test inputs exercise major function regions and corner-case regions. Transition-level criteria are proposed to measure coverage of transitions within the major function region.

TESTRNN \cite{DBLP:journals/tr/HuangSZSRMH22} targets LSTM-specific components, such as gates and hidden outputs. Boundary Coverage evaluates how many components exceed training-derived thresholds. Step-wise Coverage captures temporal dynamics by assessing whether the value change of a component between adjacent time steps surpasses a threshold. Temporal Coverage measures how well tests reflect the temporal distribution learned during training.

RNN-Test \cite{DBLP:journals/tse/GuoZZSJS22} assesses how thoroughly test inputs exercise hidden and cell states in RNNs without training-phase knowledge. Hidden State Coverage quantifies the proportion of hidden state units that reach their maximum values across time steps, layers, and batches during testing. Cell State Coverage partitions the tanh output range of cell states into intervals and computes the proportion of cell state units whose values fall into each.

\textbf{(2) Transformer-specific coverage criteria}. Transformer models rely on a self-attention mechanism to capture global dependencies within sequences, enabling parallel processing and effective contextual representation. It comprises stacked attention layers and feedforward sublayers. Targeting NLP tasks, MNCOVER \cite{DBLP:conf/naacl/SekhonJDQ22} evaluates how thoroughly a test suite activates neurons in attention layers, measuring coverage at both the word level (token embeddings) and attention level (attention weights between token pairs). It discretizes neuron outputs observed during training into equal sections and computes coverage as the proportion of activated sections. To focus on critical components, it applies importance masks to filter out less relevant tokens and attention pairs.

\textbf{(3) DRL-specific coverage criteria}.
DRL integrates DNNs with Reinforcement Learning principles to approximate value functions or policies from high-dimensional inputs. In this paradigm, an agent interacts with an environment over a series of episodes, each consisting of multiple iterations (or time steps), during which it learns optimal actions through trial-and-error guided by cumulative rewards. 

Trujillo et al. \cite{DBLP:conf/icse/TrujilloLEDC20} introduce neuron-based coverage metrics that reflect the episodic and iterative nature of DRLs. Neuron Coverage$_{RL}$ quantifies the proportion of activated neurons during a given iteration and episode relative to the total number of neurons. Neuron Layered Coverage refines this by focusing on individual layers. These metrics are extended to Cumulative Neuron Coverage and Cumulative Neuron Layered Coverage, which track neuron activations across multiple episodes and iterations.

Shi et al. \cite{DBLP:journals/jss/ShiYZ24} define the DRL coverage domain as state and action spaces, represented as the Cartesian product of their exploration ranges observed during training. At the state level, they propose State Boundary Coverage to assess whether test inputs reach state space boundaries, and $K$-Multisection State Coverage to measure the proportion of evenly partitioned state space sections that are covered. At the action level, Action Boundary Coverage and $K$-Multisection Action Coverage apply the same principles. State-Action Coverage calculates the ratio of unique state-action pairs covered by test actions relative to all possible pairs. Optimized State Coverage measures the extent to which tests cover discretized state subspaces defined by single, pairwise, or higher-order relations among state dimensions, computed as the ratio of visited to total relational subspaces.

\begin{tcolorbox}
    [colback=gray!10, rounded corners]
\ding{46} \textbf{Summary} $\blacktriangleright$ Since the first DL coverage criterion was introduced in 2017, white-box and gray-box testing strategies have gained increasing attention, evolving from basic FNNs to more complex architectures. For FNNs and CNNs, coverage criteria have been developed at various granularities, neuron-wise, layer-wise, path-wise, and connection-wise, to probe structural behaviors and internal activation patterns. Recent efforts have also explored architecture-specific designs for RNNs, Transformers, and DRLs, each defining tailored coverage domains that reflect their unique structural syntax and computational semantics. These advances highlight an ongoing trend toward designing coverage criteria that closely align with the intrinsic operational characteristics of diverse neural architectures. $\blacktriangleleft$ 
\end{tcolorbox}
 
\begin{table}
\begin{threeparttable}
  \caption{Overview of Black-box Coverage Criteria for DL Models Organized by Coverage Domain}
  \small
  \label{Tab-BlackCC}
  \fontsize{8.7}{10}\selectfont
  \setlength{\tabcolsep}{0.2 mm}{
  \begin{tabular}{c|cllcccc}
    \toprule
\textbf{Coverage} & \multirow{2}{*}{\textbf{Year}} & \multirow{2}{*}{\textbf{Study}} &  \multirow{2}{*}{\textbf{Criterion}}  & \textbf{Training} & \textbf{Hyperparameter} &  \multirow{2}{*}{\textbf{Percentage}} &  \textbf{Task} \\
\textbf{Domain} & & & &\textbf{Inps}  &\textbf{-Free} & & \textbf{-Agnostic} \\
    \midrule

\multirow{5}{*}{\textbf{Input}}  & 2020 & Byun et al. \cite{DBLP:conf/icse/ByunR20} & $K$-section Manifold Combination Coverage & \footnotesize\ding{51}& \footnotesize \ding{55}& \footnotesize \ding{51}& \footnotesize \ding{55} \\ \cline{2-8} 

& 2023  & IDC \cite{DBLP:journals/tosem/DolaDS23}& Input Distribution Coverage &  \footnotesize \ding{51}& \footnotesize \ding{55}& \footnotesize \ding{51}& \footnotesize \ding{55}  \\ \cline{2-8} 

& \multirow{3}{*}{2024}  & \multirow{3}{*}{TISA \cite{DBLP:conf/icse/NeelofarA24}} & Instance Space Coverage &  \footnotesize \ding{55}& \footnotesize \ding{51} & \footnotesize \ding{51}& \footnotesize \ding{51} \\  
&&& Area of the Instance Space & \footnotesize \ding{55} & \footnotesize \ding{51} & \footnotesize \ding{55} & \footnotesize \ding{51} \\
&&& Area of the Buggy Region & \footnotesize \ding{55} & \footnotesize \ding{51} & \footnotesize \ding{55} & \footnotesize \ding{51} \\ \midrule

\multirow{5}{*}{\textbf{Output}} & \multirow{3}{*}{2019}  & \multirow{3}{*}{Mani et al. \cite{DBLP:journals/corr/abs-1911-07309}} 
& Equivalence Partitioning & \footnotesize \ding{55} & \footnotesize \ding{51} & \footnotesize \ding{51} & \footnotesize \ding{55} \\ 
&&& Boundary Conditioning & \footnotesize \ding{55} & \scriptsize \ding{55} & \footnotesize \ding{51} & \footnotesize \ding{55} \\ 
&&& Pairwise Boundary Conditioning & \footnotesize \ding{55} & \footnotesize \ding{55} & \footnotesize \ding{51} & \footnotesize \ding{55} \\  \cline{2-8} 

& 2024  & LV-CIT \cite{DBLP:conf/issre/WangHWNNC24}& $\tau$-way label Value Combination Coverage &  \footnotesize \ding{55}& \footnotesize \ding{55} & \footnotesize \ding{51} & \footnotesize \ding{55}  \\ \cline{2-8}

& 2023  & ATOM \cite{DBLP:conf/kbse/HuWWCTJNN23}& $k$-label Combination & \footnotesize \ding{55}& \footnotesize \ding{51}& \footnotesize \ding{55}& \footnotesize \ding{55}  \\  
\bottomrule
  
\end{tabular}}
\begin{tablenotes}
\small
\item[] The symbol \ding{51} indicates support, while \ding{55} indicates lack of support. 
\end{tablenotes}
\end{threeparttable}
\end{table}

\subsection{Black-box Coverage Criteria} 
\label{BlackCC}
Black-box testing is especially relevant in scenarios where internal details of a DNN are inaccessible or impractical to obtain. This challenge is common in models built upon pre-trained backbones, which are often deployed on remote servers and accessed via inference APIs, rendering their architectures and parameters opaque to end users. In such cases, black-box criteria become essential for evaluating test adequacy, enabling testers to systematically explore the input or output space without insight into the model's internal workings.

Black-box criteria for traditional software programs rely on input specifications to partition the input space \cite{DBLP:conf/icst/KuhnMKL13}. Such specifications are unavailable for DNNs. To address this limitation, recent research leverages learned input representations to define coverage domains. Specifically, $K$-section Manifold Combination Coverage (KMCC) \cite{DBLP:conf/icse/ByunR20} and Input Distribution Coverage (IDC) \cite{DBLP:journals/tosem/DolaDS23} utilize Variational Autoencoders to learn latent representations from training inputs. Test inputs are then encoded into latent vectors, with the latent space serving as a surrogate for input specifications. Both approaches partition the latent space and apply combinatorial testing to assess how thoroughly feature combinations are exercised. The key insight is that latent dimensions encode high-level semantic features of inputs, and comprehensive coverage requires exploring interactions among these dimensions. KMCC introduced this approach in an initial study, IDC advances it by incorporating OOD detectors to filter irrelevant inputs, removing noisy latent dimensions, and focusing on feasible feature combinations.

Neelofar et al. \cite{DBLP:conf/icse/NeelofarA24} propose Test Suite Instance Space Adequacy (TISA) metrics to assess the adequacy of testing scenarios in autonomous driving systems. TISA identifies features most impactful to safety outcomes and projects scenarios onto a 2D semantic space. Within this space, it calculates three metrics: $area_{IS}$, the region spanned by the test scenarios, $area_{Bugs}$, the region containing failing cases, and $cov_{IS}$, the ratio of $area_{IS}$ to the total scenario region. These metrics quantify test coverage and diversity with respect to the input domain.

Another line of black-box criteria focuses on the output domain. Mani et al. \cite{DBLP:journals/corr/abs-1911-07309} propose adequacy metrics based on classification boundaries. Equivalence Partitioning measures the proportion of test inputs per class, ensuring a balanced distribution of test inputs across classes. Boundary Conditioning focuses on inputs near decision boundaries, identified by low-confidence predictions, where misclassifications are more likely. Pairwise Boundary Conditioning extends this to multi-class tasks by examining decision boundaries between class pairs. Since these metrics rely on observable outputs (class labels and prediction confidences), they fall under the black-box paradigm. Both LV-CIT \cite{DBLP:conf/issre/WangHWNNC24} and ATOM \cite{DBLP:conf/kbse/HuWWCTJNN23} propose label combination coverage for multi-label classification systems. Drawing inspiration from combinatorial testing, these approaches construct combinations of distinct labels from the label space to simulate real-world scenarios where multiple objects co-occur. Each label combination is regarded as a test target, enabling assessment of the model's behavior across diverse label interactions.

\begin{tcolorbox}
    [colback=gray!10, rounded corners]
\ding{46} \textbf{Summary} $\blacktriangleright$ By abstracting coverage domains from learned input representations or observable output behaviors (class label and prediction confidence), black-box criteria facilitate test adequacy analysis of DL models in scenarios where internal access is restricted or infeasible. Despite their growing advancements, existing methods remain largely confined to computer vision applications and classification tasks. There is a pressing need for the development of black-box criteria that can adapt to the unique challenges of input abstraction and output characterization across diverse application domains and tasks.
 $\blacktriangleleft$ 
\end{tcolorbox}
 
\subsection{Discussion}
\label{DiscussionCC}
This section provides a detailed analysis of some characteristics of existing coverage criteria. The `Training Inps’, `Hyperparameter-Free', `Percentage’, and `Task-Agnostic' columns in Tables \ref{Tab-CC4FNN}-\ref{Tab-BlackCC} highlight key differences among the criteria, offering a comparative overview.

\textbf{\textit{Training-Phase Knowledge}}. The decision-making logic of a DL model is inherently shaped by its training data. Accordingly, several coverage criteria leverage information acquired during the training phase to guide test adequacy evaluation. By analyzing structural behaviors observed during training, such as neuron activation patterns, output value ranges, and distributional characteristics, these criteria establish reference baselines that a test suite is expected to satisfy or surpass. Leveraging such prior knowledge enables evaluation of test adequacy within the context of the model’s learned patterns, thereby capturing the intricacies of its responses to diverse inputs. For instance, both $k$-Multisection Neuron Coverage and Neuron Boundary Coverage \cite{DBLP:conf/sigsoft/MaLLZG18} define coverage domains based on neuron outputs collected during training. Surprise Coverage \cite{DBLP:conf/icse/KimFY19, DBLP:conf/icse/KimY20, DBLP:journals/ese/GuoTH24} quantifies the novelty of test inputs by measuring deviations in activation traces from those observed during training.

\textbf{\textit{Percentage-Based Coverage Reporting}}. 
Many criteria adopt a percentage-based reporting format, providing a normalized and bounded scale (ranging from 0\% to 100\%) to quantify how thoroughly a test suite exercises a given coverage domain. This format facilitates intuitive interpretation and practical utility, enabling testers to set measurable targets (\textit{e.g.}, achieving 80\% coverage) and evaluate test generators relative to these targets.

Certain criteria, such as Top-$k$ Neuron Pattern Coverage \cite{DBLP:conf/sigsoft/MaLLZG18}, Cluster-based Coverage \cite{DBLP:conf/icml/OdenaOAG19}, Gradient Vector Coverage \cite{DBLP:journals/ijon/ParkCKK23}, and NeuraL Coverage \cite{DBLP:conf/icse/YuanPW23}, do not conform to this bounded format due to the absence of well-defined upper limits in their coverage domains. These criteria report coverage in absolute terms. For example, Top-$k$ Neuron Pattern Coverage counts the number of unique top-$k$ neuron activation patterns, while Cluster-based Coverage measures the number of output clusters identified during testing. Percentage-based reporting supports standardization, while value-based reporting offers greater expressiveness. By avoiding fixed bounds, such metrics can capture more nuanced aspects of model behavior, such as the diversity of activation patterns or responses to rare inputs. However, interpreting these results often requires domain-specific knowledge and a deeper understanding of the criterion’s design and intent. To enable fair comparisons across models and criteria, normalization strategies may be necessary to map raw values onto a comparable scale.

\textbf{\textit{Task-Agnostic}}. Given a collection of data $X$ labeled as $Y$, most DNNs can be categorized as discriminative and generative models. Discriminative models learn the conditional distribution $P(y \mid x)$ of $X$ and $Y$, mapping inputs to labels in tasks, including classification and regression. Generative models estimate the joint distribution $P(x, y)$, generating new data samples that resemble the training distribution.

Although a key advantage of many coverage criteria is their applicability across various learning tasks without the need for task-specific adaptations, such generality is not universally guaranteed. As current CGT research rarely targets generative models, `task-agnostic' in this paper refers to the applicability of a criterion to both classification and regression tasks. However, several existing criteria \cite{DBLP:conf/icse/KimFY19, DBLP:conf/icse/ByunR20, DBLP:journals/tosem/XieLWMGJL22, DBLP:journals/tosem/DolaDS23, DBLP:journals/infsof/BaiHHWXQY24} are task-specific, depending on class labels to define their coverage domains. For instance, NPC \cite{DBLP:journals/tosem/XieLWMGJL22} and Class-Critical Neuron Coverage \cite{DBLP:journals/infsof/BaiHHWXQY24} depend on class-specific training data to capture structural behaviors and quantify coverage at the class level. To generalize such criteria for regression tasks, one possible approach is to remove class dependencies and instead evaluate model behavior over the entire training distribution.

\textbf{\textit{Interpretation-Based Abstractions of DNN Internals}}. Due to the inherent black-box nature of DL models, traditional structural testing techniques based on control flow analysis are unsuitable for measuring coverage for DNNs. A central challenge lies in defining meaningful testing targets, particularly in identifying which internal components or behavioral patterns should be exercised by a test suite. To address this, interpretation-based techniques are adopted to abstract and reason about DNN internals. 

Causality-Aware Coverage \cite{DBLP:conf/icse/JiMYW23} 
applies causal inference to model the DNN as a Structural Causal Model, allowing test adequacy evaluation based on the extent to which causal relationships within the network are exercised. Besides, recent studies have leveraged Layer-wise Relevance Propagation (LRP) to identify critical neurons for coverage analysis. For instance, NPC \cite{DBLP:journals/tosem/XieLWMGJL22} and Shi et al. \cite{DBLP:journals/infsof/ShiYS25} utilize critical neurons to construct Critical Decision Paths, abstracting the internal information flows of DNNs in a manner analogous to control and data flows in traditional program analysis. DeepLID \cite{DBLP:journals/ese/GuoTH24} and IDC \cite{DBLP:conf/icse/GerasimouE0C20} define coverage domains based on outputs of critical neurons identified via LRP. Rather than merely capturing surface-level structural activation patterns, these criteria incorporate semantic importance to refine the coverage domain. By integrating interpretability into the testing process, these criteria enable the construction of more human-understandable abstractions of DNN behavior, bridging the gap between opaque internal mechanisms and testable logic.

\begin{tcolorbox}
    [colback=gray!10, rounded corners]
\ding{46} \textbf{Summary} $\blacktriangleright$ 
This section discusses existing coverage through four key perspectives. First, many utilize training-phase information, such as activation patterns or data distributions, to define or calibrate coverage objectives. Second, coverage is reported using either percentage-based metrics, which offer standardized comparison, or value-based metrics, which capture finer behavioral detail. Third, some criteria are task-agnostic and apply to both classification and regression without modification. Fourth, interpretation-based abstractions are used to make DNN internals more testable and semantically meaningful.$\blacktriangleleft$ 
\end{tcolorbox}

\section{Coverage-Guided Test Input Generation}
\label{Sec-generation}
Coverage-guided test input generation aims to generate inputs that reveal errors while maximizing test coverage. It is based on the assumption that more thoroughly exercising the model's internals increases the likelihood of exposing potential errors. Unlike adversarial attacks \cite{DBLP:journals/corr/GoodfellowSS14, DBLP:conf/iclr/KurakinGB17a, DBLP:journals/csur/ZhangYTY24}, which focus on generating adversarial examples that lead to incorrect model predictions, coverage-guided testing aims to broaden the model's behavioral coverage. These methods seek to explore rarely tested components/paths rather than only cause incorrect predictions. Among various test input generation approaches, coverage-guided fuzzing (CGF) has received significant attention in the DL testing community. Section \ref{Sec-CGF} details CGF, while Sections \ref{Sec-fal} to \ref{Sec-concolic}, cover other approaches, including coverage-guided falsification, combinatorial testing, and concolic testing.

\begin{figure}  
    \centering  
    \includegraphics[width = 15.3 cm]{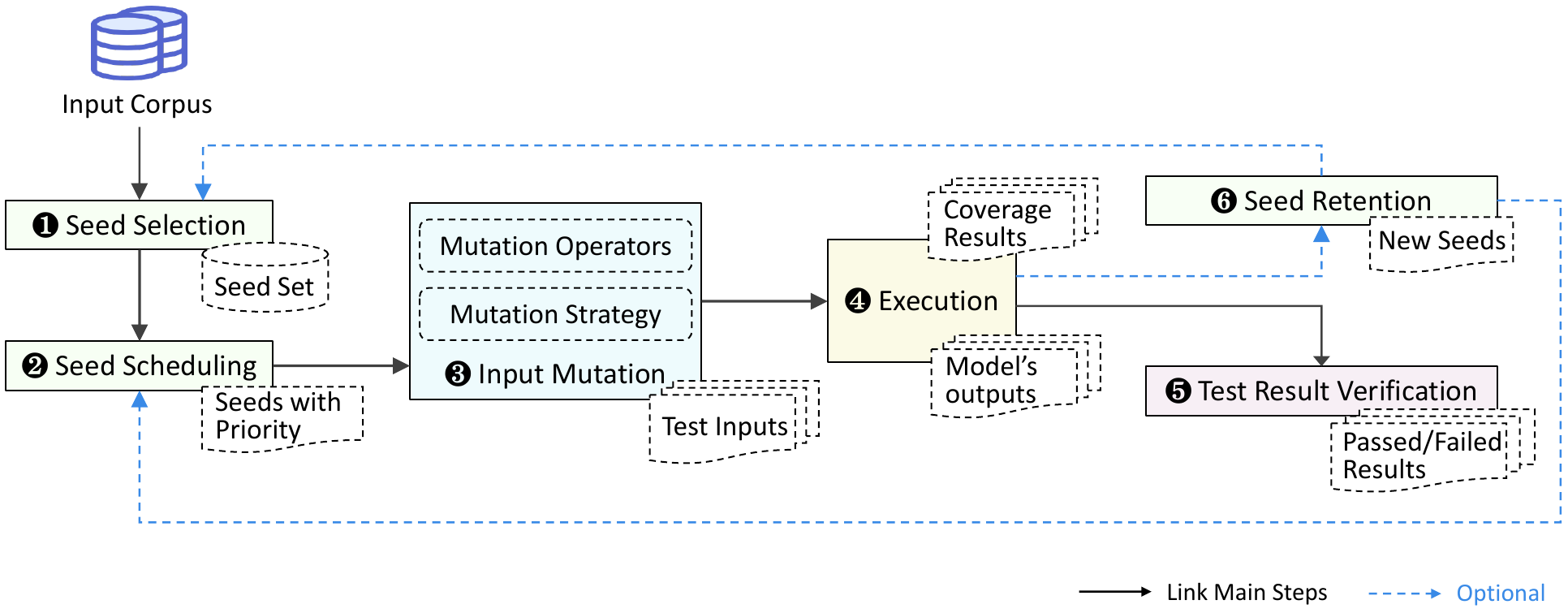} 
    \caption{Workflow of Coverage-Guided Fuzzing for DL Models}  
    \label{WorkflowCGF}
\end{figure}

\subsection{Coverage-Guided Fuzzing}
\label{Sec-CGF}
A typical fuzzer selects an initial seed input and mutates it to yield multiple new test inputs, and tests each of them, on and on. Figure \ref{WorkflowCGF} illustrates the workflow of the CGF. The CGF engine executes these steps iteratively, continuing the process until a user-defined budget is reached. Main steps are connected by solid black lines, and optional steps are connected by dashed blue lines. 

\begin{tcolorbox}
    [colback=gray!10, rounded corners]
\footnotesize
\ding{172}: \textbf{Seed Selection} samples initial test inputs from the input corpus as seeds, forming an optimal seed set.   \\
\ding{173}: \textbf{Seed Scheduling} determines the order in which seeds are processed. \\
\ding{174}: \textbf{Input Mutation} modifies seeds through mutation operators and strategies, generating new test inputs for execution. \\
\ding{175}: \textbf{Execution} is the phase where the mutated inputs are fed into the model under test, and the model’s outputs and runtime coverage are monitored. \\
\ding{176}: \textbf{Test Result Verification} validates the execution results to determine whether the generated inputs expose errors. \\
\ding{177}: \textbf{Seed Retention} evaluates whether newly generated test inputs should be retained as seeds for future fuzzing iterations.
\end{tcolorbox}

Coverage guidance could be incorporated at multiple stages of the fuzzing process. During seed scheduling, seeds with a higher potential to trigger new coverage are prioritized and allocated more `energy'. In the seed retention phase, newly generated inputs that increase coverage are retained for future mutations. While coverage is central to seed mechanisms, alternative strategies, such as class-aware seed selection, model confidence-based seed scheduling, may also be employed depending on the testing objective. In the input mutation phase, various strategies leverage coverage metrics to guide targeted exploration. 

Table \ref{Tab-CGF} summarizes CGF approaches for DL models, detailing key components including seed mechanisms, mutation strategies, and application domains. The symbols \ding{172}–\ding{180} represent application domains: \ding{172} for image classification, \ding{173} for autonomous driving, \ding{174} for text generation, \ding{175} for speech recognition, \ding{176} for video recognition, \ding{177} for malware detection, \ding{178} for RL tasks (e.g., Flappy Bird, Catcher, and Pong), \ding{179} for lipophilicity analysis, and \ding{180} for code summarization generation. The following abbreviations are used: Cov (Coverage), Xform (Transformation), Pred (Prediction), CovInc (Coverage Increase), Freq (Frequency), and DisCstr (Distance Constraint).

\subsubsection{Strategy for Seed Selection, Scheduling, and Retention}
CGF maintains a dynamic set of seeds, which is initially populated with a set of inputs for processing. The processes of seed selection, scheduling, and retention are interconnected within a feedback loop, where the execution performance of seeds in previous iterations influences their scheduling and retention in subsequent iterations. This dynamic mechanism continuously refines the seed set, steering the exploration towards maximizing model coverage and uncovering potential model errors.

\textbf{(1) Seed Selection.}
The computing resources available for testing are inherently limited, constrained by factors such as the number of seeds and the available execution time. The seed selection step samples seeds from a large-scale input corpus, and this subset serves as the basis for subsequent mutations \cite{DBLP:conf/uss/RebertCAFWGB14, DBLP:conf/ccs/KleesRCW018, DBLP:conf/issta/HerreraGMNPH21}. In traditional software testing, CGF conceptualizes this problem as corpus minimization \cite{DBLP:conf/issta/HerreraGMNPH21, DBLP:conf/uss/RebertCAFWGB14}, where the objective is to select the smallest set of seeds while disregarding redundant inputs that are not worth exploring. 

\textbf{\textit{Class-aware Strategies}.}
Most CGF studies initiate the seed set by randomly sampling inputs from the corpus, while some approaches adopt more refined strategies by incorporating class information. Specifically, GradFuzz \cite{DBLP:journals/ijon/ParkCKK23}, CriticalFuzz \cite{DBLP:journals/infsof/BaiHHWXQY24}, and the work proposed by Fabrice et al. \cite{DBLP:conf/sigsoft/Harel-CanadaWGG20} sample seed inputs from each class, ensuring that all classes are adequately represented in the seed set. DeepSmartFuzzer \cite{DBLP:conf/ijcai/DemirE020} utilizes a clustering algorithm to group similar inputs and randomly selects a cluster during each fuzzing cycle. 

\textbf{\textit{Hybrid Strategies}.}
Zhi et al. \cite{DBLP:journals/tosem/ZhiXSSZG24} introduce seed selection strategies optimizing three objectives: coverage increase, error detection, and robustness enhancement. Coverage guidance selects a minimal subset of seeds that can achieve the same level of coverage as the entire input corpus. For error detection, model uncertainty metrics such as Surprise Adequacy \cite{DBLP:conf/icse/KimFY19} and Prediction confidence Score \cite{DBLP:conf/icse/ZhangW0D0WDD20} evaluate the potential of seeds in revealing errors. To enhance robustness, seeds capable of triggering errors through loss gradients \cite{DBLP:conf/icse/WangCSMWSC21} are selected. Strategies include three single-objective and two multi-objective optimization (MOO) methods, balancing these objectives. Experiments show that optimizing coverage and error detection boosts fuzzing effectiveness. MOO-based selection achieves promising results in robustness enhancement.

\textbf{\textit{Input diversity-aware Strategies}.}
Some search-based testing studies \cite{DBLP:journals/tr/DaiSLZ24, DBLP:journals/tosem/FahmyPBS23, DBLP:conf/issta/ZohdinasabRGT21, DBLP:conf/kbse/RiccioHJT21}, though not explicitly coverage-guided, propose seed selection strategies to construct an initial population for subsequent mutations. DeepMetis \cite{DBLP:conf/kbse/RiccioHJT21} initializes the population by calculating Euclidean distances between bitmaps and greedily builds a diverse seed set. SEDE \cite{DBLP:journals/tosem/FahmyPBS23} evaluates an individual’s contribution to population diversity by measuring its distance from its nearest neighbor. Individuals with the highest fitness values are selected for further use. DFuzzer \cite{DBLP:journals/tr/DaiSLZ24} incrementally selects seeds from the input corpus that are `far away’ from each other, as quantified by difference measurement metrics such as information-based or feature-based difference measurement. Overall, these strategies employ various distance metrics to iteratively construct the seed set with diverse inputs. They can be integrated into the CGF framework to facilitate broader exploration of the input space. 

\textbf{(2) Seed Scheduling.}
Seed scheduling determines the optimal order in which seeds undergo mutations. Recent studies have proposed various scheduling strategies to ensure that seeds with a higher potential to increase model coverage or reveal errors are given higher priority and more resources for further mutation. 

\textbf{\textit{Strategies based on the historical execution of seeds}.}
Historical execution information of inputs serves as a heuristic for seed prioritization, reflecting each seed’s utility in guiding fuzzing, based on factors such as selection frequency, contribution to coverage, and time since inclusion in the seed set. Specifically, recency-aware strategies favor recently used seeds that triggered new coverage \cite{DBLP:conf/icml/OdenaOAG19, DBLP:journals/tse/ZhangRDD22}.
Frequency-aware strategies consider how often seeds are selected during fuzzing. For example, DeepHunter \cite{DBLP:conf/issta/XieMJXCLZLYS19} prioritizes less-fuzzed seeds while ensuring a minimum selection probability for frequently used ones. CriticalFuzz \cite{DBLP:journals/infsof/BaiHHWXQY24} balances the potential for coverage increase with selection frequency to avoid overusing specific seeds. DeepController \cite{DBLP:conf/seke/DaiSL22} incorporates coverage triggering, selection frequency, and the time of a seed added to the seed set into a scheduling strategy. Besides, for DRLs, DRLFuzz \cite{DBLP:journals/jss/WanLLCZ24} further incorporates reward signals, prioritizing seeds that not only lead to new coverage but also result in low episodic rewards in prior iterations.

\textbf{\textit{Strategies based on model prediction confidence}.} Several approaches \cite{DBLP:journals/infsof/TaoTGH023, DBLP:journals/apin/SunLW23, LIU2024107640} have explored model prediction probabilities to prioritize inputs with high potential for error detection. For instance, DLRegion \cite{DBLP:journals/infsof/TaoTGH023} ranks seeds based on model-predicted maximum probabilities, where lower confidence predictions suggest a higher likelihood of uncovering unexpected behaviors. DeepMC \cite{DBLP:journals/apin/SunLW23} uses the median misclassification probability as a threshold to identify inputs more likely to induce prediction errors. In addition, DeepRTest \cite{DBLP:conf/qrs/YangYW22} targets inputs near decision boundaries by generating mutant models and selecting those that produce divergent classifications from the original model, followed by perturbation to push them across the boundary.

\textbf{(3) Seed Retention.}
Seed retention determines whether newly generated inputs should be preserved as seeds, ensuring valuable inputs are maintained for future mutation cycles. Most approaches use coverage increase as the primary retention criterion, guiding fuzzing toward maximizing model coverage. Although seed retention is optional, runtime information associated with seeds (\textit{e.g.}, coverage increase, time of addition to the seed set) is leveraged to adjust mutation priorities and update the seed scheduling strategy accordingly.

\textbf{\textit{Strategies based on coverage updates}.}
Coverage-aware seed retention strategies (denoted as `CovInc' in the `Retention’ column) follow two principles. One retains an input if it increases coverage by activating previously unseen internal components \cite{DBLP:conf/sigsoft/GuoJZCS18, DBLP:conf/issta/LeeCLO20, DBLP:journals/infsof/TaoTGH023, DBLP:conf/issta/XieMJXCLZLYS19, DBLP:conf/icse/YuanPW23, DBLP:journals/tse/ZhangRDD22, DBLP:conf/icse/TianPJR18, DBLP:conf/sigsoft/DuXLM0Z19, DBLP:conf/seke/DaiSL22, DBLP:journals/infsof/BaiHHWXQY24}. The other does so if the input exhibits sufficient dissimilarity in the coverage space relative to existing seeds. The symbol \textcolor{blue}{$\dagger$} indicates that coverage updates are assessed based on coverage-space similarity, where inputs are retained if their coverage vectors differ significantly from those of existing seeds \cite{DBLP:journals/ijon/ParkCKK23, DBLP:journals/jss/WanLLCZ24, DBLP:conf/icml/OdenaOAG19}. For example, TensorFuzz \cite{DBLP:conf/icml/OdenaOAG19} uses an Approximate Nearest Neighbor search to retain inputs that exceed a distance threshold from their closest seed. 

\textbf{\textit{Strategies based on semantic constraints}.}
Semantic distance constraints (denoted as `DisCstr' in the `Retention’ column) are often integrated with coverage-based retention, especially in computer vision tasks \cite{DBLP:conf/sigsoft/GuoJZCS18, DBLP:conf/issta/LeeCLO20, DBLP:journals/infsof/TaoTGH023, DBLP:conf/issta/XieMJXCLZLYS19, DBLP:journals/tse/ZhangRDD22, DBLP:conf/kbse/MissaouiGM23}. A mutated input is retained only if its distance from the original seed remains within a predefined threshold, ensuring that it remains visually coherent and semantically consistent. 

\begin{table}
\begin{threeparttable}
   \caption{Summary of Coverage-Guided Fuzzing for DL Models}
  \small
  \label{Tab-CGF} 
  \fontsize{6.3}{8.6}\selectfont
  \setlength{\tabcolsep}{0.03 mm}{
  \begin{tabular}{lc|cll|ccc|c}
    \toprule
\multirow{2}{*} {\textbf{Year}} & \multirow{2}{*} {\textbf{Study}}  & \multicolumn{3}{c|}{\textbf{Mutation Strategy}}  &  \multicolumn{3}{c|}{\textbf{Seed Strategy}}  &   \multirow{2}{*} {\textbf{Domain}}  \\ 
\cmidrule(lr) {3-5} \cmidrule(lr) {6-8}

&& \multicolumn{1}{c}{\textbf{Info.} \customtnote{*} } & \textbf{Core Idea} & \textbf{Solution} & \multicolumn{1}{c}{\textbf{Selection}} & \multicolumn{1}{c}{\textbf{Scheduling}} & \multicolumn{1}{c|}{\textbf{Retention}} &  \\   \toprule

2017 &  DeepXplore \cite{DBLP:conf/sosp/PeiCYJ17} &  \scalebox{0.80}{\sqbox{gray}} & Joint Opt (Different Behaviors + Neuron Cov) & Gradient & Random & \scriptsize / & \tiny \ding{55}& \scriptsize \ding{172} \ding{173} \ding{177} \\ \midrule

2018 &  DLFuzz \cite{DBLP:conf/sigsoft/GuoJZCS18} & \scalebox{1.3} {\sqboxEmpty{black}}& Joint Opt (Pred Uncertainty +  Neuron Cov)&  Gradient & Random & \scriptsize / & CovInc \& DisCstr & \scriptsize \ding{172} \\  \midrule

2020 &  ADAPT \cite{DBLP:conf/issta/LeeCLO20} & \scalebox{1.3} {\sqboxEmpty{black}}& Opt (Adaptive Neuron Cov) &  Gradient  & Random & \scriptsize / & CovInc \& DisCstr &  \scriptsize \ding{172} \\ \midrule

2020 &  Fabrice et al. \cite{DBLP:conf/sigsoft/Harel-CanadaWGG20} &  \scalebox{0.80}{\sqbox{gray}} & Joint Opt (CW Attack + Neuron Output Diverg.) &  Gradient  & Class-Aware & \scriptsize / & \tiny \ding{55}  & \scriptsize\ding{172} \ding{173} \\ \midrule

2021 &  DeepCon \cite{DBLP:conf/wcre/ZhouDLZ0Y21} & \scalebox{1.3} {\sqboxEmpty{black}}& Joint Opt (Neuron Contribution + Neuron Output) &  Gradient  & \scriptsize / & \scriptsize / & \tiny  \ding{55}& \scriptsize \ding{172} \\ \midrule

2021 &  Dola et al. \cite{DBLP:conf/icse/DolaDS21} &  \scalebox{0.80}{\sqbox{gray}} & Joint Opt (Baseline Obj +Reconstruction Probability) &  Gradient  & Random & \scriptsize / & \tiny  \ding{55}&  \scriptsize \ding{172} \\ \midrule

2022 & DeepRTest \cite{DBLP:conf/qrs/YangYW22} &  \scalebox{0.80}{\sqbox{gray}} & Joint Opt (Pred Uncertainty + Neuron Cov) & Gradient & \scriptsize / & Classification Boundary & CovInc & \scriptsize \ding{172} \\ \midrule

2022 & RNN-Test \cite{DBLP:journals/tse/GuoZZSJS22} &  \scalebox{0.80}{\sqbox{gray}} \textcolor{orange}{\textbf{\normalsize \textregistered}} & Joint Opt (State Inconsistency + State Cov) & Gradient & \scriptsize / & \scriptsize / & \tiny \ding{55} & \scriptsize \ding{172} \ding{174} \ding{175} \\ \midrule

2023 & BTM \cite{DBLP:conf/qrs/ZhangZWCH23} & \scalebox{0.80}{\sqbox{gray}} & Joint Opt (Different Behaviors + Neuron Cov) & Gradient & Random & \scriptsize / & \tiny \ding{55} &  \scriptsize \ding{172} \\  \midrule

2023 & DLRegion \cite{DBLP:journals/infsof/TaoTGH023} &  \scalebox{0.80}{\sqbox{gray}} & Joint Opt (Pred Uncertainty + Region Cov) &  Gradient & Random & Pred Probability & CovInc \& DisCstr & \scriptsize \ding{172} \\  \midrule

2023 & DeepMC \cite{DBLP:journals/apin/SunLW23} &  \scalebox{0.80}{\sqbox{gray}} & Joint Opt (Pred Uncertainty + Neuron Cov) &  Gradient & \scriptsize / &  Pred Probability &\tiny \ding{55} & \scriptsize \ding{172} \\  \midrule

2023 & Test4Deep \cite{DBLP:journals/tnn/YuDY23} &  \scalebox{0.80}{\sqbox{gray}} & Joint Opt (Pred Inconsistency + Neuron Cov) &  Gradient & Random & \scriptsize / & \tiny \ding{55} & \scriptsize \ding{172} \ding{177} \\  \midrule

2024 &  DRLFuzz \cite{DBLP:journals/jss/WanLLCZ24} & \scalebox{0.80}{\sqbox{gray}} \textcolor{magenta}{\textbf{\scriptsize  \textsection}} & Opt (Differences of Q-values between States)& Gradient  & \scriptsize / & CovInc, Episodic Reward & \textcolor{blue}{\textbf{\scriptsize $\dagger$}} CovInc & \scriptsize \ding{178} \\
 \midrule 
 
 2024 & DeepCNP \cite{LIU2024107640} &  \scalebox{0.80}{\sqbox{gray}} & Joint Opt (Critical Neuron Paths Alighment) &  Gradient & Random & Pred Probability & \tiny \ding{55} & \scriptsize \ding{172} \ding{177} \\  \midrule

2024 & Themis \cite{DBLP:conf/issre/0005LXC24} &  \scalebox{0.80}{\sqbox{gray}} & Opt (Sensitivity of Neurons to Input Perturbations) &  Gradient & \scriptsize / &  \scriptsize / & \tiny \ding{55} & \scriptsize \ding{172} \ding{177} \\  \bottomrule

2018 &  DeepCruiser \cite{DBLP:journals/corr/abs-1812-05339} &  \centering \scalebox{0.7} \blackdiamond  \textcolor{orange}{\textbf{\normalsize \textregistered}} & Metamorphic Mutation & Audio Signal Xform  & Random & \scriptsize / & CovInc & \scriptsize \ding{175} \\ \midrule

2019 &  TensorFuzz \cite{DBLP:conf/icml/OdenaOAG19} & \centering \scalebox{0.7} \blackdiamond & Metamorphic Mutation &  Image Xform  & Random & Recently CovInc & \textcolor{blue}{\textbf{\scriptsize $\dagger$}} CovInc & \scriptsize \ding{172} \\ \midrule

2019 &  DeepHunter \cite{DBLP:conf/issta/XieMJXCLZLYS19} & \centering \scalebox{0.7} \blackdiamond & Metamorphic Mutation & Image Xform   & Random & Selection Freq & CovInc \& DisCstr & \scriptsize \ding{172} \\ \midrule

2019 &  DeepStellar \cite{DBLP:conf/sigsoft/DuXLM0Z19} &  \centering \scalebox{0.7} \blackdiamond \textcolor{orange}{\textbf{\normalsize \textregistered}} & Metamorphic Mutation & Image, Audio Signal Xform  & Random & \scriptsize / & CovInc & \scriptsize \ding{172} \ding{175} \\ \midrule

\multirow{2}{*} {2022} &  \multirow{2}{*} {DeepController \cite{DBLP:conf/seke/DaiSL22}} & \multirow{2}{*} {\centering \scalebox{0.70}  \blackdiamond}  & \multirow{2}{*} {Metamorphic Mutation} & \multirow{2}{*} {Image Xform}  & \multirow{2}{*} {Random} & CovInc, Selection & \multirow{2}{*} {CovInc} &  \multirow{2}{*} {\scriptsize \ding{172}}  \\ 
&&&&&& Freq, Retention Recency&& \\  \midrule

2023 & GradFuzz \cite{DBLP:journals/ijon/ParkCKK23} & \centering \scalebox{0.70}  \blackdiamond  & Metamorphic Mutation & Image Xform &  Class-Aware & Uniform Class Sampling & \textcolor{blue}{\textbf{\scriptsize $\dagger$}} CovInc&  \scriptsize \ding{172}  \\ \midrule

2024 &  CriticalFuzz \cite{DBLP:journals/infsof/BaiHHWXQY24} & \centering \scalebox{0.70}   \blackdiamond & Metamorphic Mutation & Image Xform & Class-Aware & CovInc, Selection Freq & CovInc &  \scriptsize  \ding{172}\\  \bottomrule

2018 &  DeepTest \cite{DBLP:conf/icse/TianPJR18} &  \scalebox{0.80}{\sqbox{gray}} & Search (Maximize Cov) &  GS with Image Xform & Random & \scriptsize / &\tiny \ding{55}  & \scriptsize \ding{173} \\ \midrule

2019 &  DeepEvolution \cite{DBLP:conf/icsm/BraiekK19} &  \scalebox{0.80}{\sqbox{gray}}   & Search (Fitness by Neuron Cov)  &  SM with Image Xform & Random & \scriptsize / & \tiny \ding{55}& \scriptsize \ding{172} \\ \midrule

2020 &  SENSEI \cite{DBLP:conf/icse/GaoSPR20} &  \scalebox{0.80}{\sqbox{gray}}   & Search (Fitness by Loss or Cov)  & GA with Image Xform & Random & \scriptsize / & \tiny \ding{55}& \scriptsize \ding{172} \\ \midrule

2020 &  DeepSmartFuzzer \cite{DBLP:conf/ijcai/DemirE020} &  \scalebox{0.80}{\sqbox{gray}} & Search (Fitness by Cov) &  MCTS with Image Xform & Random, Clustering & \scriptsize / & \tiny \ding{55} & \scriptsize \ding{172} \\ \midrule

2021 &  TACTIC \cite{DBLP:conf/icml/LiPZL21} &   \scalebox{0.80}{\sqbox{gray}}  & Search (Fitness by Erroneous Behavior and Cov)  & ES with MUNIT & \scriptsize / & \scriptsize / & \tiny \ding{55}& \scriptsize \ding{173} \\ \midrule

2021 &  GANC \cite{DBLP:journals/asc/Al-NimaHACW21} &   \scalebox{0.80}{\sqbox{gray}} \textcolor{magenta}{\textbf{\scriptsize  \textsection}}  & Search (Fitness by Neuron Cov)  & GA with Image Xform & \scriptsize / & \scriptsize / & \tiny \ding{55}& \scriptsize \ding{173} \\ \midrule

2021 &  FilterFuzz \cite{DBLP:conf/qrs/WeiC21} &   \scalebox{0.80}{\sqbox{gray}} & Search (Fitness by Filter Cov)  & GA with Image Xform & \scriptsize / & \scriptsize / & \tiny  \ding{55}& \scriptsize \ding{173} \\ \midrule

2022 &  TESTRNN \cite{DBLP:journals/tr/HuangSZSRMH22} &   \scalebox{0.80}{\sqbox{gray}} \textcolor{orange}{\textbf{\normalsize \textregistered}} & Search (Fitness by LSTM Cov)  & GA with Domain Xform & Random & \scriptsize / & \tiny \ding{55}&  \scriptsize  \ding{172} \ding{173} \ding{176} \ding{179}\\ \midrule

2022&  DiverGet \cite{DBLP:journals/ese/YahmedBKBZ22} &   \scalebox{0.80}{\sqbox{gray}}  & Search (Fitness by Cov Divergences)  & SM/GA with Image Xform & Random & \scriptsize / & \tiny  \ding{55}& \scriptsize \ding{172} \\ \midrule 

2023 &  CoCoFuzzing \cite{DBLP:journals/tr/WeiHYWW23} &   \scalebox{0.80}{\sqbox{gray}} & Search (Cov Improvement Heuristic) & Program Mutation Operators & \scriptsize /  & Random & \tiny \ding{55} & \scriptsize  \ding{180} \\ \midrule

2024 &  DEEPWALK \cite{DBLP:journals/tse/YuanPW24} & \centering  \scalebox{0.80}{\sqbox{gray}} & Search (Cov-Guided Latent Exploration) &  Footprint-Aware Mutation & Random & Selection Freq & CovInc & \scriptsize \ding{172} \\ \bottomrule

2022 &  CAGFuzz \cite{DBLP:journals/tse/ZhangRDD22} & \centering \scalebox{0.7} \blackdiamond & GAN & CycleGAN & \scriptsize / & Recently CovInc & CovInc \& DisCstr & \scriptsize \ding{172} \\ \midrule

2022 &  FuzzGAN \cite{DBLP:conf/hpcc/HanLTHG22} &  \scalebox{0.80}{\sqbox{gray}} & GAN & ACGAN with Cov Guidance  & \tiny \ding{55}& \tiny \ding{55}& \tiny \ding{55}& \scriptsize \ding{172} \\ \midrule

2023 &  GENFUZZER \cite{DBLP:conf/kbse/MissaouiGM23} & \centering \scalebox{0.7} \blackdiamond & Generative Models &  Text-Conditional Generative & Uniform Sampling & \scriptsize / & CovInc \& DisCstr & \scriptsize \ding{172} \\  \bottomrule
 \end{tabular}}
 
\begin{tablenotes}
\scriptsize
  \item[\customtnote{*}] It reflects the tester's knowledge of the DNN model during input mutation. \scalebox{1.5}{\sqboxEmpty{black}}: white-box. \scalebox{0.9}{\blackdiamond}: black-box. \scalebox{0.85}{\sqbox{gray}}: gray-box. The symbol \textcolor{orange}{\textbf{\normalsize \textregistered}} indicates that the target network architecture is RNN, while \textcolor{magenta}{\textbf{\scriptsize \textsection}} denotes that the target is DRL. Approaches without any symbol are designed for FNNs. 
  \item[] The symbol \ding{55} denotes a lack of support, whereas $/$ signifies support, though the specific strategy is not discussed in the literature.
  
\end{tablenotes}
\end{threeparttable}
\end{table}

\subsubsection{Strategy for Input Mutation}
Input mutation strategies of CGF fall into four categories: gradient-based optimization, metamorphic mutation, search-based mutation, and generative model-based mutation.

\textbf{(1) Gradient-Based Optimization.}
Gradient-based optimization optimizes an objective function by iteratively perturbing seeds in the direction guided by the gradient of the objective. The objective function quantifies how well an input performs relative to testing targets. Recent studies have designed optimization objectives using diverse testing metrics, such as coverage criteria and model uncertainty measures. Formally, let $J$ denote the objective function, formulated based on a set of testing metrics $M=\left\{ m_{1}, m_{2},...,m_{k}\right\}$. Given a seed $x$, the goal is to find a set of mutated inputs $T= \left\{ t_{1},t_{2},...,t_{n}\right\}$ that maximizes $J(T)$:
\begin{equation}
max \quad J(T)= \bigcup_{t_{j}\in T}^{}\sum _{i=1}^{k}\omega _{i} \cdot O\left ( m_{i},t_{j} \right ).
\end{equation}
$O\left ( m_{i},t_{j} \right )$ denotes evaluation of input $t_{j}$ with respect to the testing metric $m_{i}$, and $\omega _{i}$ is the weight assigned to $m_{i}$. The mutation process is performed iteratively for each seed until a stopping criterion is met (\textit{i.e.}, a maximum number of iterations, time limit, or a predefined threshold for the objective function). In each iteration, gradients are used as perturbations, progressively steering the seed toward maximizing the objective function.

Despite shared reliance on the gradient-based optimization framework, each approach tailors its objectives to the specific network architectures and testing goals. According to the number of metrics used in the objective function, recent strategies can be categorized into joint optimization and single-objective optimization. In Table \ref{Tab-CGF}, we refer to joint optimization strategies as `Joint Opt' and single-objective strategies as `Opt'. Furthermore, gradient-based optimization strategies can be classified into white-box and gray-box approaches based on the extent of model information utilized in their design objectives. 

\textbf{\textit{Joint Optimization}.} Joint optimization integrates multiple testing objectives into a unified function. Recent studies have settled on a common formulation that jointly optimizes for test coverage and the induction of erroneous behaviors. For example, DeepXplore \cite{DBLP:conf/sosp/PeiCYJ17}, maximizes Neuron Coverage while inducing differential behaviors across multiple models. To improve coverage, it iteratively selects inactive neurons and adjusts seeds to ensure that neurons' outputs exceed an activation threshold. To detect errors, it perturbs inputs such that at least one model's prediction diverges from those of the others. DLFuzz \cite{DBLP:conf/sigsoft/GuoJZCS18} maximizes Neuron Coverage and misclassification likelihood via targeted perturbations guided by heuristics: frequently or rarely activated neurons, neurons with large weights, and neurons near activation thresholds. DLRegion \cite{DBLP:journals/infsof/TaoTGH023} refines neuron selection by partitioning neurons into regions based on output distributions learned during training. For each input, the region with the most active neurons is chosen, and neurons are selected based on random sampling, least frequent activation, or most frequent activation in prior tests. Dola et al. \cite{DBLP:conf/icse/DolaDS21} adopt a probabilistic encoder-decoder framework that samples from the latent space to generate inputs aligned with the training data distribution. They enhance DeepXplore's joint objective \cite{DBLP:conf/sosp/PeiCYJ17} by incorporating probability density from a generative model, enabling the generation of semantically valid yet error-inducing inputs. BTM \cite{DBLP:conf/qrs/ZhangZWCH23} introduces a framework based on meta noise training and adaptive transformation. It first performs joint optimization to train a universal perturbation that misleads one surrogate model, preserves correct predictions on others, and maximizes Neuron Coverage. During input transformation, a gradient-free method iteratively updates the image, guided by prediction confidences to target decision boundaries most likely to yield misclassifications. RNN-Test \cite{DBLP:journals/tse/GuoZZSJS22} jointly optimizes state inconsistency and state coverage. The former highlights discrepancies in internal state transitions, while the latter ensures a diverse exploration of temporal behaviors through the activation of previously unseen states.

Focusing on robustness evaluation, RobOT \cite{DBLP:conf/icse/WangCSMWSC21} introduces two robustness metrics, Zero-Order Loss and First-Order Loss, to direct the fuzzing process. It formulates a joint optimization objective that combines prediction uncertainty with robustness metrics to capture input perturbations. Building on this framework, QUOTE \cite{DBLP:journals/tosem/ChenWMSSZC23} extends the application to fairness testing by generating inputs that reveal biased or inconsistent outputs across demographic groups. Although neither approach follows a CGF paradigm, both illustrate the flexibility of joint optimization strategies in supporting testing objectives beyond structural coverage.

\textbf{\textit{Single-objective optimization}.} Single-objective optimization aims to optimize a single testing metric. ADAPT \cite{DBLP:conf/issta/LeeCLO20} focuses on maximizing Neuron Coverage, by introducing 29 atomic features to characterize neurons, such as those in dense layers, with top 10\% weights, or never activated. Based on these features, it introduces a scoring mechanism and proposes adaptive neuron selection strategies to prioritize promising neurons. Gradients are computed with respect to the selected neurons. As ADAPT relies on internal weights, it is classified as a white-box fuzzer. Besides, Themis \cite{DBLP:conf/issre/0005LXC24} performs gradient ascent over the input space to find an input that maximizes the sensitivity of neurons to input perturbations. It aims to promote convergence of the sensitivity distribution toward the desired coverage level, thereby triggering erroneous behaviors.

Targeting DRL models, DRLFuzz \cite{DBLP:journals/jss/WanLLCZ24} employs a single-objective strategy by leveraging the temporal difference error as a mutation guide. It maximizes the discrepancy between predicted and target Q-values for a given state, thereby encouraging test inputs that drive the agent toward unexplored regions of the state space. 

\textbf{(2) Metamorphic Mutation.} Metamorphic mutation strategies apply domain-specific transformations to seeds, generating inputs that are syntactically valid and semantically consistent with seeds. The term `metamorphic’ originates from metamorphic testing, which constructs test oracles based on the expectation that semantically equivalent inputs should yield consistent model outputs \cite{DBLP:conf/issta/XieMJXCLZLYS19, DBLP:conf/icsm/BraiekK19}. These strategies are black-box, as they do not use coverage metrics during input mutation but incorporate coverage signals during seed selection and retention.

Among these, image-level mutations have been the most extensively researched. In computer vision tasks, various image transformations have been used as mutation operators \cite{DBLP:conf/icse/TianPJR18, DBLP:conf/icml/OdenaOAG19, DBLP:conf/issta/XieMJXCLZLYS19, DBLP:conf/seke/DaiSL22, DBLP:journals/ijon/ParkCKK23, DBLP:conf/icse/YuanPW23, DBLP:journals/infsof/BaiHHWXQY24}, including linear transformations (brightness and contrast adjustment), affine transformations (translation, scaling, shearing, and rotation), and convolutional transformations (blurring, weather effects such as fog or rain). Such image transformations are employed to simulate diverse real-world environmental conditions. In addition, DeepTest \cite{DBLP:conf/icse/TianPJR18}, designed for testing autonomous driving systems, introduces weather effects such as fog and rain to perform robustness evaluation under diverse operational conditions. DiverGet \cite{DBLP:journals/ese/YahmedBKBZ22} targets Hyperspectral image classification by applying pixel-level transformations such as salt-and-pepper noise, dropout, and striping effects, each tailored to hyperspectral image properties. DeepHunter \cite{DBLP:conf/issta/XieMJXCLZLYS19} employs conservative transformation parameters and constrains perturbations using both $L_0$ and $L_{\infty}$ norms, limiting the number of modified pixels and the maximum per-pixel deviation, respectively.

Initial progress has been made in developing mutation strategies for other input modalities, including natural language text, audio, source code, and executable binaries. In textual domains, TESTRNN \cite{DBLP:journals/tr/HuangSZSRMH22} applies four semantic-preserving operators: synonym replacement, random insertion, swapping, and deletion. In audio recognition, signal-level perturbations such as white noise injection, speed variation, and volume adjustment simulate realistic acoustic fluctuations while maintaining semantic integrity \cite{DBLP:journals/corr/abs-1812-05339, DBLP:conf/sigsoft/DuXLM0Z19}. For structured inputs like source code or executable binaries, syntax- or format-preserving modifications ensure functional validity. For example, CoCoFuzzing \cite{DBLP:journals/corr/abs-1812-05339} mutates programs using syntax-preserving transformations such as inserting unreachable control flow statements (\textit{e.g.}, if-else, for, while), renaming variables, and duplicating statements. 

\textbf{(3) Search-Based Mutation.} 
Search-based mutation employs heuristic algorithms to generate inputs that meet coverage objectives. These strategies are typically gray-box, utilizing intermediate outputs rather than internal weights or gradients. Based on their search mechanisms, existing strategies can be categorized into four types: population-based search, tree-based search, feedback-driven walk, and greedy search. The first three share a common principle, leveraging coverage metrics to guide mutations, while greedy search adopts a simpler strategy, directly selecting mutated inputs that fulfill coverage criteria.

\textbf{\textit{Population-based search}.} 
Population-based methods are inspired by biological evolution and swarm intelligence, using a group of candidate solutions to explore the input space. Through iterative cycles involving selection, crossover, and mutation, the population evolves based on fitness functions, defined by coverage metrics.

DeepEvolution \cite{DBLP:conf/icsm/BraiekK19} applies Swarm (SM)-based metaheuristics to evolve an initial set of candidate image transformations. Its fitness function combines local Neuron Coverage (neurons uniquely activated by mutated inputs compared to their seeds) and global Neuron Coverage (neurons not yet activated by any prior inputs). TESTRNN \cite{DBLP:journals/tr/HuangSZSRMH22} targets LSTM architectures by evolving input sequences through crossover and mutation, to satisfy boundary, structural, and temporal coverage targets. DiverGet \cite{DBLP:journals/ese/YahmedBKBZ22} directs mutations that reveal behavioral discrepancies between an original DNN and its quantized variant. The fitness function focuses on maximizing the difference in $K$-Multisection Neuron Coverage \cite{DBLP:conf/sigsoft/MaLLZG18} between two model variants.

Rather than directly mutating inputs, TACTIC \cite{DBLP:conf/icml/LiPZL21} explores a latent semantic space learned by generative models. Designed for autonomous driving, it employs the Multimodal Unsupervised Image-to-Image Translation (MUNIT) model \cite{DBLP:conf/eccv/HuangLBK18} to encode environmental conditions as high-dimensional style vectors. An Evolutionary Strategy (ES) then searches this space to identify styles that maximize steering angle deviations and enhance neuron-based coverage \cite{DBLP:conf/sosp/PeiCYJ17, DBLP:conf/sigsoft/MaLLZG18}. The resulting styles are used to generate realistic, safety-critical testing scenarios.

\textbf{\textit{Tree-based search}.}
Tree-based search incrementally explores the input or mutation space through branching decisions, with reward signals accumulated during search. DeepSmartFuzzer \cite{DBLP:conf/ijcai/DemirE020} models input mutation as a sequential decision-making process. It adopts a two-stage mutation: first selecting a region, then choosing an operator (pixel-level and image-level). Monte Carlo Tree Search (MCTS) \cite{DBLP:journals/tciaig/BrownePWLCRTPSC12, doi:10.1142/MCTS} is used to build a game tree, where each node represents the current input being mutated, and an edge corresponds to a mutation action. The MCTS process involves selecting a node, expanding the tree by adding new nodes for potential mutation actions, simulating the outcomes of these actions, and backpropagating the results to update rewards of mutation actions. Guided by feedback from coverage metrics, MCTS prioritizes mutation paths that yield higher coverage, steering the search toward mutation actions that maximize the exploration of the model's behavior. 

\textbf{\textit{Feedback-driven walk}.}
Feedback-driven walk explores the search space by iteratively adjusting mutation directions based on coverage feedback. DeepWalk \cite{DBLP:journals/tse/YuanPW24} employs a latent-space-centric strategy. Utilizing Generative Adversarial Network (GAN) inversion, it maps seed inputs onto class-specific latent manifolds and restricts mutations to these continuous spaces. It performs stepwise random walks, dynamically adjusting the direction and step size based on feedback from prior coverage improvements. A key innovation of DeepWalk is maintaining a footprint-aware covariance matrix that adaptively guides mutations along directions on the data manifold most likely to increase test coverage, balancing exploitation of learned behaviors and exploration of new ones.

\textbf{\textit{Greedy search}.} 
Some approaches rely on simple heuristics rather than explicit fitness-driven search. DeepTest \cite{DBLP:conf/icse/TianPJR18} adopts a greedy search to find image transformation combinations that improve coverage. Starting from a seed, it randomly applies two transformations, increasing their selection probability in subsequent iterations if the mutation leads to new coverage. CoCoFuzzing \cite{DBLP:journals/tr/WeiHYWW23} targets neural code models for code summarization tasks. It employs program mutation operators to modify seed programs, treating the resulting mutated programs as the search space. The primary objective is to discover mutated programs that increase model coverage.

\textbf{(4) Generative Model-Based Mutation.}
Recent advances in generative models, such as Generative Adversarial Networks (GANs) and Variational Autoencoders (VAEs), have enabled more effective CGF approaches for DL models. These leverage various generative networks to produce realistic inputs that resemble real data distributions while introducing subtle perturbations. Current test input generators \cite{DBLP:journals/tse/ZhangRDD22, DBLP:conf/kbse/MissaouiGM23, DBLP:conf/hpcc/HanLTHG22} can be compared along three core dimensions of the fuzzing pipeline: input mutation, semantic preservation, and coverage feedback.

Regarding input mutation, CAGFuzz \cite{DBLP:journals/tse/ZhangRDD22} and GENFUZZER \cite{DBLP:conf/kbse/MissaouiGM23} employ generative models, CycleGAN \cite{DBLP:conf/iccv/ZhuPIE17} and text-conditional inpainting diffusion, respectively, to synthesize semantically valid yet diverse inputs. FuzzGAN \cite{DBLP:conf/hpcc/HanLTHG22} injects neuron-level feedback into the generative loop, combining noise, class labels, and selected neuron locations to steer input mutations. Regarding semantic preservation, CAGFuzz and GENFUZZER rely on perceptual similarity metrics, such as cosine similarity in feature space, FID \cite{DBLP:conf/nips/HeuselRUNH17}, or CLIP \cite{DBLP:conf/icml/RadfordKHRGASAM21}. FuzzGAN ensures semantic preservation by training the generator to synthesize realistic inputs that maintain valid class labels. For coverage feedback, all three approaches incorporate neuron-level metrics to guide input selection. CAGFuzz and GENFUZZER evaluate newly generated inputs using standard Neuron Coverage \cite{DBLP:conf/sosp/PeiCYJ17} or Surprise Adequacy \cite{DBLP:conf/icse/KimFY19} and retain only those with incremental gains. FuzzGAN tracks coverage dynamically during mutation and focuses on uncovering neurons that are not activated in the previous mutations. 

\subsubsection{Test Result Verification}
In traditional software testing, test result verification compares program outputs against expected results in response to test inputs. In the context of DL models, this process presents unique challenges due to their non-deterministic nature \cite{DBLP:journals/tse/ZhangHML22, DBLP:conf/kbse/WangS20}. To address the challenges arising from the lack of explicit test oracles, metamorphic testing and differential testing have emerged as effective approaches, supporting tasks such as robustness testing \cite{DBLP:conf/issta/XieMJXCLZLYS19, DBLP:conf/icse/TianPJR18}, fairness testing \cite{DBLP:conf/ijcai/0004WL20}, quantization assessment \cite{DBLP:journals/ese/YahmedBKBZ22, DBLP:conf/ijcai/XieMWLLL19, DBLP:journals/tosem/TianZWCSMJ23}, regression testing \cite{DBLP:conf/icse/YouWCLL23}, DL library testing \cite{DBLP:conf/icse/PhamLQT19, DBLP:conf/kbse/NejadgholiY19}, and DL compiler testing \cite{DBLP:journals/pomacs/XiaoLYPW22}. For deeper insight into metamorphic testing and differential testing, readers are encouraged to refer to comprehensive surveys \cite{DBLP:journals/csur/ChenKLPTTZ18, DBLP:journals/tse/SeguraFSC16, DBLP:journals/corr/abs-2002-12543}. 

\textbf{Metamorphic Testing.} A de facto metamorphic testing framework in DL models applies semantics-preserving mutations to seeds and defines the expected output behavior of a model in response to these mutated inputs using metamorphic relations (MRs)  \cite{DBLP:conf/kbse/YuanPW22, DBLP:conf/kbse/WangS20}. This is particularly useful when explicit oracles are infeasible. The design of MRs varies by application domain, tailored to specific characteristics and testing requirements.

Recent studies \cite{DBLP:conf/issta/XieMJXCLZLYS19, DBLP:conf/icse/TianPJR18, DBLP:conf/icml/LiPZL21, DBLP:journals/tse/ZhangRDD22, DBLP:journals/tr/HuangSZSRMH22} frequently adopt an MR such that, when an input undergoes minor mutations, the output of the DL model should remain consistent or fall within a specified confidence threshold. If the output deviates significantly, this could indicate an error in the model. For example, DeepTest \cite{DBLP:conf/icse/TianPJR18}, which targets autonomous driving systems, defines an MR such that errors produced by mutated inputs should fall within a range of $\lambda$ times the Mean Squared Error produced by the seed. Here, $\lambda$ is a user-configurable parameter, allowing testers to adjust the sensitivity of error tolerance according to specific testing requirements.

\textbf{Differential Testing.} Differential testing compares the outputs of two or more DNN models, or different versions of the same model, when presented with the identical inputs or similar inputs with the same semantics. The assumption is that functionally equivalent models should produce consistent outputs for identical or semantically similar inputs \cite{DBLP:journals/ese/YahmedBKBZ22, DBLP:conf/sosp/PeiCYJ17}. The strength of differential testing lies in its capacity to utilize existing models as implicit oracles. It is particularly useful in scenarios like regression testing \cite{DBLP:conf/icse/YouWCLL23} and quantization assessment \cite{DBLP:conf/ijcai/XieMWLLL19, DBLP:journals/tosem/TianZWCSMJ23}, where multiple implementations or variants of a model exist, enabling cross-validation of outputs. 

DeepXplore \cite{DBLP:conf/sosp/PeiCYJ17} leverages the concept of differential testing to generate test inputs that trigger behavioral differences among DNN models. Both DeepEvolution \cite{DBLP:conf/icsm/BraiekK19} and DiverGet \cite{DBLP:journals/ese/YahmedBKBZ22} focus on testing quantized models, identifying test inputs that induce divergence between the original and quantized versions of a DL model. 

\begin{tcolorbox}
    [colback=gray!10, rounded corners]
\ding{46} \textbf{Summary} $\blacktriangleright$  
CGF has emerged as a powerful paradigm for testing DL models, leveraging coverage metrics as feedback signals to drive seed selection, seed scheduling, input mutation, and the retention of generated inputs. At the core of CGF lies the input mutation, which determines how new inputs are derived from seeds. Current research predominantly explores gradient-based optimization and search-based mutation strategies, both of which flexibly design testing objectives under the guidance of coverage metrics. Gradient-based strategies leverage gradients to directly steer inputs toward satisfying testing objectives. Search-based strategies explore the mutation space by selecting operators and regions. Metamorphic mutation introduces domain-specific, semantics-preserving transformations without directly utilizing coverage metrics, resulting in more random and global input alterations. Generative model-based mutation, while promising in its ability to generate diverse and realistic inputs, remains relatively underexplored, particularly in integrating coverage feedback into the input mutation process.
 $\blacktriangleleft$ 
\end{tcolorbox}

\subsection{Coverage-Guided Falsification}
\label{Sec-fal}
Falsification is a validation technique for assessing the behavior of Cyber-Physical Systems (CPSs) against formal temporal specifications. It aims to generate input signals that violate specifications, typically expressed in Signal Temporal Logic (STL) \cite{DBLP:journals/tcs/FainekosP09, DBLP:conf/formats/DonzeM10}, by minimizing robustness, a quantitative measure of how strongly a signal satisfies a given specification \cite{DBLP:journals/tcad/ZhangESAH18, DBLP:journals/tse/YamagataLADH21}. The process terminates upon finding an input with negative robustness, indicating a violation. Since falsification relies on robustness measures derived from formal specifications, it is primarily used in system-level testing and requires manually designed specifications, often focused on safety properties.

A representative example is FalsifyAI \cite{DBLP:journals/tse/ZhangLAMHZ23}, the first coverage-guided falsification approach for hybrid control systems with neural network-based controllers. It employs a dual-loop strategy that leverages coverage metrics to drive exploration while performing local exploitation around promising solutions to minimize robustness, thereby achieving a balance between exploration and exploitation. In the outer loop, a dynamically maintained seed queue is prioritized based on time-aware coverage metrics and organized in a First-In-First-Out manner. The inner loop selects seeds from the queue and applies local search algorithms, such as hill climbing, to generate neighboring signals with reduced robustness. Upon discovering a falsifying input, the process halts. Otherwise, inputs that enhance coverage are retained and prioritized to further guide robustness minimization. Through this design, FalsifyAI explores the behavioral space and exposes violations of temporal specifications.

\subsection{Combinatorial Testing}
Combinatorial testing (CT), also known as Combinatorial Interaction Testing (CIT), evaluates software behaviors influenced by the interactions among multiple input parameters. Traditional CT leverages covering arrays to systematically exercise all relevant value combinations across parameters. It generates inputs that minimize the test suite size while ensuring that combinatorial coverage targets are achieved. It often formulates test generation as a constraint-solving problem \cite{DBLP:conf/wcre/MaJXLLLZ19}. Recent studies have extended CT principles to DL models.

DeepCT \cite{DBLP:conf/wcre/MaJXLLLZ19} maintains a coverage table to monitor combinatorial coverage among neurons and incrementally generates new inputs through random selection or constraint solving to uncover untested neuron combinations. This process continues until full coverage is achieved or resource limits are met. It may incur high computational overhead for large-scale models, particularly due to the combinatorial complexity of neuron interactions.

LV-CIT \cite{DBLP:conf/issre/WangHWNNC24} and CIT4DNN \cite{DBLP:conf/icse/DolaMDS24} adopt black-box CT strategies. LV-CIT targets multi-label image classification by ensuring coverage of all $\tau$-way combinations of label values. It adaptively constructs label-value covering arrays and synthesizes composite images using predefined objects with known labels. These images are used to evaluate whether the classifier correctly predicts all relevant labels, with misclassifications revealing potential interaction-related errors. CIT4DNN applies CT to the latent space of a Variational Autoencoder. It partitions the latent space into equal-density regions and uses a Radial Constrained Covering Array to ensure coverage of region combinations. Test inputs are generated by sampling according to the covering array, aiming to produce feature-diverse inputs that stress the model across different distributional areas. The objectives of the two approaches differ: LV-CIT detects interaction errors at the label level, while CIT4DNN aims to enhance feature diversity by exploring the latent space. Although black-box methods are more scalable, they often face challenges in maintaining the semantic validity of synthesized inputs, particularly when manipulating label or latent spaces.

\subsection{Concolic Testing}
\label{Sec-concolic}
Concolic testing is a well-established testing technique that systematically explores program behaviors by combining concrete execution with symbolic analysis \cite{DBLP:conf/sigsoft/SenMA05, DBLP:conf/icse/ChaHLO18}. In traditional software testing, the process begins with executing a program using a concrete input. After execution, an unexplored path is selected and encoded symbolically \cite{DBLP:conf/osdi/CadarDE08, DBLP:conf/tacas/XieMSN05}. A constraint solver is then used to generate new concrete inputs that satisfy the symbolic conditions. This loop continues until sufficient structural coverage is achieved.

Sun et al. \cite{DBLP:conf/kbse/SunWRHKK18, DBLP:journals/tecs/SunHKSHA19} pioneered the adaptation of concolic testing for DNNs. Their framework, DeepConcolic \cite{DBLP:conf/kbse/SunWRHKK18}, iteratively generates test inputs to satisfy specific coverage criteria by alternating between concrete execution and symbolic reasoning. Beginning with an initial input, the model is executed to observe its activation behavior. If the desired coverage goals, such as Neuron Coverage \cite{DBLP:conf/sosp/PeiCYJ17} and Sign-Sign Coverage \cite{DBLP:journals/tecs/SunHKSHA19}, are not met, symbolic analysis is used to formulate an optimization problem, It is solved using techniques such as global optimization or linear programming (LP) to generate a new input that satisfies the coverage targets.

In a follow-up extension \cite{DBLP:journals/tecs/SunHKSHA19}, Sun et al. generate inputs by encoding test conditions into an LP formulation. Given a concrete input $x_1$, an LP formulation is derived to preserve $x_1$’s activation pattern (encoded via ReLU operations and node dependencies) while achieving additional coverage. Solving the LP yields a new input $x_2$ with the same neuron activation structure but satisfying specific test conditions. To overcome the scalability bottleneck of LP solvers, a heuristic gradient-based search is introduced. Initialized by adversarial perturbations (\textit{e.g.}, FGSM \cite{DBLP:journals/corr/GoodfellowSS14}), this search iteratively refines inputs using gradient ascent/descent combined with binary search, generating input pairs $(x_1,x_2)$ that satisfy testing conditions even on large models.

Current concolic testing approaches focus on optimizing coverage at the neuron activation level. Unlike traditional programs, DNNs lack explicit control flow structures, making the symbolic encoding inherently ambiguous. Consequently, the effectiveness of concolic testing is limited by its relatively shallow representation of DNN behaviors. Future research could explore more expressive abstractions that capture broader structural properties or semantic-level behaviors. 

\begin{tcolorbox}
    [colback=gray!10, rounded corners]
\ding{46} \textbf{Summary (Section \ref{Sec-fal} to \ref{Sec-concolic})} $\blacktriangleright$ Originally developed for traditional software, falsification, combinatorial testing, and concolic testing have been adapted to DL models. Coverage-guided falsification adopts coverage metrics to guide seed scheduling and retention, enhancing input space exploration. However, integrating coverage into the exploitation phase, aimed at minimizing local robustness, remains underexplored. Combinatorial testing generates inputs satisfying combinatorial coverage targets, probing diverse interactions among output features or internal representations. Expanding combinatorial coverage to diverse input modalities and capturing structural abstractions are promising research directions. Concolic testing alternates between concrete execution and symbolic reasoning to satisfy coverage targets. The lack of explicit control flow structures in DL models challenges the expressiveness and scalability of symbolic encodings, particularly for large-scale models. In summary, despite notable progress, open challenges remain in advancing falsification's exploitation mechanisms, enriching combinatorial coverage abstractions, and enabling scalable, semantically meaningful symbolic reasoning for DL models.
 $\blacktriangleleft$ 
\end{tcolorbox}

\section{Coverage-Guided Test Optimization}
\label{Sec-opt}
\begin{figure}  
    \centering  
    \includegraphics[width = 13.9 cm]{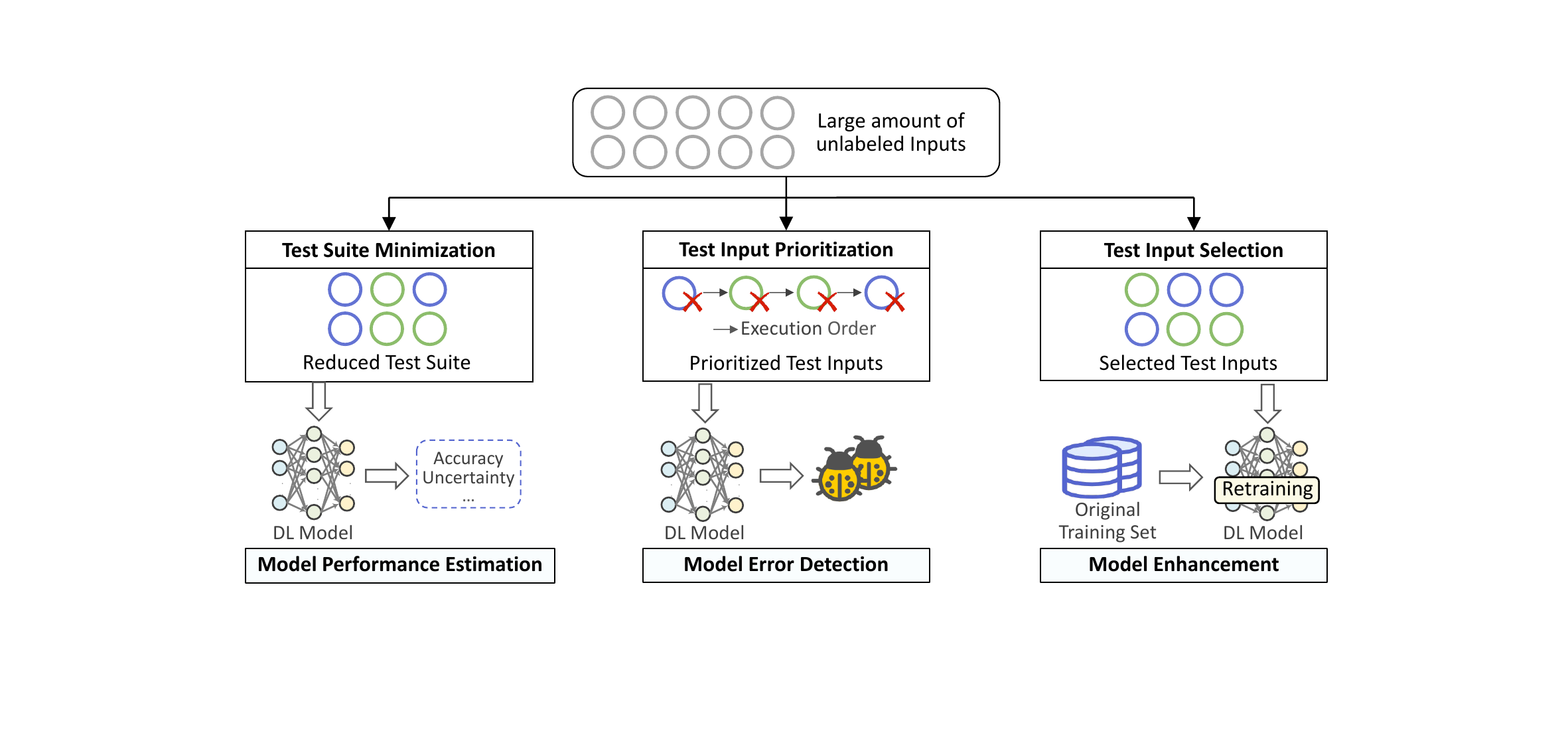} 
    \caption{Illustrations of Test Optimization Tasks for DL Models}  
    \label{TestOpt}
\end{figure}

For traditional software, test optimization is commonly used in regression testing to reduce execution time and resource consumption. Coverage-guided optimization refines the test suite based on structural coverage metrics, ensuring that the most critical code segments, particularly recent modifications, are tested effectively and efficiently \cite{DBLP:journals/stvr/NardoABL15, DBLP:journals/stvr/YooH12}. DL test optimization presents distinct objectives. DL testing is carried out by constructing a test suite derived from available labeled datasets. However, obtaining accurate labels for large-scale unlabeled data is time-consuming and labor-intensive \cite{DBLP:conf/icse/GaoFYL0X22, DBLP:conf/sigsoft/KimJFY20}. This challenge is exacerbated by the demand for high-quality annotations, which require considerable manual effort from domain experts \cite{DBLP:journals/tosem/HuGCXMPT22}, and may even necessitate collaboration among multiple specialists \cite{DBLP:journals/tosem/AghababaeyanADB24}. To address these challenges, researchers and practitioners have explored various test optimization techniques to improve the cost-effectiveness of DL testing. Figure \ref{TestOpt} illustrates three DL test optimization tasks, all based on sampling from an unlabeled dataset. 

\begin{tcolorbox}
    [colback=gray!10, rounded corners]
\small
\textbf{(1) Test suite minimization:} The goal is to select an input subset for estimating model performance, such as accuracy and uncertainty. Sampling-based minimization \cite{DBLP:journals/tosem/0003WWYZY20, DBLP:conf/sigsoft/LiM0CX019, DBLP:conf/issre/ZhouLD0H20} allocates a predefined labeling budget to annotate a limited set of inputs, and assesses model performance based on these labeled samples. Labeling-free minimization \cite{DBLP:conf/icse/HuGXCPMT23} estimates model performance without requiring manual annotation. \\
\textbf{(2) Test input prioritization:} The goal is to rank unlabeled test inputs based on their likelihood of exposing incorrect model predictions \cite{DBLP:conf/aitest/ByunSVRC19, DBLP:conf/icse/WangY0ZDZ21, DBLP:journals/tosem/DangLPKBT24, DBLP:conf/issta/FengSGWF020}. 
It ensures that error-revealing test inputs are identified, labeled, and processed earlier in the testing process. Recent studies \cite{DBLP:conf/issta/FengSGWF020, DBLP:journals/tosem/MaPTCT21, DBLP:conf/icse/LiuFY022, DBLP:journals/kbs/WuSZLSS23} have employed various model uncertainty quantification metrics to evaluate the error detection capabilities of test inputs.\\
\textbf{(3) Test input selection:} A subset of test inputs is sampled from an unlabeled dataset, which is labeled and utilized for model retraining. This process not only facilitates performance evaluation but also improves model quality through targeted data augmentation. \cite{DBLP:conf/kbse/ShenLCHZX20, DBLP:journals/tosem/AghababaeyanADB24, DBLP:conf/issta/LiXJP0WL24, DBLP:conf/wcre/HaoHGS23, DBLP:journals/tosem/AttaouiFPB23, DBLP:conf/issre/GuoTH23}. Many prioritization techniques can also be applied to identify error-revealing inputs that contribute to model enhancement.
\end{tcolorbox}

The research community has advanced the field by conducting comprehensive surveys \cite{DBLP:journals/tosem/HuGXCMPT24, 10562069} and empirical studies \cite{DBLP:journals/tosem/MaPTCT21, DBLP:conf/qrs/ShiYZL21, DBLP:conf/euromicro/MosinSDNPK22, Weiss-issta, DBLP:journals/infsof/ZhaoMCZJW22} that investigate various test optimization techniques. We encourage interested readers to refer to these references for a more comprehensive understanding. Table \ref{Tab-Opt} summarizes studies that propose new coverage-guided optimization approaches and introduce coverage-based methods as baselines.

\subsection{Coverage-Guided Test Optimization}
In terms of test suite minimization, DeepReduce \cite{DBLP:conf/issre/ZhouLD0H20} optimizes the test suite by reducing its size while preserving test adequacy and output distribution similarity. It operates in two phases. First, a greedy algorithm is applied to select a minimal subset of inputs that achieves the same Neuron coverage as the full test set. Second, a heuristic optimization process iteratively refines this subset by minimizing the relative entropy between the output distributions of the subset and the original set. HeatC \cite{DBLP:conf/setta/SunLLS23} partitions the benchmark dataset into heat feature buckets and selects inputs that cover previously uncovered buckets while maximizing input diversity. Both approaches aim to minimize the test suite size while maintaining diversity and adequacy, thereby ensuring that the selected test suite explores the state or feature space. 

Regarding test selection, HashC \cite{DBLP:journals/jsa/SunXLZS23} partitions neuron outputs into discrete intervals (buckets), and greedily selects inputs that maximize the number of independent bucket combinations, aiming to enhance the diversity and adequacy of feature interactions. Furthermore, Shi et al. \cite{DBLP:journals/jss/ShiYZ24} observed a negative correlation between state space coverage and DRL rewards. Based on this insight, they sample diverse state sequences to maximize state space coverage, enhancing error detection effectiveness. In follow-up work, starting with an initial seed set, Shi et al. \cite{DBLP:journals/infsof/ShiYS25} iteratively sample multiple candidate subsets from the test pool and greedily select the one that yields the highest coverage improvement. This process continues until the desired test set size is reached. 

\begin{table}
\begin{threeparttable}
  \caption{Summary of Research Related to Coverage-Guided Test Optimization}
  \scriptsize
  \label{Tab-Opt}
  \fontsize{8.5}{10}\selectfont
  \setlength{\tabcolsep}{0.1 mm}{
  \begin{tabular}{c|llll}
    \toprule
{}& \textbf{Year} & \textbf{Study} &  \textbf{Task}  & \textbf{Core Idea} \\
    \midrule

&2020 & DeepReduce \cite{DBLP:conf/issre/ZhouLD0H20}  &  Minimization & Test suite minimization with coverage preservation\\
\textbf{New}
&2024 & Shi et al. \cite{DBLP:journals/jss/ShiYZ24} & Selection & Select state sequence to maximize coverage of state space \\  
&2023 & HeatC \cite{DBLP:conf/setta/SunLLS23} & Minimization & Sample inputs evenly across different heat feature buckets \\ 
\textbf{Technique} &2023 & HashC \cite{DBLP:journals/jsa/SunXLZS23} & Selection & Greedy input selection to maximize bucket combination coverage \\ 

&2025 & Shi et al. \cite{DBLP:journals/infsof/ShiYS25}  &  Selection & Greedy test subset selection to maximize DL coverage \\
\midrule

\multirow{6}{*}{\textbf{Benchmark}} &2020 & Deepgini \cite{DBLP:conf/issta/FengSGWF020}  & Prioritization & Incorporate DL coverage with CTM and CAM\\

&2022 & Weiss et al. \cite{Weiss-issta}  & Prioritization, Selection &  Incorporate DL coverage with CTM and CAM  \\  

&2022 & ATS \cite{DBLP:conf/icse/GaoFYL0X22}  & Selection & Greedy input selection to maximize DL coverage \\

&2023 & RTS \cite{10319282}  & Selection & Greedy input selection to maximize DL coverage \\
  
& 2024 & Wang et al. \cite{10.1145/3672446}  & Prioritization & Incorporate DL coverage with CAM \\ 

& 2025 & DeepVec \cite{deepvec}  & Selection & Incorporate DL coverage with CTM and CAM \\ \bottomrule
  
\end{tabular}}
\end{threeparttable}

\end{table}

\subsection{Benchmark}
Recent studies have proposed coverage-guided test input prioritization approaches by combining coverage criteria with traditional strategies such as the Coverage-Total Method (CTM) and the Coverage-Additional Method (CAM) \cite{DBLP:conf/issta/YooH07}. CTM ranks inputs based on their overall contribution to coverage, selecting those that collectively maximize coverage metrics. CAM adopts a greedy strategy, iteratively selecting inputs that cover the most uncovered structures until a coverage threshold is reached. For example, DeepGini \cite{DBLP:conf/issta/FengSGWF020} employs both CAM and CTM strategies, adopting criteria such as Neuron Coverage \cite{DBLP:conf/sosp/PeiCYJ17}, $k$-Multisection Neuron Coverage, and Strong Neuron Activation Coverage \cite{DBLP:conf/kbse/MaJZSXLCSLLZW18}. Building on these strategies, Weiss et al. \cite{Weiss-issta} examine the error detection effectiveness of coverage-guided, uncertainty-based, and Surprise Adequacy-based prioritization approaches. Wang et al. \cite{10.1145/3672446} investigate the effectiveness of coverage criteria in detecting ID errors, adopting CAM for input prioritization. Their findings suggest that Surprise Adequacy metrics outperform neuron output-based criteria in error detection, revealing the limitations of coverage-guided prioritization in uncovering erroneous behaviors in DL models. Furthermore, ATS \cite{DBLP:conf/icse/GaoFYL0X22} and RTS \cite{10319282} iteratively select inputs using a greedy strategy that maximizes the coverage gain with respect to a predefined criterion. At each step, the next input is chosen based on its capability to cover the largest portion of previously uncovered regions, leveraging feedback from the previously selected set.

\begin{tcolorbox}
    [colback=gray!10, rounded corners]
\ding{46} \textbf{Summary} $\blacktriangleright$ Test optimization for DL models remains an active area of research. While many approaches rely on internal behaviors and output prediction confidence of DL models for test optimization, they rarely incorporate coverage metrics directly. This is primarily due to two factors: first, in test suite minimization, previous coverage metrics are insufficient for capturing the diversity and representativeness of the original test set, including aspects such as class distribution and latent input representations. Second, current coverage metrics often exhibit weak correlations with error detection, rendering them less effective than alternative strategies, such as uncertainty-based or Surprise Adequacy-based methods. A promising future direction is to develop hybrid strategies that integrate coverage signals with task-specific properties, such as class distribution and error-proneness, to enhance test optimization.
 $\blacktriangleleft$ 
\end{tcolorbox}

\section{Empirical Evaluation}
\label{Sec-evaluation}
\subsection{Datasets and Models Used in CGT Studies}

\begin{figure}  
    \centering  
    \includegraphics[width = 15.8 cm]{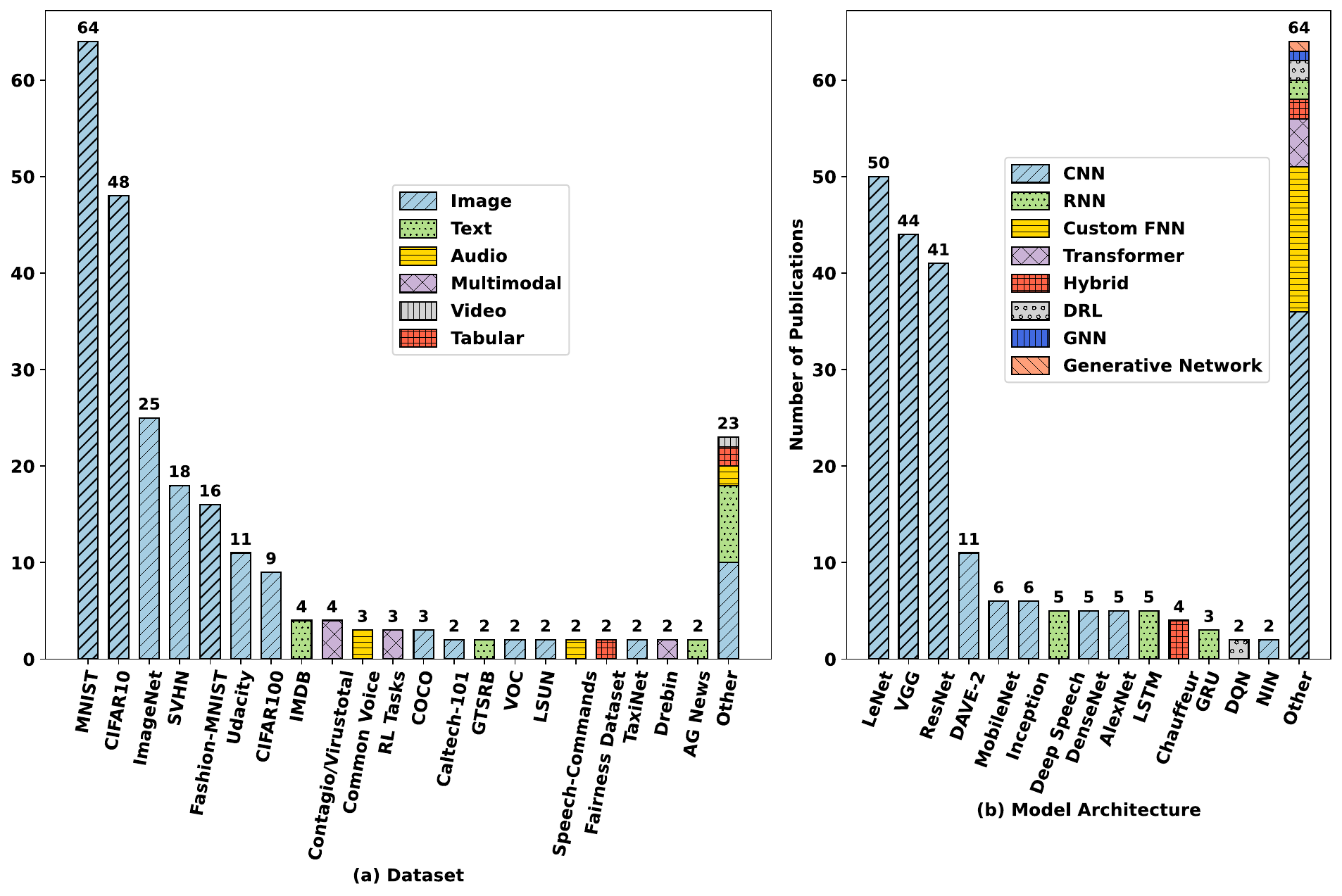} 
    \caption{Overview of Model Architectures and Datasets in Coverage-Guided Testing Studies}  
    \label{model-dataset}
\end{figure}

\subsubsection{Datasets}
Figure \ref{model-dataset}(a) provides an overview of datasets employed in CGT studies, illustrating their distribution across different modalities. The vertical axis lists datasets used in more than one publication, while those utilized in only a single study are grouped under the `Other’ category. The horizontal axis represents the number of publications utilizing each dataset, with datasets further categorized into six types according to their modalities, \textit{i.e.}, image, text, audio, multimodal, video, and tabular, as indicated in the legend. It is worth noting that the text category contains natural language, source code, and symbolic sequences. The multimodal category includes datasets with heterogeneous data formats, such as PDF documents and mobile applications.

The figure reveals a strong preference for image datasets in CGT studies, with MNIST (64 studies), CIFAR-10 (48 studies), and ImageNet (25 studies) standing out as the most frequently utilized benchmarks. Additional image datasets such as SVHN (16 studies), Fashion-MNIST (11 studies), and Udacity (9 studies) further reinforce the field’s predominant focus on computer vision tasks. However, other dataset types, such as audio, multimodal, and tabular, have received relatively limited attention, accounting for only a small fraction of the studies. This indicates that CGT research remains largely constrained to vision-centric domains. The `Other' category, encompassing 23 studies, suggests that while a diverse array of datasets has been examined, few have gained widespread adoption. These observations underscore the need for future research to diversify dataset selection to enhance the generalizability of CGT methodologies across a broader range of DL applications.

\subsubsection{DL Models}
Figure \ref{model-dataset}(b) presents a statistical overview of model architectures examined in CGT research. The vertical axis lists architectures referenced in more than one publication, while those cited only once are aggregated under the `Other' category. The horizontal axis indicates the number of studies in which each architecture has been employed. Architectures are further categorized into seven types: CNN, RNN, Custom FNN, Transformer, Hybrid, DRL, GNN, and Generative Network.

The analysis reveals a predominant focus on CNNs, with LeNet (50 studies), VGG (44 studies), and ResNet (41 studies) being the most frequently adopted architectures. Additional CNN-based models, such as DAVE-2, MobileNet, and Inception, further underscore the community’s emphasis on CNNs, particularly in vision-centric tasks.
Alternative architectures, such as RNNs (LSTM, GRU), DRLs, Transformers, GNNs, and generative networks, have received comparatively limited attention. The `Other' category, comprising 64 studies, suggests that while a wide range of architectures has been explored, many lack sustained or widespread adoption. Given the growing development of diverse architectural paradigms in DL, future research may benefit from broadening the architectural scope to enhance its applicability across a broader set of tasks and domains.

\begin{tcolorbox}
    [colback=gray!10, rounded corners]
\ding{46} \textbf{Summary} $\blacktriangleright$ CGT research is primarily centered on image datasets, with benchmarks such as MNIST, CIFAR-10, and ImageNet driving much of the research. Datasets from other modalities, such as text, audio, and tabular data, remain relatively underexplored, indicating a potential gap in the application of CGT across a wider range of domains. Model architectures are predominantly centered around CNNs, while other architectures, such as RNNs, generative networks, and DRL models, have received less attention. As DL continues to evolve, the growing prominence of transformer-based models and DRL presents an opportunity for future research. To sum up, future work should expand the scope of CGT by incorporating a broader range of datasets and investigating emerging architectures to enhance the generalizability of testing methodologies across diverse DL domains. $\blacktriangleleft$ 
\end{tcolorbox}

\subsection{Experimental Design of Evaluating Coverage Criteria}
Table \ref{EmpiricalCC} provides an overview of the experimental design employed to evaluate coverage criteria, organizing them into six main categories: model quality properties, error detection, input/output diversity, coverage profile, testing cost, and criteria setting. Each category has specific evaluation aspects, supported by relevant references and the corresponding dataset treatments.

\subsubsection{Dataset Treatments}
\begin{figure}  
    \centering  
    \includegraphics[width = 16 cm]{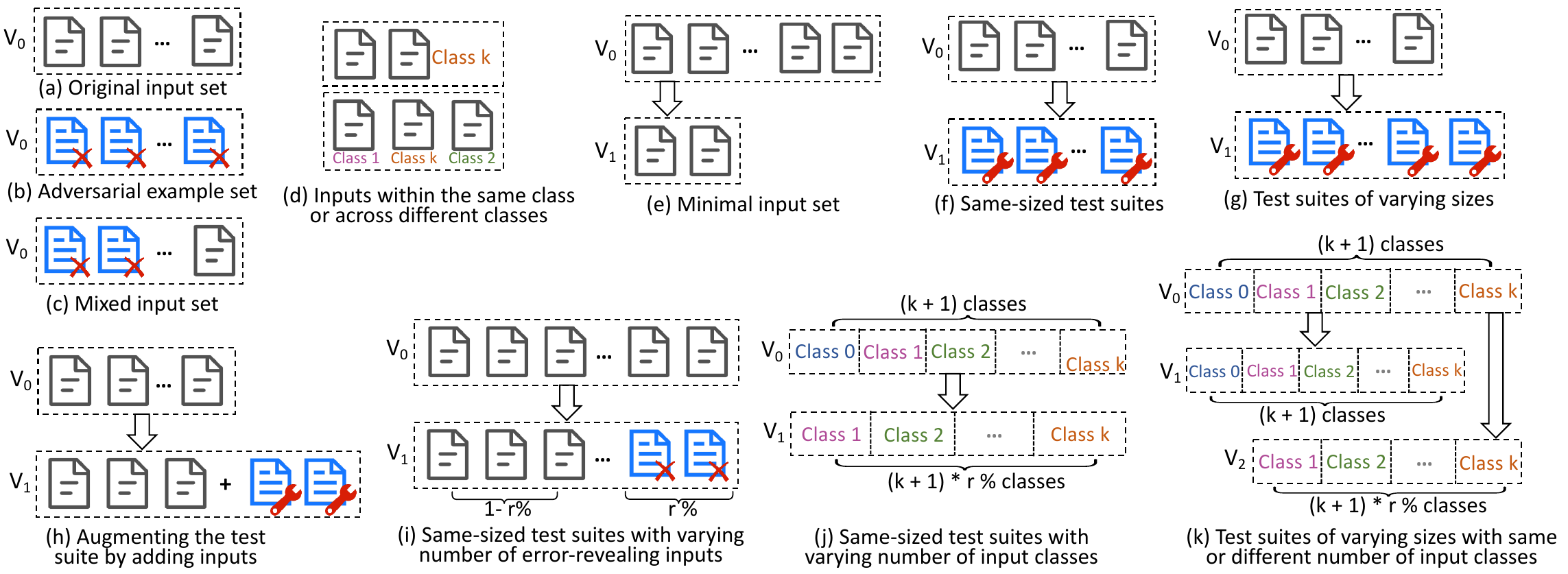} 
    \caption{Illustrations of Dataset Treatments for Coverage Criteria Evaluation}  
    \label{DataTreatment}
\end{figure}

Various dataset treatments are designed to simulate different testing scenarios by varying factors such as test suite size, input/output diversity, class distributions, and error rates. Figure \ref{DataTreatment} provides high-level illustrations of dataset treatments used in coverage criteria evaluation. For instance, starting with an original test suite denoted as $V_0$, an input set $V_1$ can be derived from $V_0$ through various treatments. The coverage of $V_0$ and newly generated input sets is analyzed from multiple perspectives.

\textbf{(a) Original input set.}
Several studies have adopted inputs directly from the original test set or training set without any modification to evaluate coverage criteria.

\textbf{(b) Adversarial example set.}
This set comprises adversarial examples crafted by applying adversarial attack techniques to the original test set.

\textbf{(c) Mixed Input set.}
This set is constructed by mixing original test inputs with adversarial examples according to a predefined proportion.

\textbf{(d) Inputs within same or across classes.}
This treatment organizes test inputs according to their respective categories, aiming to investigate how coverage varies under different input configurations. It is adopted to examine the effects of input diversity both within individual classes and across multiple classes.

\textbf{(e) Minimal input set.}
Starting from an original test suite $V_0$, a minimal input set $V_1$ is derived by reducing test inputs. It is performed with the constraint that removing any additional test input from $V_1$ would result in a coverage decrease. It ensures that $V_1$ retains maximal coverage while minimizing the number of test inputs.

\textbf{(f) Same-sized test suites.}
New input sets are generated with the same size as the original set $V_0$, employing techniques such as adversarial attack strategies or metamorphic mutations. It ensures consistency in the size of the test input sets while varying other factors, such as input diversity and error rates.

\textbf{(g) Test suites of varying sizes.}
New input sets are derived from the original set $V_0$, with varying sizes. Despite these variations, the differences in coverage across input sets are analyzed by using the average coverage.

\textbf{(h) Augmenting the test suite by adding inputs.}
The original input set $V_0$ is augmented by incorporating additional inputs, such as original inputs, adversarial examples, backdoor inputs, and mutated inputs. Recent studies have employed a progressive strategy, incrementally increasing the size of $V_0$ by constructing a series of test suites. This treatment assesses how incremental additions to a test suite influence the coverage.

\textbf{(i) Same-sized test suites with varying number of error-revealing inputs.}
New input sets are created by replacing a portion of the inputs in $V_0$ with error-revealing inputs, such as adversarial examples or naturally occurring error-inducing inputs. While the size of each test suite remains constant, the proportion of error-revealing inputs is varied. It aims to examine how the inclusion of these inputs influences coverage.

\textbf{(j) Same-sized test suites with varying number of input classes.}
For an original test suite $V_0$ with $k+1$ input categories, new test sets are generated by varying the number of input categories within $V_0$. Specifically, the number of input categories is adjusted by a factor of $r\%$, while maintaining the total number of inputs consistent with  $V_0$. The inputs within $V_1$ are uniformly distributed across the available categories. This treatment explores how changes in input category distribution affect coverage performance.

\textbf{(k) Test suites of varying sizes with the same or different number of input classes.}
For an original test suite $V_0$ with $k+1$ input categories, two distinct treatments are applied: 1) The number of categories is fixed while adjusting the number of inputs within each category, ensuring equal inputs per category. 2) A proportion of $(k+1) *r\%$ of the original categories is retained, leading to new test suites with different total input counts compared to $V_0$. Both treatments are adopted to investigate the effect of input diversity on coverage. 

\begin{table}
\begin{threeparttable}
  \caption{Overview of Experimental Design for Evaluating Coverage Criteria}
  \small
  \label{EmpiricalCC}
  \fontsize{7.1}{8.2}\selectfont
  \setlength{\tabcolsep}{0.1 mm}{
  \begin{tabular}{cllcc}
    \toprule
\multirow{2}{*}{\textbf{Category}} & \multirow{2}{*}{\textbf{Evaluation Aspect} \customfootnote{$\dagger$}} & \multirow{2}{*}{\textbf{Reference}} & \multirow{2}{*}{\textbf{\# studies}} & \textbf{Dataset}  \\ 
&&& & \textbf{Treatment}  \\ \midrule
    
& Corr. between test coverage and model robustness  & \cite{DBLP:conf/iceccs/DongZWLSH0WDD20, DBLP:conf/sigsoft/YanTLZMX020} & 2 & (a)  \\ 
 \textbf{Model}& Corr. between test coverage and model fairness  & \cite{DBLP:journals/tse/ZhengLWC24, DBLP:conf/esem/DuC24} & 2 & (a, h)  \\  
 \textbf{Quality}& Corr. between test coverage and DRL rewards & \cite{DBLP:conf/icse/TrujilloLEDC20, DBLP:journals/jss/ShiYZ24} & 2  &  (a) \\  
\textbf{Property} & Model robustness variations with coverage-increasing inputs & \cite{DBLP:conf/wcre/YangSAL22, DBLP:conf/iceccs/DongZWLSH0WDD20} & 2 &  (h) \\  
& Coverage Comp. between the original model and after applying bias mitigation & \cite{DBLP:journals/tse/ZhengLWC24} & 1 & N.A  \\  
\midrule   
 
& Coverage variations with augmenting test suites via adding error-revealing inputs & \cite{DBLP:conf/sigsoft/YanTLZMX020, DBLP:conf/kbse/MaJZSXLCSLLZW18, DBLP:conf/aitest/WangWFCC19, DBLP:conf/icse/KimFY19, DBLP:conf/icse/JiMYW23, DBLP:journals/tosem/DolaDS23, DBLP:conf/sigsoft/DuXLM0Z19, DBLP:journals/tr/HuangSZSRMH22, DBLP:journals/infsof/ShiYS25} & 9 & (h)\\ 

& Coverage variations in same-sized test suites with increasing error-revealing inputs & \cite{DBLP:conf/icse/LiM0C19, DBLP:journals/ese/GuoTH24, DBLP:journals/tosem/XieLWMGJL22, DBLP:journals/corr/abs-1812-05339,  DBLP:journals/sttt/UsmanSGDMP23} & 5 & (i)\\ 

\textbf{Error} & Corr. between test coverage and error detection & \cite{10.1145/3672446, DBLP:journals/smr/YanCCWKW24, DBLP:journals/sqj/ZhouPCQZZ25, DBLP:conf/icse/NeelofarA24, DBLP:journals/infsof/ShiYS25} & 5 & (a, i) \\
 
\textbf{Detection} & Coverage Comp. between benign/original inputs and error-revealing inputs & \cite{DBLP:journals/tse/ZhangLAMHZ23, 10.1145/3672446, DBLP:journals/tse/ZhengLWC24, DBLP:conf/icse/YuanPW23, DBLP:conf/iclr/LiuTLM024, DBLP:conf/wcre/ZhouDLZ0Y21} & 6 & (a, f, g)\\

& Error detection rate Comp. between original and reduced input sets  & \cite{DBLP:conf/naacl/SekhonJDQ22} & 1 & (e)\\ 

& Error detection capability & \cite{DBLP:conf/icse/KimFY19, DBLP:journals/tse/RossoliniBB23} & 2 & (b, c)\\ \midrule

& Coverage variations across test suites with varying levels feature diversity & \cite{DBLP:conf/icse/YuanPW23, DBLP:journals/tosem/DolaDS23, DBLP:conf/icse/GerasimouE0C20, DBLP:conf/icse/JiMYW23, DBLP:conf/safecomp/AbrechtAGGHHW20, DBLP:journals/jsa/SunXLZS23} & 6 & (f, g, h) \\ 
 
& Coverage variations across test suites with varying levels of class diversity & \cite{DBLP:conf/icse/JiMYW23, DBLP:conf/icse/YuanPW23, DBLP:journals/tr/HuangSZSRMH22, DBLP:journals/sttt/UsmanSGDMP23, DBLP:conf/ijcnn/SunLS21, DBLP:journals/infsof/ShiYS25} & 6 & (j, k)\\ 

\textbf{Input/}& Coverage of test suites within the same class and across different classes & \cite{DBLP:conf/sosp/PeiCYJ17, DBLP:conf/icse/ByunR20, 10.1145/3672446, DBLP:conf/safecomp/AbrechtAGGHHW20, DBLP:journals/corr/abs-1911-07309} & 5 & (d)\\

\textbf{Output}& Comp. between test coverage and mutation testing w.r.t feature diversity & \cite{DBLP:journals/tosem/DolaDS23} & 1 & (h)\\  
  
\textbf{Diversity}& Corr. between test coverage and the number of input classes & \cite{10.1145/3672446, DBLP:journals/smr/YanCCWKW24} & 2 & (j)\\

& Corr. between coverage metrics and input diversity metrics & \cite{DBLP:conf/icse/NeelofarA24} & 1 & (i)\\

& Corr. between test coverage and model output impartiality &  \cite{DBLP:journals/tosem/XieLWMGJL22, DBLP:conf/sigsoft/Harel-CanadaWGG20} & 2 & (f) \\   \midrule
 
& Comp. or Corr. between coverage criteria across test suites& \cite{DBLP:conf/sosp/PeiCYJ17, DBLP:conf/icse/KimFY19, DBLP:conf/icse/GerasimouE0C20, DBLP:conf/sigsoft/YanTLZMX020, DBLP:journals/tse/ZhengLWC24, DBLP:conf/wcre/ZhouDLZ0Y21, DBLP:journals/tr/HuangSZSRMH22}  & 7 & (a, h, e) \\
& Coverage variations among test suites of increasing sizes & \cite{DBLP:journals/tse/ZhengLWC24, DBLP:conf/sigsoft/YanTLZMX020, DBLP:conf/naacl/SekhonJDQ22, DBLP:journals/tr/HuangSZSRMH22, DBLP:conf/setta/SunLLS23, DBLP:journals/sttt/UsmanSGDMP23, DBLP:conf/ijcnn/SunLS21} & 7 & (a, h, k) \\  
\textbf{Coverage} & Corr. between test coverage and test suite size & \cite{DBLP:journals/tse/ZhengLWC24} & 1 & (g) \\ 
\textbf{Profile}  & Corr. between test coverage and model outputs & \cite{DBLP:conf/icse/TianPJR18} & 1 & (a) \\  
& Coverage differences across various models & \cite{DBLP:journals/jss/ShiYZ24} & 1 & (a) \\  
& Corr. between test coverage and mutation scores & \cite{DBLP:conf/icst/JahangirovaT20} & 1 & (h) \\  
\midrule

\textbf{Testing} & Execution time overhead & \cite{DBLP:journals/ese/GuoTH24, DBLP:conf/icse/JiMYW23, DBLP:journals/tse/ZhangLAMHZ23, DBLP:journals/tosem/XieLWMGJL22, DBLP:journals/tosem/DolaDS23, DBLP:conf/icse/YuanPW23, DBLP:conf/icse/NeelofarA24, DBLP:journals/jsa/SunXLZS23, DBLP:journals/tse/RossoliniBB23} & 9 & / \\
\textbf{Cost}  & GPU memory usage & \cite{DBLP:conf/icse/JiMYW23, DBLP:journals/tse/RossoliniBB23} & 2 & /  \\ \midrule
     
\textbf{Criteria} & \multirow{2}{*}{Effects of hyper-parameters} & \cite{DBLP:conf/wcre/MaJXLLLZ19, DBLP:conf/icse/LiM0C19, DBLP:journals/infsof/BaiHHWXQY24, DBLP:conf/sigsoft/Harel-CanadaWGG20, DBLP:conf/kbse/MaJZSXLCSLLZW18, DBLP:conf/sosp/PeiCYJ17, DBLP:conf/icse/GerasimouE0C20, DBLP:conf/iclr/LiuTLM024, DBLP:conf/issre/0005LXC24} & \multirow{2}{*}{16} & \multirow{2}{*}{/} \\
&&\cite{10.1145/3672446, DBLP:conf/ijcnn/YeWZK22, DBLP:journals/tosem/XieLWMGJL22, DBLP:journals/tse/ZhangLAMHZ23, DBLP:conf/sigsoft/YanTLZMX020, DBLP:conf/aitest/WangWFCC19, DBLP:journals/smr/YanCCWKW24} &  &  \\ \cmidrule(lr) {2-5} 
\textbf{Setting}& Effects of layer selection & \cite{DBLP:conf/icse/KimFY19, DBLP:conf/icse/GerasimouE0C20, DBLP:conf/wcre/ZhouDLZ0Y21, DBLP:conf/icse/YuanPW23, DBLP:journals/tecs/SunHKSHA19, DBLP:conf/wcre/MaJXLLLZ19, DBLP:conf/icse/JiMYW23, DBLP:conf/icse/TrujilloLEDC20, DBLP:conf/safecomp/AbrechtAGGHHW20, DBLP:conf/iclr/LiuTLM024} & 10 & / \\ \bottomrule
  
\end{tabular}}
\begin{tablenotes}
\footnotesize
  \item[\customtnote{$\dagger$}] Corr.: Correlation, Comp.: Comparison.
\end{tablenotes}
\end{threeparttable}
\end{table}

\subsubsection{Evaluation Aspects}
The following section details the aspects adopted for coverage criteria evaluation.

\textbf{(1) Model Quality Property.} This aspect explores the relationships between test coverage and model quality properties, including robustness, fairness, and DRL rewards. It aims to investigate whether test coverage observations could provide implications for model quality evaluation and enhancement. Correlation analysis is typically performed using statistical methods such as Pearson and Kendall correlation coefficients. Trujillo et al. \cite{DBLP:conf/icse/TrujilloLEDC20} and Shi et al. \cite{DBLP:journals/jss/ShiYZ24} explored the correlation between test coverage and DRL rewards, which serve as a key measure for capturing an agent’s ability to achieve predefined objectives. Regarding fairness enhancement, Zheng et al. \cite{DBLP:journals/tse/ZhengLWC24} investigated variations in coverage before and after applying model bias mitigation strategies. Regarding robustness, Yang et al. \cite{DBLP:conf/wcre/YangSAL22} augmented the training set with coverage-improving inputs for model retraining and investigated how robustness metrics evolve with increasing test coverage.

\textbf{(2) Error Detection.} Traditional CGT practices operate on the assumption that testing a broader range of code or functionality increases the probability of detecting errors. In DL models, this principle is extended by observing coverage variations to distinguish input sets with varying error-revealing capabilities.

The sensitivity of coverage metrics to errors is assessed by examining whether increasing the number of error-revealing inputs leads to higher coverage. Prior work \cite{DBLP:conf/kbse/MaJZSXLCSLLZW18, DBLP:conf/sigsoft/DuXLM0Z19, DBLP:journals/smr/YanCCWKW24, DBLP:journals/tr/HuangSZSRMH22, DBLP:conf/sigsoft/YanTLZMX020} examined coverage variations across test suites of varying sizes as additional error-revealing inputs are introduced. To mitigate the effect of test suite size, multiple test suites of equal size but with varying error rates are constructed \cite{DBLP:journals/ese/GuoTH24, DBLP:journals/tosem/XieLWMGJL22, DBLP:journals/corr/abs-1812-05339, DBLP:conf/icse/LiM0C19}. When an original test suite contains a certain proportion of erroneous inputs, MNCOVER \cite{DBLP:conf/naacl/SekhonJDQ22} generates a minimal test set by filtering out inputs that do not contribute to coverage increases. The underlying assumption is that test coverage can help identify and eliminate redundant inputs while retaining those with error-revealing capability.

Prior studies have assessed whether coverage metrics could identify novel or unexpected inputs likely to trigger errors. Coverage comparisons are made between original inputs or inputs that pass the test against those that fail, such as original test inputs versus adversarial examples \cite{DBLP:conf/icse/YuanPW23}, ID inputs versus OOD inputs \cite{10.1145/3672446}, or original inputs versus fairness-violating instances \cite{DBLP:journals/tse/ZhengLWC24}. Moreover, SADL \cite{DBLP:conf/icse/KimFY19} and the method introduced by Rossolini et al. \cite{DBLP:journals/tse/RossoliniBB23}directly utilize test coverage information as an indicator for identifying adversarial samples.

\textbf{(3) Input/Output Diversity.} Recent studies have investigated whether higher coverage reflects greater input diversity. These studies have analyzed coverage variations across test suites with differing levels of feature and class diversity. Feature diversity refers to variation in characteristics or attributes of data representations. For instance, DeepImportance \cite{DBLP:conf/icse/GerasimouE0C20} applies pixel-level perturbations to seed inputs, whereas NLC \cite{DBLP:conf/icse/YuanPW23} produces feature-diverse input sets by injecting white noise into seeds. Class diversity quantifies the number of unique labels in a test suite and examines how evenly they are distributed. For example, CC \cite{DBLP:conf/icse/JiMYW23} evaluated coverage variations across test suites of equal size but with differing proportions of input classes. NLC \cite{DBLP:conf/icse/YuanPW23} examined how coverage metrics respond to test diversity by comparing input sets varying in class count and label distribution. Additionally, Dola et al. \cite{DBLP:journals/tosem/DolaDS23} compared coverage metrics with mutation testing in assessing feature diversity, analyzing mutation scores and their proposed coverage metric. While mutation testing is effective for evaluating test adequacy in traditional software, existing DL mutation operators often fail to account for feature diversity. 

Previous studies \cite{DBLP:conf/sigsoft/Harel-CanadaWGG20, DBLP:journals/tosem/XieLWMGJL22} have explored the correlation between test coverage and output impartiality, a metric that quantifies the extent to which model predictions exhibit bias toward specific class labels. Coverage metrics sensitive to output diversity could guide test input generators to produce diverse inputs. The output impartiality is defined as: ${output\_impartiality}(T) = \frac{\sum_{t \in C_k} P_{t=C_k} \log P_{t=C_k}}{\log |C|}$, where $|C|$ denotes the number of distinct classes, and $P_{t=C_{k}}$ represents the proportion of test samples predicted to belong to class $C_{k}$. It evaluates the distribution of predictions across classes, with higher impartiality reflecting a more balanced distribution.

\textbf{(4) Coverage Profile.} Additional characteristics of coverage criteria have been analyzed, including the effects of test suite size, the correlation or comparison between different coverage metrics, the relationship between test coverage and model outputs, and the coverage differences across various models. For example, DeepImportance \cite{DBLP:conf/icse/GerasimouE0C20} and SADL \cite{DBLP:conf/icse/KimFY19} leveraged input sets with progressively increasing sizes to examine whether the growth trends of coverage remain consistent, providing insights into the correlations between different criteria. Concerning coverage profiles on the same test suites, DeepXplore \cite{DBLP:conf/sosp/PeiCYJ17} compared code coverage against Neuron Coverage using 10 randomly selected inputs from the original test set. TestRNN \cite{DBLP:journals/tr/HuangSZSRMH22} compared neuron output-based coverage metrics with RNN-specific coverage metrics on minimal test sets. Furthermore, DeepTest \cite{DBLP:conf/icse/TianPJR18} performed statistical correlation analysis to explore the relationship between test coverage metrics and the outputs of autonomous driving systems, such as steering angle and steering direction. 

\textbf{(5) Testing Cost.} Recent research \cite{DBLP:journals/tse/ZhangLAMHZ23, DBLP:conf/icse/JiMYW23, DBLP:journals/tosem/DolaDS23} has evaluated the execution cost of coverage analysis, considering practical factors such as execution time overhead and GPU memory requirements. This evaluation focuses on conducting coverage analysis on a single test suite, without the need to generate additional test sets. Consequently, treatments related to dataset manipulation are not presented in Table \ref{EmpiricalCC}.

\textbf{(6) Criteria Setting.} Recent studies \cite{DBLP:conf/icse/LiM0C19, DBLP:conf/sigsoft/Harel-CanadaWGG20} have explored the sensitivity of test coverage metrics to hyperparameters and investigated whether the findings can guide the selection of appropriate values across different testing scenarios. In addition, the choice of the network layer to which the coverage criterion is applied affects the evaluation outcomes \cite{DBLP:conf/icse/TrujilloLEDC20, DBLP:journals/tecs/SunHKSHA19, DBLP:conf/wcre/MaJXLLLZ19}. Similar to testing cost, dataset treatments are not included in Table \ref{EmpiricalCC}.

\begin{tcolorbox}
    [colback=gray!10, rounded corners]
\ding{46} \textbf{Summary} $\blacktriangleright$ Recent studies on coverage criteria have explored various dataset treatments to evaluate their effectiveness across diverse application scenarios, including employing natural inputs with different levels of diversity, simulating data distribution shifts, and generating adversarial examples. The evaluation of coverage criteria is conducted from multiple perspectives, such as their correlation with model quality properties, ability to detect errors, sensitivity to input/output diversity, and the influence of configuration settings. Despite growing research efforts, there is still no consensus on how to systematically evaluate coverage criteria. In particular, determining appropriate dataset treatments for specific testing contexts remains a non-trivial challenge due heterogeneous nature of real-world scenarios. Therefore, it is crucial to develop standardized guidelines on experimental design and dataset preparation. Such standards would support the reproducibility of empirical evaluations and enhance comparability across studies. $\blacktriangleleft$ 
\end{tcolorbox}

\subsection{Experimental Design for Evaluating Coverage-Guided Test Input Generation}
\label{SecEmp-CGT}
\begin{table}
\begin{threeparttable}
  \caption{Overview of Experimental Design for Evaluating Coverage-Guided Test Input Generation}
  \small
  \label{EmpiricalCGF}
  \fontsize{7.4}{8.7}\selectfont
  \setlength{\tabcolsep}{0.1 mm}{
  \begin{tabular}{cllc}
    \toprule
\textbf{Category} & \textbf{Evaluation Aspect} & \textbf{Reference} & \textbf{\# Studies} \\ \midrule
    
\textbf{Coverage} & Frequency of test suite satisfying a coverage target & \cite{DBLP:journals/tr/HuangSZSRMH22} & 1 \\ 
& \multirow{2}{*}{Coverage (improvement) achieved by newly generated inputs} & \cite{DBLP:conf/safecomp/AbrechtAGGHHW20, DBLP:conf/wcre/MaJXLLLZ19,  DBLP:conf/sigsoft/GuoJZCS18, DBLP:conf/issta/LeeCLO20,  DBLP:journals/corr/abs-1812-05339, DBLP:conf/sigsoft/DuXLM0Z19, DBLP:conf/icml/LiPZL21, DBLP:conf/ijcai/DemirE020, DBLP:journals/tr/HuangSZSRMH22, DBLP:conf/seke/DaiSL22, DBLP:journals/infsof/BaiHHWXQY24, DBLP:conf/icsm/BraiekK19, LIU2024107640, DBLP:conf/sigsoft/Harel-CanadaWGG20, DBLP:conf/icse/DolaMDS24,  DBLP:conf/icse/DolaDS21,  DBLP:journals/tse/GuoZZSJS22} &  \multirow{2}{*}{29} \\
\textbf{Analysis} &&\cite{DBLP:journals/ijon/ParkCKK23, DBLP:conf/icse/TianPJR18, DBLP:conf/kbse/MissaouiGM23, DBLP:journals/tr/WeiHYWW23, DBLP:conf/wcre/ZhouDLZ0Y21, DBLP:conf/issta/XieMJXCLZLYS19, DBLP:journals/tse/ZhangRDD22, DBLP:conf/sigsoft/YanTLZMX020, DBLP:conf/kbse/SunWRHKK18, DBLP:journals/tnn/YuDY23, DBLP:conf/issre/WangHWNNC24, DBLP:conf/sosp/PeiCYJ17} & \\
    \midrule

& \multirow{3}{*}{Number or proportion of (error-revealing) inputs generated} & \cite{DBLP:conf/sigsoft/GuoJZCS18, DBLP:conf/issta/LeeCLO20, DBLP:journals/tse/GuoZZSJS22, DBLP:journals/infsof/BaiHHWXQY24, DBLP:conf/icsm/BraiekK19, DBLP:conf/icml/LiPZL21, DBLP:conf/seke/DaiSL22,  DBLP:journals/tr/HuangSZSRMH22, DBLP:conf/ijcai/DemirE020,  DBLP:conf/sigsoft/DuXLM0Z19, DBLP:conf/hpcc/HanLTHG22,  DBLP:conf/wcre/MaJXLLLZ19, DBLP:conf/sigsoft/Harel-CanadaWGG20,  DBLP:conf/icse/DolaMDS24,  DBLP:conf/icse/DolaDS21, DBLP:conf/issre/0005LXC24} & \multirow{3}{*}{36}\\
&&\cite{DBLP:journals/infsof/TaoTGH023,  DBLP:conf/kbse/SunWRHKK18, DBLP:conf/sosp/PeiCYJ17, DBLP:journals/ijon/ParkCKK23, DBLP:journals/jss/WanLLCZ24, DBLP:journals/apin/SunLW23, DBLP:conf/icse/TianPJR18,   DBLP:conf/qrs/WeiC21, DBLP:journals/ese/YahmedBKBZ22, DBLP:conf/sigsoft/YanTLZMX020, DBLP:conf/wcre/YangSAL22} & \\ 
&& \cite{DBLP:journals/tse/YuanPW24, DBLP:journals/tnn/YuDY23, DBLP:conf/issta/Yuan0024, DBLP:journals/tse/ZhangLAMHZ23, DBLP:conf/wcre/ZhouDLZ0Y21, DBLP:conf/ijcnn/YeWZK22, DBLP:conf/qrs/ZhangZWCH23, DBLP:conf/icse/YuanPW23, DBLP:journals/tse/ZhangRDD22} \\
\textbf{Error} & Number of seeds from which error-revealing inputs are generated & \cite{LIU2024107640, DBLP:conf/issta/LeeCLO20, DBLP:journals/infsof/TaoTGH023} & 3\\
& Evaluation of model predictions on newly generated inputs & \cite{DBLP:journals/tse/GuoZZSJS22, DBLP:journals/corr/abs-1812-05339, DBLP:journals/tr/WeiHYWW23, DBLP:conf/qrs/YangYW22} & 4\\ 
\textbf{Detection} & Diversity of triggered errors & \cite{DBLP:conf/issta/LeeCLO20, DBLP:conf/issta/XieMJXCLZLYS19, DBLP:journals/ijon/ParkCKK23, DBLP:conf/icse/YuanPW23, DBLP:journals/infsof/BaiHHWXQY24, DBLP:journals/infsof/TaoTGH023, LIU2024107640, DBLP:journals/tse/YuanPW24, DBLP:conf/qrs/ZhangZWCH23}  & 9 \\
& Number of errors found \textit{w.r.t} the number of transformations & \cite{DBLP:conf/issta/Yuan0024} & 1\\
& Number of detected erroneous properties \textit{w.r.t} the number of epochs & \cite{DBLP:conf/issta/Yuan0024} & 1\\
\midrule

 & \multirow{2}{*}{Automated validity of the generated inputs} & \cite{ DBLP:conf/hpcc/HanLTHG22, DBLP:conf/icse/RiccioT23,  DBLP:conf/sigsoft/Harel-CanadaWGG20, DBLP:journals/tse/GuoZZSJS22, LIU2024107640,  DBLP:journals/asc/Al-NimaHACW21, DBLP:conf/kbse/MissaouiGM23, DBLP:journals/tr/HuangSZSRMH22, DBLP:journals/tecs/SunHKSHA19, DBLP:conf/icse/DolaMDS24,  DBLP:conf/icse/DolaDS21} & 18 \\
\textbf{Input} &&\cite{ DBLP:conf/icse/YuanPW23, DBLP:conf/kbse/SunWRHKK18, DBLP:conf/sigsoft/YanTLZMX020,  DBLP:journals/tse/YuanPW24,  DBLP:journals/tnn/YuDY23, DBLP:journals/ese/YahmedBKBZ22, DBLP:conf/qrs/ZhangZWCH23} \\ 
\textbf{Validity} & \multirow{2}{*}{Manual verification of the validity preservation in generated inputs} & \cite{DBLP:conf/ijcai/DemirE020, DBLP:conf/icml/LiPZL21, DBLP:conf/sigsoft/GuoJZCS18, DBLP:conf/issta/LeeCLO20, DBLP:journals/tse/GuoZZSJS22, DBLP:journals/tr/HuangSZSRMH22, DBLP:conf/hpcc/HanLTHG22, DBLP:journals/asc/Al-NimaHACW21, DBLP:conf/sosp/PeiCYJ17, DBLP:conf/sigsoft/Harel-CanadaWGG20, DBLP:conf/icse/RiccioT23, DBLP:conf/icse/DolaMDS24} & \multirow{2}{*}{18} \\ 
&& \cite{DBLP:conf/qrs/ZhangZWCH23, DBLP:conf/icse/YuanPW23, DBLP:journals/tse/YuanPW24, DBLP:conf/icse/TianPJR18, DBLP:conf/issta/XieMJXCLZLYS19, DBLP:journals/infsof/TaoTGH023}\\
\midrule

\textbf{Input}& Average distance/similarity between newly generated inputs & \cite{DBLP:conf/sosp/PeiCYJ17, DBLP:journals/jss/WanLLCZ24, DBLP:journals/tnn/YuDY23, DBLP:conf/qrs/ZhangZWCH23} & 4 \\  
\textbf{Diversity}& Internal pattern diversity of newly generated inputs & \cite{DBLP:conf/qrs/WeiC21, DBLP:conf/icse/DolaMDS24} & 2 \\
 \midrule

\textbf{Model} & Retraining the model with newly generated inputs & \cite{DBLP:journals/infsof/TaoTGH023, DBLP:journals/jss/WanLLCZ24, DBLP:journals/ijon/ParkCKK23, DBLP:conf/icse/TianPJR18, DBLP:journals/asc/Al-NimaHACW21, DBLP:conf/qrs/YangYW22, DBLP:conf/issta/Yuan0024, DBLP:conf/issre/0005LXC24} & 8  \\ 
\textbf{Enhancement} & Retraining the model with training set and newly generated inputs & \cite{DBLP:conf/sosp/PeiCYJ17, DBLP:journals/tse/GuoZZSJS22, DBLP:journals/tr/WeiHYWW23, DBLP:journals/tse/ZhangRDD22, DBLP:conf/hpcc/HanLTHG22, DBLP:conf/sigsoft/YanTLZMX020, DBLP:conf/wcre/YangSAL22, DBLP:conf/icse/GaoSPR20, DBLP:journals/tnn/YuDY23, DBLP:journals/apin/SunLW23, DBLP:conf/ijcnn/SunLS21} & 11 \\ \midrule

\multirow{5}{*}{\textbf{Efficiency}} & Execution time to achieve coverage targets & \cite{DBLP:conf/sosp/PeiCYJ17, DBLP:conf/issre/0005LXC24} & 2\\
& Coverage achieved by newly generated inputs under stopping criteria & \cite{DBLP:journals/infsof/BaiHHWXQY24, DBLP:journals/asc/Al-NimaHACW21, DBLP:conf/issta/LeeCLO20, DBLP:journals/tnn/YuDY23, DBLP:conf/seke/DaiSL22, DBLP:conf/icse/DolaMDS24, DBLP:journals/ijon/ParkCKK23} & 7 \\
& Average execution time or iterations to generate test inputs & \cite{DBLP:conf/sigsoft/GuoJZCS18, DBLP:journals/tse/ZhangLAMHZ23, DBLP:conf/sosp/PeiCYJ17, DBLP:journals/ijon/ParkCKK23, DBLP:journals/ese/YahmedBKBZ22, DBLP:journals/tse/GuoZZSJS22, LIU2024107640, DBLP:journals/tnn/YuDY23, DBLP:journals/apin/SunLW23, DBLP:conf/issre/WangHWNNC24, DBLP:conf/qrs/ZhangZWCH23} & 11 \\
& Execution time for seed selection, input mutation, or input execution & \cite{DBLP:conf/seke/DaiSL22, DBLP:journals/tr/WeiHYWW23} & 2\\
& Number of newly generated inputs within a fixed time limit & \cite{DBLP:conf/ijcai/DemirE020, DBLP:conf/ijcnn/YeWZK22, DBLP:journals/infsof/BaiHHWXQY24} & 3\\
 \midrule

\multirow{5}{*}{\textbf{Other}} & Effects of mutation strategies and their hyperparameters & \cite{DBLP:conf/sosp/PeiCYJ17, DBLP:conf/sigsoft/GuoJZCS18, DBLP:conf/issta/LeeCLO20, DBLP:journals/infsof/TaoTGH023, DBLP:journals/ese/YahmedBKBZ22, DBLP:journals/tr/WeiHYWW23, DBLP:journals/tecs/SunHKSHA19, DBLP:conf/icse/GaoSPR20, DBLP:journals/tnn/YuDY23, DBLP:journals/apin/SunLW23} & 10 \\
 & Effects of coverage guidance in test input generation & \cite{DBLP:conf/sosp/PeiCYJ17, DBLP:journals/tse/GuoZZSJS22, DBLP:journals/jss/WanLLCZ24, DBLP:conf/issta/XieMJXCLZLYS19, DBLP:conf/icml/LiPZL21, DBLP:journals/ese/YahmedBKBZ22, DBLP:journals/tr/WeiHYWW23, DBLP:conf/icse/GaoSPR20} &  8 \\ 
& Effects of hyper-parameters in test oracle configurations & \cite{DBLP:conf/icse/TianPJR18} & 1\\
 & Effects of different seed selection strategies & \cite{DBLP:journals/tr/HuangSZSRMH22} & 1 \\
& Effects of the number of iterations  & \cite{DBLP:conf/sosp/PeiCYJ17, DBLP:journals/jss/WanLLCZ24, DBLP:conf/issta/XieMJXCLZLYS19} & 3 \\
 \bottomrule
  
\end{tabular}}
\end{threeparttable}
\end{table}

Existing coverage-guided test input generation studies primarily rely on the original seed inputs and the newly generated inputs for evaluation. Table \ref{EmpiricalCGF} summarizes the experimental design for evaluating generators, categorizing them into seven aspects: coverage analysis, error detection, input validity, input diversity, model enhancement, efficiency, and other. 

\textbf{(1) Coverage Analysis.} Coverage-guided generators produce inputs that maximize coverage of specific test conditions defined by a given adequacy criterion. Prior research has assessed the extent to which newly generated inputs improve or satisfy these coverage targets, often comparing them to original test seeds or adversarial examples. Such analysis highlights the potential of CGT in improving test adequacy. For example, DeepStellar \cite{DBLP:conf/sigsoft/DuXLM0Z19} evaluated test generation effectiveness by comparing coverage results obtained from initial seeds and those generated through testing. TestRNN \cite{DBLP:journals/tr/HuangSZSRMH22} introduced the metric of coverage times, quantifying how frequently a test condition is satisfied during execution.

\textbf{(2) Error Detection.} For discriminative models, testing involves evaluating output consistency between original seeds and their mutated variants. A test is deemed to fail if the model produces divergent outputs where consistency is expected. The most commonly used metric is the number of error-revealing test inputs generated within an allocation of testing resources. Alternative terms, such as the number of erroneous behaviors, number of errors, proportion of error-revealing inputs, adversary rate, and attack success rate, are also employed in studies. For generative models, defining criteria for `passing' or `failing' a test is more complex due to the subjective and diverse nature of their outputs. As such, domain-specific metrics are employed to assess the quality and relevance of outputs, such as the Perplexity score for text generation \cite{DBLP:journals/tse/GuoZZSJS22}, Word Error Rate (WER) for speech-to-text tasks \cite{DBLP:journals/tse/GuoZZSJS22, DBLP:journals/corr/abs-1812-05339}, and BLEU scores for code summarization \cite{DBLP:journals/tr/WeiHYWW23}. Furthermore, several studies \cite{DBLP:journals/infsof/TaoTGH023, DBLP:conf/issta/LeeCLO20, LIU2024107640} reported the number of seeds from which errors are revealed, highlighting the potential of seeds in revealing errors.

Beyond the quantity of errors, recent studies have also highlighted error diversity, assuming that a diverse set of errors allows for more thorough exploration of the error space. Common metrics include the number of seeds that trigger errors \cite{DBLP:conf/issta/XieMJXCLZLYS19} and the number of classes covered by error-revealing inputs \cite{DBLP:conf/icse/YuanPW23, DBLP:conf/issta/LeeCLO20, DBLP:journals/infsof/BaiHHWXQY24}. When multiple test suites cover the same number of classes, NLC introduced an additional measure using scaled entropy, defined as: $ \# CoveredClasses = -\frac{1}{|C|}\sum p_{c}log p_{c}$, where $p_c$ denotes the ratio of incorrect predictions of class $c$, and $|C|$ denotes the total number of classes. Higher entropy indicates a more diverse distribution of errors.

\textbf{(3) Input Validity.} Input validity is a fundamental concept in software testing, ensuring that the test inputs belong to the input domain and can be executed by the software under test. In the context of DL testing, input validity extends across multiple dimensions. Beyond syntactic correctness, valid inputs must also be semantically meaningful, \textit{e.g.}, visually recognizable by human practitioners and maintain naturalness even after undergoing mutations. Recent studies have adopted both automated and manual validation strategies to assess input quality.

Automated validation employs quantitative metrics, such as semantic similarity between newly generated inputs and their seeds, and domain-specific indicators, to assess input quality. For instance, Yan et al. \cite{DBLP:conf/sigsoft/YanTLZMX020} calculated average $l_{p}$-norm distances between benign and generated inputs. RNN-Test \cite{DBLP:journals/tse/GuoZZSJS22} and TestRNN \cite{DBLP:journals/tr/HuangSZSRMH22} reported the average perturbation magnitude to quantify the distortion introduced by mutations. GANC \cite{DBLP:journals/asc/Al-NimaHACW21} adopted the Structural Similarity Index, Mean Absolute Percentage Error, and Difference Entropy to evaluate differences between generated inputs and their seeds. These metrics collectively examine whether core input features are preserved despite variations. In addition, domain-specific metrics are widely adopted. DiverGet \cite{DBLP:journals/ese/YahmedBKBZ22} assessed the quality of synthetic hyperspectral images (HSIs) using Peak Signal-to-Noise Ratio, a standard indicator of information loss caused by distortions in HSI processing. Image-domain studies \cite{DBLP:conf/hpcc/HanLTHG22, DBLP:conf/icse/YuanPW23, DBLP:conf/sigsoft/Harel-CanadaWGG20} leveraged the Inception Score \cite{DBLP:conf/nips/SalimansGZCRCC16} and Fr\'{e}chet Inception Distance \cite{DBLP:conf/nips/HeuselRUNH17} to the naturalness of generated images.

Manual Inspection involves human practitioners reviewing generated inputs to ensure semantic coherence and verifying adherence to domain-specific requirements. It is common practice to manually inspect whether the generated inputs are visually recognizable and can be confidently labeled according to the input domain. Many studies \cite{DBLP:conf/icse/TianPJR18, DBLP:conf/sosp/PeiCYJ17, DBLP:journals/infsof/TaoTGH023, DBLP:journals/tse/GuoZZSJS22} present visualizations of generated inputs to demonstrate their naturalness and relevance, and report the number or proportion of inputs deemed valid through manual analysis \cite{DBLP:conf/icse/RiccioT23}.

\textbf{(4) Input Diversity.} To quantify input diversity, recent studies employed various distance metrics to measure the dissimilarity among newly generated inputs. These metrics provide insights into how broadly the inputs cover the input space. For example, DRLFuzz \cite{DBLP:journals/jss/WanLLCZ24} utilized the Mean Pairwise Distance, calculated as the average Euclidean distance between all pairs of inputs, to quantify diversity. CIT4DNN \cite{DBLP:conf/icse/DolaMDS24} and FilterFuzz \cite{DBLP:conf/qrs/WeiC21} utilized the diversity of internal patterns within DL models as an indirect measure of input diversity. Specifically, CIT4DNN measured feature diversity of test sets using Filled Cells and Coverage Sparseness metrics computed over feature maps. FilterFuzz evaluated diversity based on activated filter patterns elicited by the inputs.

\textbf{(5) Model Enhancement.} Current studies have incorporated newly generated inputs into model retraining or fine-tuning processes. One strategy is to update DL models with new inputs, without retraining from scratch on the entire dataset \cite{DBLP:journals/infsof/TaoTGH023, DBLP:journals/jss/WanLLCZ24, DBLP:journals/ijon/ParkCKK23, DBLP:conf/icse/TianPJR18, DBLP:journals/asc/Al-NimaHACW21, DBLP:conf/issta/Yuan0024}. Alternatively, to alleviate catastrophic forgetting, other studies \cite{DBLP:journals/tr/WeiHYWW23, DBLP:journals/tse/ZhangRDD22, DBLP:conf/hpcc/HanLTHG22, DBLP:conf/sigsoft/YanTLZMX020, DBLP:conf/wcre/YangSAL22} augment the original training set with generated inputs and retrain the model.

For discriminative models, model accuracy before and after retraining is measured. For example, DeepTest \cite{DBLP:conf/icse/TianPJR18} assessed whether retraining DL-based autonomous driving systems with synthetic images improves robustness, using Mean Squared Error as the metric. For generative models, task-specific metrics are used. RNN-Test \cite{DBLP:journals/tse/GuoZZSJS22}, for instance, evaluated the Perplexity changes of the language model after retraining.

\textbf{(6) Efficiency.} This aspect primarily evaluates the efficiency of test input generation by examining the execution time or the number of iterations required to achieve specific testing objectives \cite{DBLP:journals/infsof/BaiHHWXQY24, DBLP:conf/seke/DaiSL22, DBLP:journals/jss/WanLLCZ24, DBLP:journals/ese/YahmedBKBZ22}. It also considers the extent to which these testing objectives can be met under fixed constraints of time and iterations \cite{DBLP:journals/ijon/ParkCKK23, DBLP:conf/ijcai/DemirE020}. In this context, testing objectives refer to achieving predefined coverage targets with the newly generated inputs or generating a sufficient number of error-revealing inputs. 

\textbf{(7) Other.} This aspect explores how various configurations, such as mutation hyperparameters, seed selection, coverage guidance, test oracle design, and iteration count, impact the effectiveness of test input generation and helps identify optimal settings. ADAPT \cite{DBLP:conf/issta/LeeCLO20} and DLRegion \cite{DBLP:journals/infsof/TaoTGH023} investigated how different neuron selection strategies, used to define optimization objectives, affect input mutation. CoCoFuzzing \cite{DBLP:journals/tr/WeiHYWW23} compared mutation operators by evaluating their coverage improvements. Regarding coverage guidance, RNN-Test \cite{DBLP:journals/tse/GuoZZSJS22} evaluated test generators under RNN-specific coverage and Neuron Coverage to assess their effectiveness in fault detection.

\begin{tcolorbox}
    [colback=gray!10, rounded corners]
\ding{46} \textbf{Summary} $\blacktriangleright$ Among the aspects for evaluating coverage-guided test input generation, error detection capability has received the most attention, followed by input validity and coverage analysis. To support these evaluations, various domain-specific and task-dependent evaluation metrics are employed. However, research has predominantly focused on adversarial-oriented testing. The actual impact of coverage guidance on the input generation remains insufficiently explored. The potential of CGT as a fundamental paradigm for DL quality assurance remains an open question. Therefore, further investigation is necessary to clarify the specific role of coverage in guiding the generation process. $\blacktriangleleft$ 
\end{tcolorbox}

\subsection{Experimental Design for Evaluating Coverage-Guided Test Optimization}
\label{SecEmp-Opt}

\begin{table}
\begin{threeparttable}
  \caption{Overview of Experimental Design for Evaluating Coverage-Guided Test Optimization}
  \small
  \label{EmpiricalOpt}
  \fontsize{7.3}{9.7}\selectfont
  \setlength{\tabcolsep}{0.1 mm}{
  \begin{tabular}{clllc}
    \toprule
\textbf{Category} & \textbf{Task} \customfootnote{$\dagger$} &\textbf{Evaluation Aspect} & \textbf{Reference} & \textbf{\# Studies} \\ \midrule
    
\textbf{Model Performance}& \multirow{2}{*}{Minimization} & Accuracy difference achieved between the reduced and the original set &\cite{DBLP:conf/issre/ZhouLD0H20, DBLP:journals/infsof/ZhaoMCZJW22} & 2 \\
\textbf{Estimation}& & Correlation between accuracy estimation error and coverage difference & \cite{DBLP:journals/infsof/ZhaoMCZJW22} & 1 \\ \midrule

& Prioritization & Average Percentage of Fault Detection (APFD) & \cite{DBLP:conf/issta/FengSGWF020, Weiss-issta, 10.1145/3672446} & 3 \\
\textbf{Error} & Prioritization, Selection & Number or ratio of the detected errors / error types & \cite{DBLP:conf/icse/GaoFYL0X22, 10319282, deepvec, DBLP:journals/infsof/ShiYS25} & 4 \\
\textbf{Detection} & Prioritization, Selection &  Correlation between coverage-based metrics and error detection & \cite{DBLP:journals/tosem/MaPTCT21} & 1 \\
& Selection & Average rewards achieved by the selected state sequences & \cite{DBLP:journals/jss/ShiYZ24} & 1 \\
 \midrule

\textbf{Model Enhancement}  & Selection & Model retraining on both the training set and selected input set &  \cite{DBLP:journals/tosem/MaPTCT21, Weiss-issta, DBLP:conf/icse/GaoFYL0X22, 10319282, deepvec} & 5 \\ \midrule

\textbf{Input Diversity} & Minimization & Average distance between inputs  & \cite{DBLP:conf/setta/SunLLS23} & 1 \\ \midrule

\textbf{Efficiency} & / & Execution time overhead & \cite{DBLP:conf/issta/FengSGWF020, Weiss-issta, deepvec} & 3 \\ \midrule

\multirow{2}{*}{\textbf{Other}} & / & Effects of coverage guidance & \cite{DBLP:conf/issre/ZhouLD0H20} & 1 \\
& / & Effects of stopping criteria & \cite{DBLP:conf/issre/ZhouLD0H20} & 1 \\
 \bottomrule
  
\end{tabular}}
\begin{tablenotes}
\small
  \item[\customtnote{$\dagger$}] This column outlines the optimization tasks corresponding to each category, while `/' denotes task-agnostic evaluation.
\end{tablenotes}
\end{threeparttable}
\end{table}
Table \ref{EmpiricalOpt} presents the experimental design for evaluating coverage-guided test optimization, categorizing them into six aspects: model performance estimation, error detection, model enhancement, input diversity, efficiency, and other. Among these categories, efficiency and other configurations are task-agnostic. For the remaining categories, we specify their corresponding optimization tasks.

\textbf{(1) Model Performance Estimation.} 
This evaluation is designed for test suite minimization, measuring how well a reduced subset estimates model performance. Existing approaches \cite{DBLP:journals/infsof/ZhaoMCZJW22, DBLP:conf/issre/ZhouLD0H20} primarily assess accuracy estimation by comparing model accuracy on the reduced set versus the full test set. A small accuracy gap indicates that the reduced subset retains the critical characteristics of the full set necessary for accuracy estimation. Besides, Zhao et al. \cite{DBLP:journals/infsof/ZhaoMCZJW22} explored the correlation between accuracy estimation error and test coverage differences. While accuracy difference provides a basic metric, existing studies often overlook other aspects, such as class distribution, model confidence calibration, and other quality properties, which are essential for ensuring the representativeness of the minimized set.

\textbf{(2) Error Detection.} Test input prioritization is commonly evaluated using the Average Percentage of Fault Detection (APFD) metric, a widely adopted indicator in traditional software testing. Higher APFD values indicate that a greater proportion of errors are detected earlier in the testing process \cite{DBLP:journals/stvr/YooH12}. Given their shared objective of identifying error-revealing inputs from an unlabeled input corpus, prioritization and selection tasks often adopt similar evaluation aspects related to error detection. Common evaluation metrics include the number of detected errors or error types \cite{DBLP:conf/icse/GaoZMSCW22}, and the correlation between coverage-based metrics and error detection \cite{DBLP:journals/tosem/MaPTCT21}. In addition, Shi et al. \cite{DBLP:journals/jss/ShiYZ24} observed a negative correlation between test coverage and DRL rewards. They analyzed the rewards of coverage-improving state sequences to assess their error detection capability.

\textbf{(3) Model Enhancement.} This aspect evaluates the effectiveness of test input selection in model enhancement. Two retraining strategies are employed: retraining the model using the selected subset or augmenting the original training set with selected inputs. Recent coverage-guided selection studies \cite{DBLP:journals/tosem/MaPTCT21, Weiss-issta, DBLP:conf/icse/GaoFYL0X22} favor the latter. Effectiveness is typically evaluated by comparing model accuracy on a held-out test set before and after retraining.

\textbf{(4) Input Diversity.} HeatC \cite{DBLP:conf/setta/SunLLS23} evaluates the diversity of the minimized test set by measuring the average distance between inputs. Higher diversity implies that the optimized test set covers a wide range of input features, potentially maintaining the test adequacy.

\textbf{(5) Efficiency.} Current research \cite{DBLP:conf/issta/FengSGWF020, Weiss-issta} reported execution time overheads associated with optimizing test suites. Considering the context of DL test optimization was introduced, efficiency is inherently tied to label annotation costs \cite{DBLP:conf/icse/HuGXCPMT23}, highlighting the need to evaluate annotation efficiency.

\textbf{(6) Other.} Zhou et al. \cite{DBLP:conf/issre/ZhouLD0H20} investigated how various configurations affect optimization effectiveness, including the role of coverage guidance and stopping criteria.

\begin{tcolorbox}
    [colback=gray!10, rounded corners]
\ding{46} \textbf{Summary} $\blacktriangleright$ The design of experimental evaluations in test optimization varies according to the specific optimization objectives. Studies on test suite minimization assess whether the reduced set preserves the key properties of the original set. Research on prioritization and selection focuses on evaluating error detection effectiveness and accuracy improvement capabilities. Despite considerable efforts in evaluating optimization techniques, a major limitation remains the reliance on a narrow set of metrics. To advance the field, future research should extend beyond accuracy-centric evaluations and incorporate additional metrics, such as class-wise performance, robustness assessments, and more comprehensive diversity measures. Besides, given the unique challenges of DL testing, particular attention must be paid to the cost-effectiveness of label annotation, an essential consideration that sets DL test optimization apart from traditional regression testing paradigms. $\blacktriangleleft$ 
\end{tcolorbox}

\subsection{Empirical Study}
\label{empirical}
\begin{figure}  
    \centering  
    \includegraphics[width = 15 cm]{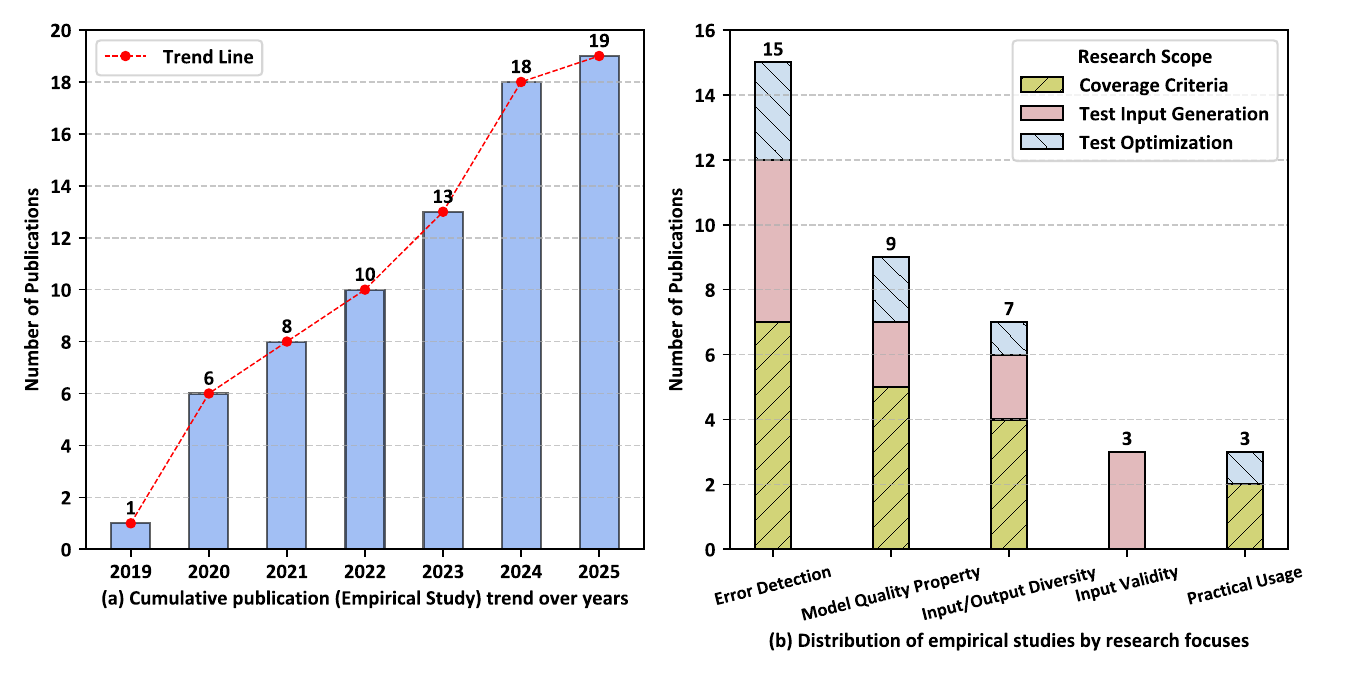} 
    \caption{Publication Trends and Research Focuses of Empirical Studies on Coverage-Guided Testing}  
    \label{PublicationTrends-Emp}
\end{figure}

The CGT field has benefited from a growing body of empirical research, investigating various aspects, such as the sensitivity of different coverage criteria to error detection and input diversity, the correlations between test coverage and model quality properties, and the effectiveness of coverage-guided input generators. Figure \ref{PublicationTrends-Emp}(a) presents the publication trends of empirical studies from 2019 to 2025. By May 2025, the total number of publications reached 19, reflecting a growing interest in investigating the actual performance of studies within this field. Figure \ref{PublicationTrends-Emp}(b) illustrates the distribution of empirical studies across different research focuses. Among these, `Practical Usage' refers to studies that explore the impact of various factors on CGT, taking practical implementations into account. As shown, empirical studies primarily focus on error detection, followed by model quality property and input/output diversity. Table \ref{Tab-Empirical} provides a detailed summary of empirical studies. 

\begin{table}
\begin{threeparttable}
  \caption{Summary of Empirical Studies Related to Coverage-Guided Testing Organized by Publication Year}
  \footnotesize
  \label{Tab-Empirical}
  \fontsize{8.6}{12.2}\selectfont
  \setlength{\tabcolsep}{0.15 mm}{
  \begin{tabular}{llcl}
    \toprule
\textbf{Year} & \textbf{Study} & \textbf{Scope} \customfootnote{$\dagger$} & \multicolumn{1}{c}{\textbf{Research Focus}} \\ \midrule

2019 & Li et al. \cite{DBLP:conf/icse/LiM0C19} &  \textcolor{teal} {\uppercase\expandafter{\romannumeral1}} & Correlation between test coverage and error detection. \\

2020 & Jahangirova et al. \cite{DBLP:conf/icst/JahangirovaT20} & \textcolor{teal} {\uppercase\expandafter{\romannumeral1}} & Comparison of test coverage and mutation scores for error detection.\\

2020 & Dong et al. \cite{DBLP:conf/iceccs/DongZWLSH0WDD20} & \textcolor{teal} {\uppercase\expandafter{\romannumeral1}} & Correlation between test coverage and model robustness. \\

2020 & Fabrice et al. \cite{DBLP:conf/sigsoft/Harel-CanadaWGG20}  & \textcolor{orange} {\uppercase\expandafter{\romannumeral2}} & Effectiveness of CGF in error detection, input naturalness, and output impartiality.  \\

\multirow{2}{*}{2020} & \multirow{2}{*}{Yan et. al. \cite{DBLP:conf/sigsoft/YanTLZMX020}} & \multirow{2}{*}{\textcolor{teal} {\uppercase\expandafter{\romannumeral1}}, \textcolor{orange} {\uppercase\expandafter{\romannumeral2}}}& Correlation between test coverage and model robustness, and effectiveness of \\ 
&&& CGF in error detection and robustness improvement. \\

2020 & Yang et al. \cite{DBLP:conf/wcre/YangSAL22} &  \textcolor{orange} {\uppercase\expandafter{\romannumeral2}} & Comparison of CGF and gradient-based adversarial attack for robustness enhancement. \\

2021 & Ma et al. \cite{DBLP:journals/tosem/MaPTCT21} & \textcolor{magenta} {\uppercase\expandafter{\romannumeral3}} & Comparison of coverage-based and uncertainty-based metrics for test input selection. \\

2021 & Sun et al. \cite{DBLP:conf/ijcnn/SunLS21} & \textcolor{teal} {\uppercase\expandafter{\romannumeral1}}  & Effectiveness of coverage criteria in distinguishing test set quality and enhancing robustness.\\

2022 & Weiss et al. \cite{Weiss-issta} & \textcolor{magenta} {\uppercase\expandafter{\romannumeral3}} & Comparison of coverage-based and uncertainty-based metrics for test input prioritization. \\

2022 & Zhao et al. \cite{DBLP:journals/infsof/ZhaoMCZJW22} & \textcolor{magenta} {\uppercase\expandafter{\romannumeral3}} & Effectiveness of test suite minimization in preserving test diversity.\\

2023 & Yuan et al. \cite{DBLP:conf/icse/YuanPW23} & \textcolor{teal} {\uppercase\expandafter{\romannumeral1}}, \textcolor{orange} {\uppercase\expandafter{\romannumeral2}} & Comparison of coverage criteria considering distribution properties and practical concerns.\\

\multirow{2}{*}{2023} & \multirow{2}{*}{Usman et al. \cite{DBLP:journals/sttt/UsmanSGDMP23}} & \multirow{2}{*}{\textcolor{teal} {\uppercase\expandafter{\romannumeral1}}} & Evaluation of coverage criteria \textit{w.r.t} input class diversity and error detection capabilities\\
&&& under adversarial attacks and data poisoning scenarios. \\

2023 & Riccio et al. \cite{DBLP:conf/icse/RiccioT23} & \textcolor{orange} {\uppercase\expandafter{\romannumeral2}} & Comparison of coverage-guided and black-box generators \textit{w.r.t} input validity. \\

2024 & Zheng et al. \cite{DBLP:journals/tse/ZhengLWC24} & \textcolor{teal} {\uppercase\expandafter{\romannumeral1}} & Correlation between test coverage and model fairness. \\

2024 & Yan et al. \cite{DBLP:journals/smr/YanCCWKW24} & \textcolor{teal} {\uppercase\expandafter{\romannumeral1}} & Correlation of test coverage with error detection and input diversity. \\

\multirow{2}{*}{2024} & \multirow{2}{*}{Wang et al. \cite{10.1145/3672446}} & \multirow{2}{*}{\textcolor{teal} {\uppercase\expandafter{\romannumeral1}}, \textcolor{magenta} {\uppercase\expandafter{\romannumeral3}}} & Correlation between test coverage and input diversity, impact of criteria \\
&&& parameters, and effectiveness in prioritization of erroneous ID inputs, and OOD detection. \\ 

2024 & Yuan et al. \cite{DBLP:conf/issta/Yuan0024} & \textcolor{orange} {\uppercase\expandafter{\romannumeral2}} & Limitations of error-triggering inputs produced by white-box and black-box generators.  \\

\multirow{2}{*}{2024} & \multirow{2}{*}{Du et al. \cite{DBLP:conf/esem/DuC24}} & \multirow{2}{*}{\textcolor{teal} {\uppercase\expandafter{\romannumeral1}}} & Impact of context settings on fairness testing, particularly the correlation between test\\
&&& coverage and fairness metrics under hyperparameter changes. \\

2025 & Zhou et al. \cite{DBLP:journals/sqj/ZhouPCQZZ25} & \textcolor{teal} {\uppercase\expandafter{\romannumeral1}} & Correlation between test coverage and error detection based on metamorphic testing. \\
\bottomrule
  
\end{tabular}}
\begin{tablenotes}
\small
\item[\customtnote{$\dagger$}] The scope of the empirical study: `\textcolor{teal} {\uppercase\expandafter{\romannumeral1}}' represents the coverage criteria evaluation, `\textcolor{orange} {\uppercase\expandafter{\romannumeral2}}' refers to the assessment of coverage-guided test input generators, and `\textcolor{magenta} {\uppercase\expandafter{\romannumeral3}}' denotes the analysis of coverage-guided optimization approaches. 
\end{tablenotes}
\end{threeparttable}
\end{table}

\subsubsection{Empirical Study on Coverage Criteria}
Early coverage criteria for DL models were motivated by traditional software testing principles, where higher coverage correlates with stronger error detection. However, empirical studies have challenged this assumption. Li et al. \cite{DBLP:conf/icse/LiM0C19} found that the effectiveness of high-coverage tests is largely due to the adversarial-oriented nature of the generated inputs, rather than the strength of coverage criteria. Results also reveal no significant correlation between neuron output-based coverage and misclassification rates on natural inputs. Jahangirova et al. \cite{DBLP:conf/icst/JahangirovaT20} compared test coverage with mutation scores (mutation testing) across test suites of varying error rates and found mutation operators to be more sensitive to error detection than Neuron Coverage \cite{DBLP:conf/sosp/PeiCYJ17} and Surprise Adequacy \cite{DBLP:conf/icse/KimFY19}. To mitigate the effect of test suite size, Yan et al. \cite{DBLP:journals/smr/YanCCWKW24} analyzed coverage variations across test suites of equal size but differing proportions of adversarial examples. Results indicate that Surprise Coverage \cite{DBLP:conf/icse/KimFY19} correlates positively with error detection, whereas neuron output-based criteria \cite{DBLP:conf/sosp/PeiCYJ17, DBLP:conf/wcre/MaJXLLLZ19} do not. Wang et al. \cite{10.1145/3672446} examined how test coverage responds to ID errors and OOD samples. Neuron-based criteria are shown to be less indicative of natural errors compared to Surprise Coverage. Zhou et al. \cite{DBLP:journals/sqj/ZhouPCQZZ25} introduce the violation rate (VR) to quantify error detection via metamorphic testing. They analyzed the correlation between VR and coverage metrics, finding that while some coverage metrics correlate with VR, the correlations lack robustness across varying test set sizes, metamorphic relations (MRs), and datasets.

As traditional structural testing assumes that higher coverage indicates greater test diversity, recent studies \cite{DBLP:conf/ijcnn/SunLS21, DBLP:journals/sttt/UsmanSGDMP23, DBLP:journals/smr/YanCCWKW24, 10.1145/3672446} have explored whether DL coverage reflect test diversity. Yan et al. \cite{DBLP:journals/smr/YanCCWKW24} analyzed the correlation between test coverage and input class diversity by examining coverage variations across multiple test suites of equal size but with differing input class proportions. Results demonstrate that Neuron Coverage and $k$-Multisection Neuron Coverage increase with class diversity, particularly when the number of classes in the test suite is relatively small. Surprise Coverage and Neuron Boundary Coverage exhibit limited correlation. Similar observations can be drawn by Wang et al. \cite{10.1145/3672446}. Their results suggest that earlier proposed criteria struggle to achieve both strong error detection capability and sensitivity to test diversity. 

Previous studies have investigated the correlation between test coverage and model quality by comparing coverage metrics on test suites with corresponding quality measures. However, the empirical evidence has generally been discouraging. Regarding robustness, Dong et al. \cite{DBLP:conf/iceccs/DongZWLSH0WDD20} investigated the correlation between coverage metrics and robustness indicators (\textit{e.g.}, Lipschitz constant \cite{DBLP:journals/ml/XuM12}, CLEVER score \cite{weng2018evaluating}) on the original test set. They also compared changes in coverage and robustness before and after retraining. Results reveal a weak correlation between test coverage and robustness, with similar trends in both coverage and robustness changes. Yan et al. \cite{DBLP:conf/sigsoft/YanTLZMX020} further found that adding coverage-improving inputs does not necessarily enhance robustness. In terms of fairness, Zheng et al. \cite{DBLP:journals/tse/ZhengLWC24} investigated the correlation between test coverage and model fairness, using group and individual evaluation indicators. Their results reveal no strong positive correlation. After applying bias mitigation, coverage metrics exhibit inconsistent behaviors. Additionally, Du et al.\cite{DBLP:conf/esem/DuC24} investigated how contextual factors (\textit{e.g.}, model-level hyperparameters and data-level label bias) impact fairness testing, observing weak correlations between test adequacy and fairness metrics under different contextual settings.

Yuan et al. \cite{DBLP:conf/icse/YuanPW23} proposed eight requirements for coverage criteria, accounting for both distributional characteristics and practical considerations. Among these, four factors are particularly relevant to real-world usage: training-phase knowledge, matrix-form computation, incremental updates, and user-friendliness. Based on these requirements, they conducted a qualitative comparison of various criteria. They also evaluated their sensitivity to test diversity, error detection capability, and the effectiveness of input generation. In addition, Wang et al. \cite{10.1145/3672446} investigated how hyperparameter configurations and architectural characteristics affect coverage performance.

\subsubsection{Empirical Study on Coverage-guided Test Input Generation}
Empirical studies on test input generation primarily evaluate various generators along dimensions such as error detection, input diversity, model enhancement, output impartiality, and input validity. Fabrice et al. \cite{DBLP:conf/sigsoft/Harel-CanadaWGG20} conducted an early investigation into the utility of Neuron Coverage as a target for test generation, evaluating its effectiveness in detecting errors, preserving input naturalness, and maintaining output impartiality. The findings raise skepticism about whether Neuron Coverage is a meaningful test target, suggesting that increasing coverage may compromise other important properties.

Yan et al. \cite{DBLP:conf/sigsoft/YanTLZMX020} compared a coverage-guided generator (DeepHunter \cite{DBLP:conf/issta/XieMJXCLZLYS19}) with a gradient descent-based adversarial generator (PGD \cite{madry2018towards}) in terms of error detection and model enhancement. Yang et al. \cite{DBLP:conf/wcre/YangSAL22} extended this study by incorporating additional datasets, DNN architectures, and baseline approaches. Their findings show that model quality does not improve with increased coverage. Gradient-based methods outperform coverage-driven ones in error detection, and retraining with one generator's inputs does not mitigate errors found by the other. Additionally, Yuan et al. \cite{DBLP:conf/icse/YuanPW23} evaluated generated inputs from different generators along dimensions including the number of triggered errors, diversity of erroneous behaviors, and input naturalness. 

Riccio et al. \cite{DBLP:conf/icse/RiccioT23} assessed the validity of inputs produced by various generators, using both automated and human validators. The study investigates four aspects: the validity of misbehavior-triggering inputs assessed by automated validators, the label preservation of misclassified inputs, and the consistency between automated and human validation. Results indicate that although generators often produce syntactically valid inputs, they frequently fail to preserve correct ground-truth labels. Automated validation generally aligns with human assessments but struggles with complex datasets and semantically valid yet atypical inputs.

Yuan et al. \cite{DBLP:conf/issta/Yuan0024} discussed limitations in existing DL testing pipelines, highlighting two challenges: the overwhelming number of error-triggering inputs and the issue of mutually exclusive input properties. The findings reveal that existing generators tend to produce redundant error-revealing inputs using limited transformations, and fine-tuning with such inputs inadvertently increases model fragility due to conflicting input properties. Therefore, they propose a property-oriented testing framework that shifts the focus from individual error-revealing inputs to the input transformations responsible for mispredictions.

\subsubsection{Empirical Study on Coverage-guided Test Optimization}
This section reviews empirical studies on coverage-guided test optimization, excluding those not involving coverage-guided methods. Weiss et al. \cite{Weiss-issta} compared the uncertainty-based metric DeepGini \cite{DBLP:conf/issta/FengSGWF020} with coverage-guided metrics in terms of execution time, input prioritization, and retraining effectiveness, finding that DeepGini performs better overall. Ma et al. \cite{DBLP:journals/tosem/MaPTCT21} evaluated various test input selection metrics, finding that uncertainty-based metrics correlate moderately to strongly with misclassification and outperform coverage-based metrics in enhancing model accuracy through retraining.

Wang et al. \cite{10.1145/3672446} adopted a coverage-addition strategy for ID input prioritization. Starting from an empty set, inputs are incrementally selected based on the largest coverage increase, remaining inputs are then ranked by descending coverage. Using the APFD metric to evaluate prioritization, results show that Distance-based Surprise Coverage outperforms both neuron output-based metrics and uncertainty-based metrics.

Zhao et al. \cite{DBLP:journals/infsof/ZhaoMCZJW22} examined whether test suite minimization methods preserve test diversity, with a focus on maintaining adequate class label coverage. Their findings reveal that prior approaches overlook certain classes, leading to reduced accuracy in performance estimation. Therefore, they propose a multi-objective optimization-based approach that maximizes class coverage while minimizing data distribution differences across clusters. Experimental results demonstrate improved class diversity and estimation accuracy over existing baselines.

\begin{tcolorbox}
    [colback=gray!10, rounded corners]
\ding{46} \textbf{Summary} $\blacktriangleright$ Empirical studies have evaluated the effectiveness of CGT approaches from multiple perspectives, employing various dataset treatments. Concerns have emerged regarding the insensitivity of some early methods to error detection and input/output diversity, weak correlation with model quality properties, and limited feasibility in real-world applications. These critiques offer valuable insights, challenging initial assumptions and informing the direction of future research. Given the diverse objectives of DL testing, there is a pressing need for more comprehensive empirical investigations to illuminate the landscape of CGT methodologies. Future work should broaden their applications to more domains, diversify evaluation metrics, and assess recent advancements. These issues are discussed in Section \ref{Sec-challenges}.
  $\blacktriangleleft$ 
\end{tcolorbox}

\section{Open Challenge and Future Direction}
\label{Sec-challenges}
This section highlights current challenges and promising directions for advancing coverage-guided testing.

\textbf{\ding{182} \textit{Correlation between Structural Coverage and Testing Objectives}.} Certain coverage criteria focus on low-level syntactic behaviors of DL models, but fail to capture higher-order semantics that are critical for testing objectives, such as robustness to input perturbations, fairness across sensitive attributes, and generalization to unseen distributions. This mismatch between structural exploration and semantic testing objectives has been highlighted by recent empirical studies, which reveal weak correlations between structural coverage and model quality properties (robustness, fairness). Increasing structural coverage does not necessarily improve error detection effectiveness. These investigations challenge a long-standing assumption, borrowed from traditional software testing, that broader structural exploration enhances testing adequacy. As a result, the practical utility of structural coverage as a reliable indicator of test adequacy in DL contexts has come under serious scrutiny.

These insights call for a paradigm shift in how coverage is conceptualized and utilized in DL model testing. Future research should develop coverage metrics that are intrinsically aligned with testing objectives rather than purely structural exploration. From white-box and gray-box perspectives, explainability techniques and semantic abstractions within models could be leveraged to capture regions that are particularly sensitive to input perturbations, sensitive attributes, or data distribution shifts, thereby redefining the coverage domain. From a black-box perspective, it is preferable to target input and output spaces, capturing regions in the input space where the model exhibits low confidence or ensuring that the model’s possible decisions are adequately explored.

\textbf{\ding{183} \textit{Generalizability Across Network Architectures and Tasks}.}
A majority of coverage criteria are designed for FNNs and standard CNNs, with limited efforts targeting RNNs, Transformers, and DRLs. CGT approaches for other architectures, such as Graph Neural Networks, Vision Transformers, and language models like GPT-series, remain largely unexplored. In addition, current research predominantly focuses on discriminative models, while generative models, such as GANs and diffusion models, have received little attention. This lack of CGT techniques for diverse architectures and learning paradigms reveals an important gap and calls for broader exploration.

Looking ahead, a comprehensive evaluation of the generalizability of existing CGT methods across different network architectures and tasks is essential. It helps identify which techniques are universally applicable and which require adaptation to specific models and learning paradigms. In addition, extending CGT to emerging network architectures and learning paradigms presents both opportunities and challenges. A key direction involves designing CGT approaches that capture the distinctive structural and functional properties of these models. Furthermore, task diversity demands rethinking CGT notions, particularly for generative models, whose complex, high-dimensional outputs require testing not only for correctness but also for diversity and creativity.

\textbf{\ding{184} \textit{Customization of CGT for Diverse Domains}.}
Most CGT research remains concentrated on computer vision tasks. While several studies have explored other domains, such as Natural Language Processing, speech recognition, malware detection, and DRL-based sequential decision-making, these efforts remain relatively insufficient. Different domains involve diverse input modalities (\textit{e.g.}, natural language texts, audio waveforms, syntax trees, environmental state-action trajectories, or 3D structures like point clouds and meshes). They often exhibit domain-specific characteristics, such as temporal dependencies in time-series data, hierarchical structure in code, or dynamic interactions in sequential decision-making. These intrinsic differences pose fundamental challenges for CGT, especially in designing modality-aware black-box coverage criteria, developing semantically valid input mutation strategies, and constructing optimization methods that generalize across domains.

To bridge the gap between general-purpose CGT frameworks and the specialized requirements of real-world DL-enabled applications, future research should focus on domain-specific customization along three directions. First, a key direction involves designing black-box coverage criteria that reflect the structural and semantic characteristics of domain inputs, \textit{e.g.}, syntax- or grammar-aware criteria for natural language, control- and data-flow-sensitive metrics for source code, or segment-level abstraction for temporal signals. Second, input mutations must be developed with domain constraints to preserve the semantic integrity of the generated inputs. Third, there is a pressing need for modular and adaptable CGT architectures that support plug-and-play integration of domain-specific components. Such flexible architectures would empower practitioners to assemble customized testing pipelines that align with the unique requirements of diverse domains.

\textbf{\ding{185} \textit{Overhead and Scalability Concerns}.}
White-box and gray-box CGT approaches require fine-grained analysis of internal states, incurring significant computational and memory overhead, especially when applied to large-scale DL models with billions of intermediate activations. Although black-box CGT avoids internal inspection, it faces scalability challenges due to the vast input and output spaces of modern models. Moreover, most existing coverage implementations do not support incremental updates, necessitating full recomputation of coverage information with each new input. These limitations hinder the practical application of CGT in large-scale or online testing pipelines. As DNN architectures continue to grow in depth and architectural complexity, the lack of scalable coverage analysis mechanisms becomes a critical bottleneck.

To address these limitations, coverage estimation techniques, such as sampling-based strategies and surrogate modeling, offer a promising way to reduce overhead while preserving the informativeness of coverage signals. In addition, incremental measurement mechanisms that update coverage dynamically, without recomputing from scratch, are critical for efficient testing. As large-scale models become increasingly common, there is a clear need for coverage frameworks tailored to their complexity. Instead of relying on exhaustive internal exploration, future efforts should prioritize scalable abstractions, critical state selection, and hierarchical analysis to manage resource costs while maintaining test effectiveness.

\textbf{\ding{186} \textit{Practical Concerns towards Contextual Factors}.}
In real-world deployments, DL models with identical architectures often diverge due to differences in hyperparameter configurations, optimization techniques, and training data distributions. These contextual factors can impact the effectiveness of CGT. However, existing studies have paid little attention to the sensitivity of coverage criteria and CGT approaches to contextual factors, limiting the generalizability of current empirical findings.

Empirical studies are needed to systematically analyze how diverse contextual factors affect the performance of coverage criteria and CGT methods. Besides, exploring context-agnostic coverage metrics and testing strategies offers a promising path toward improving their robustness against contextual changes. Such efforts would enhance the practical applicability of CGT across diverse real-world deployment settings. 

\textbf{\ding{187} \textit{Standardized Evaluation Protocols}.}
A major challenge in CGT research lies in the lack of standardized benchmarks and evaluation protocols. Inconsistent experimental setups, dataset treatments, and metric designs influence cross-study comparability and industrial deployment. Besides, evaluation metrics for test input optimization often focus narrowly on coverage improvement or error detection, overlooking key practical factors such as the annotation cost-effectiveness and the diversity of selected inputs at the class and feature levels.

We advocate for the development of standardized evaluation protocols informed by systematic analyses of how different dataset treatments affect coverage criteria. Evaluation datasets should mirror real-world deployment scenarios, including distribution shifts and adversarial perturbations. Moreover, coverage-oriented and adversarial-oriented metrics should be complemented with other dimensions. For example, test input prioritization and selection methods could be evaluated with measures capturing class bias and feature diversity. Practical considerations of labeling costs for test optimization tasks should also be incorporated. These multidimensional metrics are essential for guiding the development of CGT techniques.

\textbf{\ding{188} \textit{Tool Support for CGT}.}
Although many studies release available repositories for the proposed approaches, these implementations are often built on different DL frameworks (\textit{e.g.}, TensorFlow, PyTorch), leading to inconsistencies in coding practices, such as gradient computation, intermediate state monitoring, and input transformation encoding. This fragmentation hampers reproducibility and limits community-wide adoption. Currently, the field lacks toolkits that unify CGT support across frameworks.

To fill this gap, high-quality, framework-agnostic toolkits are urgently needed to facilitate the adoption of CGT. Such toolkits should fulfill the following key requirements. First, the toolkits should be capable of interfacing with models built using different DL frameworks, minimizing the need for reimplementation. Second, a modular design is essential for users to flexibly define, extend, and replace core components such as coverage criteria, seed mechanisms, and input mutation strategies. Customization of core hyperparameters (\textit{e.g.}, mutation rates, sampling thresholds, coverage targets) should also be supported to accommodate diverse research needs. Third, the toolkits should provide standardized implementations for diverse dataset treatments and evaluation metrics, promoting rigorous comparison and reproducibility across CGT studies.

\section{Conclusion}
\label{Sec-conclusion}
CGT has rapidly emerged as a cornerstone of quality assurance for DL models. To illuminate the current landscape and inspire future research, this article presents the first comprehensive survey dedicated to CGT. It systematically reviews recent advancements across three fundamental aspects: coverage analysis, coverage-guided test input generation, and coverage-guided test optimization. We propose refined taxonomies to categorize existing approaches based on their core methodologies and application contexts. Moreover, this article also summarizes experimental designs commonly adopted in the literature. Despite the significant progress achieved, several open challenges remain. This article pinpoints key limitations in current CGT studies and outlines promising directions for future research. We hope that this work serves as both a comprehensive reference for researchers seeking an in-depth understanding of CGT and a foundation to support the continued evolution of trustworthy DL models.

\begin{acks}
The authors would like to thank the anonymous reviewers for their insightful comments and valuable suggestions. This work is supported by the Joint Funds of the National Natural Science Foundation of China (No. U2241216), the National Natural Science Foundation of China (No. 62202223), the Natural Science Foundation of Jiangsu Province (No. BK20220881), the Open Fund of the State Key Laboratory for Novel Software Technology (No. KFKT2024B27), and the Fundamental Research Funds for the Central Universities (No. NT2024020).
\end{acks}

\bibliographystyle{ACM-Reference-Format}
\bibliography{main}

\end{document}